%% file: M.tex
\documentstyle[epsf%
,preprint%
,eqsecnum,prd,aps]{revtex}

\def\0{\over } \def\2{{1\over2}} \def\4{{1\over4}}
\def\5{\hat } \def\6{\partial }

\def\({\left(} \def\){\right)} \def\<{\langle } \def\>{\rangle }

\newcommand{\nn}{\nonumber\\ }
\newcommand{\be}{\begin{eqnarray}}
\newcommand{\ee}{\end{eqnarray}}
\def\labe{\label}

\def\simge{\mathrel{%
   \rlap{\raise 0.511ex \hbox{$>$}}{\lower 0.511ex \hbox{$\sim$}}}}
\def\simle{\mathrel{
   \rlap{\raise 0.511ex \hbox{$<$}}{\lower 0.511ex \hbox{$\sim$}}}}
\def\bigs{\mathrel{
   \rlap{\raise 0.531ex \hbox{$>$}}{\lower 0.531ex \hbox{$<$}}}}

\def\grad{\nabla}                               
\def\del{\partial}                              

\def\frac#1#2{{#1 \over #2}}

\def\half{\ifinner {\scriptstyle {1 \over 2}}
   \else {1 \over 2} \fi}

\def\simge{\mathrel{%
   \rlap{\raise 0.511ex \hbox{$>$}}{\lower 0.511ex \hbox{$\sim$}}}}
\def\simle{\mathrel{
   \rlap{\raise 0.511ex \hbox{$<$}}{\lower 0.511ex \hbox{$\sim$}}}}
\def\bigs{\mathrel{
   \rlap{\raise 0.531ex \hbox{$>$}}{\lower 0.531ex \hbox{$<$}}}}

\def\slashchar#1{\setbox0=\hbox{$#1$}           
   \dimen0=\wd0                                 
   \setbox1=\hbox{/} \dimen1=\wd1               
   \ifdim\dimen0>\dimen1                        
      \rlap{\hbox to \dimen0{\hfil/\hfil}}      
      #1                                        
   \else                                        
      \rlap{\hbox to \dimen1{\hfil$#1$\hfil}}   
      /                                         
   \fi}                                         %

\def\subrightarrow#1{
  \setbox0=\hbox{
    $\displaystyle\mathop{}
    \limits_{#1}$}
  \dimen0=\wd0
  \advance \dimen0 by .5em
  \mathrel{
    \mathop{\hbox to \dimen0{\rightarrowfill}}
       \limits_{#1}}}                           

\begin{document}
\tighten
\preprint{Saclay-T01/085, BNL-NT-01/21, hep-ph/0109115}

\title{ Nonlinear Gluon Evolution in the Color Glass Condensate: II}
\author{Elena Ferreiro}
\address{Departamento de F\'{\i}sica de Part\'{\i}culas,
15706 Santiago de Compostela, Spain}
\author{Edmond Iancu}
\address{Service de Physique Th\'eorique, CE Saclay,
        F-91191 Gif-sur-Yvette, France}
\author{Andrei Leonidov}
\address{P. N. Lebedev Physical Institute, Moscow, Russia}
\author{Larry McLerran}
\address{ Physics Department, Brookhaven National Laboratory,
                 Upton, NY 11979, USA}
\date{\today }
\maketitle
\begin{abstract}
We complete the construction of the renormalization group equation 
(RGE) for the Color Glass Condenstate begun in Paper I. This is 
the equation which governs the evolution with rapidity
of the statistical weight function for the color glass field.
The coefficients in this equation --- one-loop real and virtual
contributions --- are computed explicitly, to all orders in the
color glass field. The resulting RGE can be interpreted as the
imaginary-time evolution equation, with rapidity as the 
``imaginary time'', for a quantum field theory in two spatial dimensions.
In the weak field limit it reduces to the BFKL equation.
In the general non-linear case, it is equivalent to an
equation by Weigert which summarizes in functional form
the evolution equations for Wilson line operators previously derived 
by Balitsky and Kovchegov.

\end{abstract}

\newpage

\input{S1.tex}
\input{S2.tex}
\input{S3.tex}

\input{S4.tex}

\input{S5.tex}

\input{S6.tex}

\appendix
\setcounter{equation}{0}
\input{A1.tex}

\input{A2.tex}

\input{A3.tex}
\input{A4.tex}

\end{document}

%% file: S1.tex
\section{Introduction}

Hadronic scattering at high energy, or small Bjorken's $x$, uncovers
a novel regime of QCD where the coupling is small ($\alpha_s\ll 1$) but 
the parton densities are so large that conventional perturbation theory 
breaks down, via strong non-linear effects 
\cite{GLR,MQ,BM87,AM0,MV94,AGL,AM2,AM1,Larry01,Levin}. 
In a previous paper \cite{PI}, to be referred to as ``Paper I'' 
throughout this text, we have outlined the construction of an 
effective theory which is well suited to study the non-linear phenomena
at small $x$ and has a transparent physical interpretation: It portrays 
the gluon component of the hadron wavefunction (the relevant
component at small $x$) as a {\it Color Glass Condensate} (CGC).
This is a multiparticle quantum state with 
high occupation numbers, but to the accuracy of interest it can be 
represented as a classical stochastic color field 
with a probability law determined by a functional Fokker-Planck equation.

The latter is a renormalization group equation (RGE) which shows how to
construct the effective theory by integrating out quantum fluctuations
in the background of a strong color field (the CGC).
The non-linear effects included in the RGE via this background field
describe interactions among the gluons 
produced in the quantum evolution towards small $x$. 
This leads to a non-linear 
generalization of the BFKL equation \cite{BFKL} whose 
general structure has been originally identified in Refs. \cite{JKLW97}, 
in an effort to give a field-theoretical justification to
the McLerran-Venugopalan model for the gluon distribution
of a large nucleus \cite{MV94,K96,JKMW97,KM98}. But previous attempts 
\cite{JKW99} to compute the coefficients in this equation 
(beyond the linear, or BFKL, approximation \cite{JKLW97})
have suffered from technical complications and, moreover, appear 
to be inconsistent 
with the results obtained from other perturbative approaches 
\cite{B,K,W}.

In this paper, we continue the 
analysis in Paper I and compute explicitly the coefficients 
in the RGE alluded to above. As recently reported in Ref. \cite{SAT}, 
the non-linear effects encoded in these coefficients lead to 
{\it gluon saturation}, that is, to a limitation on the maximum
gluon density in the hadron wavefunction as $x\to 0$. 
In contrast to the linear BFKL
equation, which predicts the exponential growth of the gluon distribution 
function with $\tau\equiv \ln(1/x)$, our corresponding result 
grows only linearly, and thus respects the unitarity bounds.
This is consistent with a previous result by Mueller \cite{AM2},
and also with some recent analyses \cite{LT99,B00,LL,Levin} 
of the Balitsky--Kovchegov equations \cite{B,K}, 
which are encoded too in our RGE \cite{RGE} 
(see also Sect. \ref{sec:BK} below).

This paper is quite technical and relies heavily on the results of 
Paper I. To facilitate its reading and prepare the calculations to follow,
it is convenient to summarize, in the remaining part of this Introduction, 
the general structure of the effective theory and its quantum evolution. 
In this introductory discussion, we shall follow the presentation 
in Paper I, to which the reader may refer for more details, but we shall also 
anticipate some of the results to be obtained later in this paper.
The remaining sections are organized as follows: 
In Section II we present the explicit calculation of the
``real correction', i.e., the charge-charge correlator $\chi$
which plays the role of the diffusion kernel in the functional RGE.
In Section III we do the same for the ``virtual correction'', i.e.,
the one-point function $\sigma$ which plays the role of a force term. 
In Section IV, we present the final
result for the RGE, and discuss its general structure and 
some of its remarkable properties.
We emphasize, in particular, the Hamiltonian structure of this
equation, and its relation with a similar equation by Weigert \cite{W}.
In Section V, we derive evolution
equations for observables from the general RGE. In the weak field
limit, we thus recover the BFKL equation for the unintegrated gluon
distribution. In the general non-linear case, we obtain the
coupled evolution equations for Wilson-line correlators originally
derived by Balitsky and (in the large $N_c$ limit) also by Kovchegov.
In Section VI we present conclusions and perspectives.
Some explicit calculations, as well as the presentation of
the background field gluon propagator, are deferred to the four 
Appendices.

\subsection{The effective theory for the CGC}
\label{sec:EFT}

The effective theory applies to gluon correlations in the hadron 
 wavefunction as measured in deep inelastic 
scattering at small Bjorken's $x$. It is formulated in the hadron
infinite momentum frame, where small $x$ corresponds to soft
longitudinal momenta\footnote{Throughout, we
use light-cone vector notations, that is,
$v^\mu=(v^+,v^-,{\bf v}_\perp)$, with
$v^+\equiv (1/\sqrt 2)(v^0+v^3)$,
$v^-\equiv (1/\sqrt 2)(v^0-v^3)$, and ${\bf v}_\perp
\equiv (v^1,v^2)$. The dot product reads $k\cdot x = k^- x^+ + k^+ x^-
- {\bf k}_\perp \cdot {\bf x}_\perp$. The hadron four-momentum
reads $P^\mu=(P^+,0,0_\perp)$, with $P^+\to \infty$.}
$k^+=xP^+$, with $P^+$ the hadron momentum, and $x\ll 1$.
The main physical assumption, which is a posteriori verified in
the construction of the effective theory, is that the ``fast''
partons in the hadron wavefunction (i.e., the excitations with
relatively large longitudinal momenta $p^+\gg k^+$) can be replaced,
as far as their effects on the soft correlation functions are 
concerned, by a classical random {\it color source} $\rho^a(x)$, 
whose gross properties are determined by the kinematics.

The separation of scales in longitudinal momenta ($p^+\gg k^+$) 
implies a corresponding separation in (light-cone) energies
($p^-\sim p_\perp^2/2p^+\,\ll\,k^- \sim k_\perp^2/2k^+$),
and therefore also in (light-cone) time scales: The lifetime
$\Delta x^+ \sim 1/k^- \propto k^+$ of the
soft gluons is much shorter than the characteristic time scale
$\sim 1/p^- \propto p^+$ for the dynamics of the fast
partons. Thus, the latter appear to the soft gluons
as nearly on-shell colored particles which propagate on 
the light-cone ($z\simeq t$, or $x^-\simeq 0$)
with large $p^+$ momenta. The associated color charge density 
$\rho^a(x)$ is therefore
{\it static} (i.e., independent of $x^+$), {\it localized} near the 
light-cone (within a small distance $\Delta x^- \sim 1/p^+ \ll 1/k^+$), 
and {\it random} (since this is the instantaneous color charge in the
hadron ``seen'' by the soft gluons at the arbitrary time of their
emission). The correlations of $\rho$ are encoded in a
gauge-invariant {\it weight function} $W_\tau[\rho]$. This is the 
probability density for having a color charge distribution with 
density $\rho_a(x^-,{\bf x}_{\perp})$, normalized as:
\be\label{norm}
\int {\cal D}\rho\, \,W_\tau[\rho]\,=\,1\,.\ee
Note that we use the momentum-space rapidity 
$\tau\equiv\ln(P^+/k^+) = \ln(1/x)$ to indicate the
dependence of the weight function upon the soft scale $k^+$.

Thus, in this effective theory, the
equal-time\footnote{Only equal-time correlators are needed, since these
are the only ones to be measured by a small-$x$ external probe, which
is absorbed almost instantaneously by the hadron. In eq.~(\ref{clascorr}),
it is understood that the fields $A^i_a(x^+,\vec x)$ involve only Fourier
modes with longitudinal momenta $k^+=xP^+$.}
gluon correlation functions at the scale $k^+ =
xP^+$ are obtained as (with ${\vec x}\equiv (x^-,{\bf x}_{\perp})$):
\be\label{clascorr}
\langle A^i_a(x^+,\vec x)A^j_b(x^+,\vec y)
\cdots\rangle_\tau\,=\,
\int {\cal D}\rho\,\,W_\tau[\rho]\,{\cal A}_a^i({\vec x})
{\cal A}_b^j({\vec y})\cdots\,,\ee
where ${\cal A}_a^i\equiv {\cal A}_a^i[\rho]$ is the 
solution to the classical Yang-Mills equations with source
$\rho_a$ :
\be
(D_{\nu} F^{\nu \mu})_a(x)\, =\, \delta^{\mu +} \rho_a(x)\,,
\label{cleq0}
\ee
in the light-cone (LC) 
gauge $A^+_a=0$, which is the gauge which allows for the most
direct contact with the gauge-invariant physical quantities 
\cite{AM1,PI}. For instance, 
the gluon distribution function
($\equiv$ the total number of gluons per unit rapidity
with longitudinal momentum $k^+=xP^+$ and transverse momentum
$k_\perp^2 \le Q^2$) is obtained as \cite{AM1,PI}
\be\label{GCL}
x G(x,Q^2)&=&\frac{1}{\pi}
\int {d^2k_\perp \over (2 \pi)^2}\,\Theta(Q^2-
k_\perp^2)\,\Bigl\langle\,
|{\cal F}^{+i}_a(\vec k)|^2\Bigr\rangle_\tau\,.\ee
where ${\cal F}^{+i}_a=\partial^+ {\cal A}^i_a$ is the 
electric field associated to the classical solution
${\cal A}_a^i[\rho]$, and $\vec k \equiv (k^+,{\bf k}_\perp)$
with $k^+=xP^+=P^+{\rm e}^{-\tau}$.

To deduce explicit expressions for these classical fields,
it is preferable to express the LC-gauge solution
${\cal A}_a^i$ in terms of color source $\tilde\rho_a$
in the {\it covariant} gauge $\partial^\mu \tilde A_\mu =0$
(COV-gauge). This is possible since both the measure and the
weight function in the functional integral (\ref{clascorr})
are gauge-invariant, so that the classical average can be done 
equally well by integrating over the COV-gauge $\tilde\rho$.
In terms of this latter, the classical solution 
${\cal A}_a^\mu[\tilde\rho]$
is known explicitly \cite{PI}, and reads: 
${\cal A}_a^+=0$ (the gauge condition), ${\cal A}^-_a=0$, and 
(in matrix notations appropriate for the adjoint representation:
$\tilde\rho\equiv \tilde\rho_a T^a$, etc.)
\begin{mathletters}\be\label{Aclass}\,
{\cal A}^i\,
(\vec x) &=&{i \over g}\, U(\vec x) \,\partial^i  U^\dagger(\vec x),\\
U^{\dagger}(x^-,x_{\perp})&=&
 {\rm P} \exp
 \left \{
ig \int_{-\infty}^{x^-} dz^-\,{\alpha}(z^-,x_{\perp})
 \right \},\label{Udef}\\
- \nabla^2_\perp \alpha({\vec x})&=&\tilde\rho(\vec x).
\label{alpharho}\ee\label{Atilde}\end{mathletters}
This is the gauge-transform, with gauge function $U(\vec x)$,
of the corresponding solution in the COV-gauge, which has only
one non-trivial component: $\tilde{\cal A}^\mu_a 
=\delta^{\mu +}\alpha_a({\vec x})$.
All the fields above are static, i.e., independent of $x^+$.
Since $\tilde\rho$, and therefore $\alpha$, are localized at
$\Delta x^- < 1/k^+$, the associated field ${\cal A}^i$ appears 
effectively as a $\theta$--function :
\be\labe{APM}
{\cal A}^i(x^-,x_\perp)\,\approx\,\theta(x^-)\,
\frac{i}{g}\,V(\del^i V^\dagger)
\,\equiv\,\theta(x^-){\cal A}^i_\infty(x_\perp),\ee
to any probe with a much lower longitudinal resolution
(i.e., with momenta $q^+\ll k^+$). On the same resolution scale:
\be\labe{UTAF}
U^{\dagger}(x^-,x_{\perp})\,\approx \,
\theta(x^-)\,V^\dagger(x_{\perp}) + \theta(-x^-),\qquad\,
{\cal F}^{+i}(\vec x) \,\approx\,\delta(x^-)\,
{\cal A}^{i}_\infty(x_\perp).\ee
In the equations above, $V$ and $V^\dagger$ are
the asymptotic values of the respective
gauge rotations as $x^-\to\infty$ :
\be\labe{v}
V^\dagger(x_{\perp})\,\equiv\,{\rm P} \exp
 \left \{
ig \int_{-\infty}^{\infty} dz^-\,\alpha (z^-,x_{\perp})
 \right \}.\ee
In practice, $U(x^-,x_{\perp})=V(x_{\perp})$ for any
$x^-\gg 1/k^+$.

Note that the solution to the classical non-linear equation
is known exactly, which makes this approach particularly convenient
to study the non-linear physics at small $x$. The classical 
approximation should be indeed appropriate in this regime, which
is characterized by weak coupling and large occupation numbers.
On the other hand, the weight function $W_\tau[\rho]$ 
is obtained via a quantum calculation in which the quantum 
fluctuations with momenta $k^+\simle p^+ \simle P^+$ are
integrated out in layers of $p^+$, and in the background of
the classical fields ${\cal A}^i$ \cite{JKMW97,JKLW97,PI}. 
This calculation captures the basic 
mechanism leading to large gluon densities --- namely, the BFKL-type 
of evolution towards small $x$ ---, while also including non-linear 
effects (i.e., rescatterings among the produced gluons) 
via the background fields. In the saturation regime, where one
expects color fields as strong as ${\cal A}^i\sim 1/g$
(corresponding, via eq.~(\ref{GCL}),
to gluon densities of order $1/\alpha_s$
\cite{AM1,JKMW97,AM2,SAT}), the background field calculation
must be carried out {\it exactly}, i.e., to all orders in ${\cal A}^i$. 

\subsection{The quantum evolution of the effective theory}
\label{QEVOL}

To describe the quantum evolution, it is convenient to introduce
an arbitrary separation scale $\Lambda^+$ such as 
$k^+ \simle \Lambda^+ \ll P^+$, and assume that the
``fast'' quantum modes with momenta $p^+ \gg\Lambda^+$ have been 
already integrated out and replaced, to the accuracy of interest, by
a classical color source $\rho_a(\vec x)$ with weight function
$W_\Lambda[\rho]$. On the other hand, the ``soft'' gluons with momenta
$p^+ <\Lambda^+$ are still explicitly present in the theory.
The correlation functions at the soft scale
$k^+$ are then obtained from the following generating functional :
\be\labe{PART}
{Z}[J]\,=\,\int {\cal D}\rho\,\,W_\Lambda[\rho]
\,\,\left\{\frac{\int^\Lambda {\cal D}A_a^\mu\,
\delta(A^+_a)\,\,{\rm e}^{\,iS[A,\,\rho]-i\int J\cdot A}}
{\int^\Lambda {\cal D}A_a^\mu\,\delta(A^+_a)\,\,{\rm e}^{\,iS[A,\,\rho]}}
\right\}.\ee
where the ``external current'' $J^\mu_a$ is just a device to generate
Green's functions via functional differentiations,
and should not be confused with the physical source $\rho$.
Eq.~(\ref{PART}) is written fully in the LC gauge (in particular, $\rho$
is the LC-gauge color source), and involves
two functional integrals: ({\it a}) a quantum path integral
over the soft gluon fields $A^\mu$, which generates ($\rho$-dependent)
quantum expectation values at fixed $\rho$, e.g.,
\be\labe{2point} 
\langle {\rm T}\,A^\mu(x)A^\nu(y)\rangle [\rho]
\,=\,\frac{\int^\Lambda {\cal D}A\,\delta(A^+)
\,\,A^\mu(x)A^\nu(y)\,\,{\rm e}^{\,iS[A,\,\rho]}}
{\int^\Lambda {\cal D}A\,\delta(A^+)\,\,{\rm e}^{\,iS[A,\,\rho]}}
,\ee
and ({\it b}) a classical average over $\rho$.
As discussed in Paper I, eq.~(\ref{PART}) has the typical structure
to describe correlations in a glass.

The action $S[A,\,\rho]$ in eqs.~(\ref{PART}) and (\ref{2point})
reads as follows \cite{JKLW97} :
\be\label{ACTION}
S[A,\rho]\,=\,- \int d^4x \,{1 \over 4} \,F_{\mu\nu}^a F^{\mu\nu}_a
\,+\,{i \over {gN_c}} \int d^3 \vec x\, {\rm Tr}\,\Bigl\{ \rho(\vec x)
\,W[A^-](\vec x)\Bigr\}\,\equiv\,S_{YM}\,+\,S_W,\,\,
\ee
where $W[A^-]$ is a Wilson line in the temporal direction:
\be\label{WLINE}
W[A^-](\vec x)\, =\,{\rm P}\, \exp\left[\,ig\int dx^+ A^-(x) \right].
\ee
In the classical, or saddle point, approximation $\delta S/\delta A^\mu
=0$, the action (\ref{ACTION}) generates the desired equations
of motion, that is, eqs.~(\ref{cleq0}) with $A^-=0$. This shows that
the classical field in eqs.~(\ref{Atilde}) is the tree-level background
field in which propagate the quantum fluctuations. 
The general non-linear structure of $S_W$ in eq.~(\ref{WLINE}),
which plays a role only for the quantum corrections (in that it
generates new interaction vertices; cf. Sect. \ref{FRULES} 
below), reflects our eikonal approximation for the interaction 
between fast particles moving in the plus direction (here represented
by $\rho$) and the comparatively slow gluon fields.

In the above formulae, the intermediate scale $\Lambda^+$
enters at two levels: as an upper cutoff on the
longitudinal momenta $p^+$ of the quantum gluons, and in
the weight function $W_\Lambda[\rho]$ for the classical source.
Of course, the
final results for correlation functions at the scale $k^+$
must be independent of this arbitrary scale :
\be
\Lambda^+ \frac{\del {Z}[J]}{\del \Lambda^+}\,=\,0.\ee
This constraint can be formulated as a
renormalization group equation (RGE) for $W_\Lambda[\rho]$ which 
governs its evolution with decreasing $\Lambda^+$.

The initial conditions for this evolution are determined by the
properties of the hadronic matter at $\Lambda^+ \sim P^+$.
These are not really under control in perturbation theory,
but one can try to rely on some non-perturbative model, 
like the valence quark model.  The initial conditions might be under better 
control in the high density environment of a very large nucleus,
since we expect the coupling constant to be weaker at high density.
The indeterminancy of evolution equations due to initial conditions is hardly a new problem in QCD, since the DGLAP and BFKL equations are both
limited by such uncertainty.  In these cases, and we also hope
 here\footnote{The approximate solutions to the RGE recently
found in Ref. \cite{SAT} appear to confirm this expectation.}, 
the solution of the evolution equation for arbitrarily high energies 
is universal and its generic properties are largerly independent of the initial conditions.

Starting with these initial conditions, one then proceeds with the
quantum evolution down to the soft scale of interest
 $\Lambda^+ \sim xP^+\ll P^+$.
In this process, the original source at the scale 
$P^+$ gets dressed by the quantum fluctuations with momenta 
$\Lambda^+ < |p^+| < P^+$. We treat this process in perturbation theory, in the ``leading logarithmic approximation'' (LLA)\footnote{Indeed, it is only 
to this accuracy that the assumed separation of scales in the problem is
maintained by quantum corrections.}
--- i.e., by retaining only the terms 
enhanced by the large logarithm $\ln(1/x)$, to all orders in
$(\alpha_s\ln(1/x))^n$) ---,
which, by itself, is also the accuracy of the BFKL equation, 
but we go beyond BFKL in that we resum also finite density effects, 
which are expected to become increasingly
important as we go to smaller and smaller $\Lambda^+$ (or Bjorken's $x$). 
We do that by performing a background field calculation, that is,
by integrating out the quantum fluctuations at one step 
in the background of the classical field generated by the color
source at the previous step, with the background field simulating
(via its correlations) the finite density effects in the system.

To be more explicit, 
let us describe one step in this renormalization group procedure
in some detail. Assume that we know the effective theory at
the scale $\Lambda^+$ --- as specified by the
corresponding weight function $W_\Lambda[\rho]\equiv W_\tau[\rho]$,
with $\tau=\ln(P^+/\Lambda^+)$ ---, and we are interested
in correlations at the softer scale
$k^+ \sim b\Lambda^+$ with $b\ll 1$, but such as $\alpha_s\ln(1/b)< 1$, for
perturbation theory in powers of $\alpha_s\ln(1/b)$ to make sense.
Our purpose is to construct the new weight function 
$W_{b\Lambda}[\rho]\equiv  W_{\tau+\Delta\tau}[\rho]$,
with $\Delta\tau\equiv \ln(1/b)$, which would determine the 
gluon correlations at this softer scale. As compared to 
$W_\tau[\rho]$, this new weight function must include also the 
quantum effects induced by the ``semi-fast'' gluons 
with longitudinal momenta in the strip
\be\labe{strip}\,\,
 b\Lambda^+ \,\,\ll\,\, |p^+|\,\, \ll\,\,\Lambda^+\,.\ee
To compute these effects, it is convenient to decompose the gluon field 
in eq.~(\ref{PART}) as follows 
\be
A^\mu_c\,=\,{\cal A}^\mu_c[\rho]+a^\mu_c+\delta A^\mu_c,\ee
where ${\cal A}^\mu_c$ is the tree-level field,
$a^\mu_c$ represents the semi-fast fluctuations, and
$\delta A^\mu_c$ refers to the remaining modes with
$|p^+| \le b\Lambda^+$.
By integrating out the fields $a^\mu$, some
new correlations are induced at the soft scale $b\Lambda^+$ 
via the coupling $\delta A^-_c\delta\rho_c$,
where $\delta\rho_c$ 
is the color charge of the semi-fast gluons. 
These correlations have to be computed to lowest order
in $\alpha_s\ln(1/b)$, but to all orders in the 
background fields ${\cal A}^i$. 
This is essentially an one-loop calculation, but with the exact
background field propagator of the semi-fast gluons.
The new correlations are eventually absorbed 
into the functional change $\Delta W \equiv W_{\tau+\Delta\tau} - W_\tau$
in the weight function. Since $\Delta W\propto \Delta \tau$,
this evolution is most conveniently formulated as a (functional) 
renormalization group equation for $W_\tau[\rho]$, which reads
\cite{JKLW97,PI}
\be\label{RGE}
{\del W_\tau[\rho] \over {\del \tau}}\,=\,\alpha_s
\left\{ {1 \over 2} {\delta^2 \over {\delta
\rho_\tau^a(x_\perp) \delta \rho_\tau^b(y_\perp)}} [W_\tau\chi_{xy}^{ab}] - 
{\delta \over {\delta \rho_\tau^a(x_\perp)}}
[W_\tau\sigma_{x}^a] \right\}\,,
\ee 
in compact notation where repeated color indices 
(and coordinates) are understood to be summed (integrated) over. The
 coefficients $\sigma_{x}^a\equiv \sigma_a(x_\perp)$ and
$\chi_{xy}^{ab}\equiv\chi_{ab}(x_\perp,y_\perp)$ in the above
equation are related to the 1-point and 2-point functions
of the color charge $\delta\rho_a(x)$
of the semi-fast gluons via the following relations:
\be\label{sigperp}
\alpha_s\ln{1\over b}\,\sigma_a ({x}_\perp)&\equiv &
\int dx^- \,\langle\delta \rho_a(x)\rangle\,,\nn
\alpha_s\ln{1\over b}\,\chi_{ab}(x_\perp, y_\perp)&\equiv &
\int dx^- \int dy^-\,
\langle\delta \rho_a(x^+,\vec x)\,
\delta \rho_b(x^+,\vec y)\rangle\,,\ee
where the brackets denote the average over quantum fluctuations
in the background of the tree-level color fields ${\cal A}^i$,
as shown in eq.~(\ref{2point}).

In writing eq.~(\ref{RGE}), we have also anticipated the longitudinal
structure of the quantum evolution, which will become manifest only
after performing the quantum calculations in the next sections. Specifically,
we shall see that the {\it induced source} $\langle\delta \rho_a(x)\rangle$
($\equiv$ the correction to $\rho$ generated by the gluons 
with $p^+$ in the strip (\ref{strip})) has support at
\be\label{stripx-}
1/\Lambda^+\,\,\simle\,\,x^-\,\,\simle\,\,1/(b\Lambda^+)\,.\ee
By induction, we shall deduce that $\rho_a(\vec x)$ ($\equiv$ the color
source generated by the quantum evolution down to 
$\Lambda^+=P^+{\rm e}^{-\tau}$) has support 
at\footnote{The fact that the source has support at {\it
positive} $x^-$, rather than at generic  $x^-$ with $|x^-|\simle x^-_\tau$
will be seen to be related to our specific gauge-fixing prescription;
cf. Sect. \ref{sec:sigmaA}.} 
$0\le x^-\simle x^-_\tau$ with 
\be\label{xtau}
 x^-_\tau\,\equiv\, 1/\Lambda^+\,=\,x^-_0{\rm e}^{\tau},\qquad
x^-_0\,\equiv\, 1/P^+\,.\ee
Thus, the color source is built in layers of $x^-$, with a one-to-one
correspondence between the $x^-$--coordinate of a given layer and the 
$p^+$--momenta of the modes that have been integrated to generate that layer. 
When some new quantum modes, with rapidities
$\tau <\tau' <\tau+\Delta\tau$, are integrated out, 
the preexisting color source at
$0 < x^-< x^-_\tau$ 
remains unchanged, but some new source is added in
the interval (\ref{stripx-}). 
Because of that, 
$\Delta W \equiv W_{\tau+\Delta\tau} - W_\tau$ involves only the
change in $\rho_a$ within that last interval. In the
continuum limit, this generates the functional derivatives of
$W_\tau$ with respect to $\rho_a(\vec x)$ at $x^-=x^-_\tau$
{\it only}, that is, the derivatives with respect to
\be\label{rhotau}
\rho_{\tau}^a(x_\perp)
\,\equiv\, \rho^a(x^-=x^-_\tau,x_\perp),\ee
as shown in eq.~(\ref{RGE}).

On the other hand, given the separation of scales in the problem,
the detailed longitudinal structure of the quantum effects should 
not be too important, and this too has been anticipated in writing
eqs.~(\ref{RGE})--(\ref{sigperp}): In the same way as the original
source with support at $x^-\simle 1/\Lambda^+$ appears effectively
as a $\delta$-function to the semi-fast gluons (with 
wavelenghts $\Delta x^-\sim 1/p^+ \gg 1/\Lambda^+$), 
the induced source $\langle\delta \rho\rangle$,
although relatively delocalized as compared
to $\rho$ (cf. eq.~(\ref{stripx-})), appears to the
soft gluons with $k^+\simle b\Lambda^+$ as a rather
sharp color distribution around $x^-\simeq x^-_\tau$. 
Thus, the soft gluons can probe only
the {\it two-dimensional} color charge distribution 
in the transverse plane, as obtained after integrating out $x^-$, 
cf. eq.~(\ref{sigperp}). 
This explains the 2-dimensional structure of the coefficients
$\sigma$ and $\chi$ in the RGE (\ref{RGE}).

Eq.~(\ref{RGE}) has the structure of the diffusion equation: It
is a second-order (functional) differential equation whose r.h.s.
is a total derivative (as necessary to conserve the total
probability; cf. Sect. \ref{sect:PROP}). Thus, this equation describes
the quantum evolution as a diffusion (with diffusion ``time'' $\tau$)
of the probability density $W_\tau[\rho]$
in the functional space spanned by $\rho_a(x^-,x_\perp)$.
For this equation to be useful, its coefficients $\sigma$
and $\chi$ must be known explicitly as functionals of $\rho$.
It is therefore more convenient to use the COV-gauge source 
$\tilde\rho_a$, or the associated Coulomb field $\alpha_a$,
as the functional variable to be averaged over. Indeed,
$\sigma$ and $\chi$ depend upon the color source
via the classical field ${\cal A}^i$, which is most directly
related to $\alpha_a$, cf. eqs.~(\ref{Atilde}).

As explained in Ref. \cite{PI}, when going from the (background)
LC-gauge to the COV-gauge, the induced color charge $\sigma$ acquires 
a new contribution, in addition to the color rotation with matrix
$U^\dagger$. This is so because the transformation between
the two gauges depends upon the charge content in the problem: this
was $\rho$ at the initial scale $\Lambda^+$, thus giving a gauge
transformation $U^\dagger[\rho]$, but it becomes $\rho+\delta\rho$,
with fluctuating $\delta\rho$, at the new scale $b\Lambda^+$, 
thus inducing a fluctuating component in the corresponding 
gauge function $U^\dagger[\rho+\delta\rho]$.
After averaging out the quantum fluctuations, one is left with a RGE for
$W_\tau[\tilde\rho]$ which is formally similar to eq.~(\ref{RGE}), but
with $\rho \to \tilde\rho$,
$\chi\to \tilde\chi$, and $\sigma\to \tilde\sigma$, where:
\begin{mathletters}
\label{tildesc}
\be\label{tildechi}
\tilde \chi_{ab}(x_\perp,y_\perp)&\equiv&
V^{\dagger}_{ac}(x_\perp)\,
\chi_{cd}(x_\perp,y_\perp)\,V_{d b}(y_\perp),\\
\tilde\sigma_a (x_\perp)&\equiv&
 V^{\dagger}_{ab}(x_\perp)\,\sigma_b(x_\perp)- \delta\sigma_a(x_\perp),
\label{tildesig} \\ 
\delta\sigma_a(x_\perp)&\equiv&
{g\over 2}\,f^{abc}\int d^2y_\perp\,\,
\tilde\chi_{cb}(x_\perp,y_\perp)\,
\langle y_\perp|\,\frac{1}{-\grad^2_\perp}\,|
x_\perp\rangle\,,\label{sigclas}
\ee\end{mathletters}
with $V$ and $V^\dagger$ as defined in eq.~(\ref{v}).
The correction $-\delta\sigma$, to be referred to as the
``classical polarization'', is the result of the quantum evolution
of the gauge transformation itself.

It turns out that the RGE is most conveniently written as
an equation for $W_\tau[\alpha]\equiv
W_\tau[\tilde\rho= - \nabla^2_\perp \alpha]$, in which case it reads:
\be\labe{RGEA0}
{\del W_\tau[\alpha] \over {\del \tau}}\,=\,\alpha_s
\left\{ {1 \over 2} {\delta^2 \over {\delta
\alpha_\tau^a(x_\perp) \delta \alpha_\tau^b(y_\perp)}} 
[W_\tau\eta_{xy}^{ab}] - 
{\delta \over {\delta \alpha_\tau^a(x_\perp)}}
[W_\tau\nu_{x}^a] \right\}\,,
\ee
where $\alpha^a_\tau(x_\perp)\equiv
\alpha^a(x^- = x^-_\tau,x_\perp)$ (cf. eq.~(\ref{rhotau})), and
\begin{mathletters}
\be\label{nudef}
\nu^a (x_\perp)&\equiv&\int d^2z_\perp\,
\langle x_\perp|\,\frac{1}{-\grad^2_\perp}\,|z_\perp\rangle\,
\tilde\sigma^a (z_\perp),\\
\eta^{ab}(x_\perp,y_\perp)&\equiv&\int d^2z_\perp \int d^2u_\perp\,
\langle x_\perp|\,\frac{1}{-\grad^2_\perp}\,|z_\perp\rangle\,
\tilde\chi^{ab}(z_\perp,u_\perp)\,
\langle u_\perp|\,\frac{1}{-\grad^2_\perp}\,|y_\perp\rangle\,.
\label{etadef}\ee
\label{nueta}\end{mathletters}
This form has, in particular, the advantage to eliminate any explicit
reference to the arbitrary infrared cutoff $\mu$ which enters
the solution to eq.~(\ref{alpharho}) for $\alpha$ :
\be\labe{alpha}
\alpha ({\vec x})
\,=\,-\int \frac{d^2y_\perp}{2\pi}\,
\ln\Bigl(|{\bf x}_\perp - {\bf y}_\perp|\mu\Bigr)\,
\tilde\rho (x^-,{\bf y}_\perp).\ee
It is our ultimate goal in this paper to compute explicitly the
quantities $\eta$ and $\nu$ as functionals of the field
$\alpha$, and thus completely specify the RGE (\ref{RGEA0}).

\subsection{Feynman rules for $\chi$ and $\sigma$}
\label{FRULES}

The basic quantum calculation is that of the quantities
$\chi$ and $\sigma$ in the LC gauge.
The necessary Feynman rules can be
readily derived from the action (\ref{ACTION}) \cite{JKLW97,PI}, and are
summarized here for later reference. Eqs.~(\ref{sigperp}) involve the color
charge $\delta\rho_a$ of the semi-fast fields $a^\mu_c$. To
the order of interest, this reads
\begin{equation}\label{delta12}
\delta \rho_a (x) =\delta \rho_a^{(1)} (x)+\delta \rho_a^{(2)} (x),
\end{equation}
where $\delta \rho^{(1)}$ is linear in the fluctuations 
$a^\mu$, while $\delta \rho^{(2)}$ is quadratic\footnote{To compare
with the corresponding formulae in Paper I (see, e.g., eqs. (4.41)--(4.42)
in Ref. \cite{PI}), please note a change in our
normalization conventions: Here, the various fields preserve their 
natural normalization fixed by the action (\ref{ACTION}). 
By contrast, in Paper I we have rescaled the classical fields 
and sources by a factor $1/g$.}: 
\be\label{rho10}
 \delta \rho_a^{(1)} (x) & = & 
-2i g{\cal F}^{+i}_{ac} (\vec x) a^{ic} (x) + \nonumber \\
& & +g\rho^{ac} (\vec x)
\int dy^+ \langle x^+ |{\rm PV}\,{1 \over i\partial^-} |y^+ \rangle
 a^{c-}(y^+,\vec x), \\
\delta \rho_a^{(2)}(x)& = & 
 g f^{abc} [\partial^+ a^{b}_{i}(x)
  ]a^{c}_{i}(x) 
 \nonumber\\ &{-}& (g^2/N_c) \,\rho^{b}({\vec x})
  \int dy^+ a^{-c}(y^+,{\vec x}) \int dz^+ a^{-d}(z^+,{\vec x})
\nonumber\\ &{}& \nonumber\,\,\,\, \times\,
  \biggl\{\theta (z^+ -y^+)
  \theta (y^+ -x^+) {\rm Tr} \,(T^a T^c T^d T^b)
\\ &{}&\nonumber 
  \qquad+\ \theta (x^+ -z^+) \theta (z^+ -y^+) {\rm Tr} \,(T^a T^b T^c T^d)
\\ &{}&\qquad
+\ \theta (z^+ -x^+) \theta (x^+ -y^+) {\rm Tr}\,(T^a T^d T^b T^c) \biggr\}.
\label{rho2}
\ee
In the right hand sides of these equations, the terms involving
$a^i_c$ come from the three-gluon vertex in
$S_{YM}$, while the terms involving $a^-_c$
come from the two- and three-point vertices in $S_W$.
It is understood here that only the soft modes with
$k^+\simle b\Lambda^+$ are to be kept in the products of fields.

Consider first the following charge-charge 
correlator\footnote{Our notations are such that
$\hat\chi$ and $\chi$ or $\hat\sigma$ and $\sigma$ differ
just by a factor of $\alpha_s\ln(1/b)$.} which enters
eq.~(\ref{sigperp}) for $\chi$ :
\be\label{chi00}
{\hat \chi}_{ab}({\vec x},{\vec y}) \, \equiv \,
\langle\delta \rho_a(x^+,\vec x)\,
\delta \rho_b(x^+,\vec y)\rangle\,.\ee
By time homogeneity (recall that the background fields
${\cal A}^i$ are static), this equal-time 2-point function 
is actually independent of time. When evaluating $\hat\chi$ 
within our present accuracy,
it is sufficient to keep only the terms of first order in fluctuations:
$\delta \rho \sim \delta \rho^{(1)}$. These are given by 
eq.~(\ref{rho10}) which implies (in condensed notations,
where the PV prescription in $1/p^-$ and the condition 
$x^+=y^+$ are implicit):
\be\label{chi0}
\hat\chi_{ab}(\vec x,\vec y)\,=\,g^2
\left\langle \left(-2i{\cal F}^{+i}
a^i + \,\rho {1 \over i\partial^-} a^- \right)_x^a
\left(
2i a^i {\cal F}^{+i} +a^- {1 \over 
i\partial^-} \rho \right)_y^b \right\rangle.
\ee
This involves the propagator
\be\label{delAcorr}
iG^{\mu\nu}_{ab}(x,y)[{\cal A}]&\equiv&
\langle {\rm T}\,a_a^\mu(x) a_b^\nu(y)\rangle\ee
of the semi-fast gluons in the background of the tree-level fields 
${\cal A}^i$, and in the LC-gauge ($a^+=0$).
Given the specific structure of the background field (\ref{Atilde}),
it has been possible to construct this
propagator {\it exactly}, i.e., to all orders in ${\cal A}^i$
\cite{AJMV95,HW98,PI}. The resulting expression, to be
extensively used in what follows, is presented in Appendix A.

The construction of the propagator in Refs.
\cite{AJMV95,HW98,PI} has relied in an 
essential way on the separation of scales in the problem: 
Because of their low $p^+$ momenta ($p^+\ll \Lambda^+$),
the semi-fast gluons $a^\mu$ are unable to discriminate the
longitudinal structure of the source, but rather ``see'' this 
as a $\delta$-function at $x^-=0$. One is thus reduced to the
problem of the scattering off a $\delta$-type potential, 
whose solution is known exactly. But this also means that,
strictly speaking, the propagator (\ref{delAcorr}) is known 
only far away from the support of the source
(for $|x^-| \gg 1/\Lambda^+$), where ${\cal A}^i$ 
takes the approximate form in eq.~(\ref{APM}).
It is thus an important self-consistency check on the 
calculations to follow to verify that $\chi$ and $\sigma$
are indeed insensitive to the internal structure of the source.

Still for the construction of the propagator, it has been
more convenient to impose the strip restriction (\ref{strip})
on the LC-energy $p^-$, rather than on the longitudinal
momentum $p^+$ \cite{PI}. Indeed, to LLA, the quantum effects are 
due to nearly on-shell gluons, for which the constraint (\ref{strip})
is equivalent to the following constraint on $p^-$ :
\be\labe{strip-}\,
\Lambda^- \,\ll\, |p^-| \,\ll\, \Lambda^-/b\,,\ee
where $\Lambda^-\equiv Q_\perp^2/2\Lambda^+$, and 
$Q_\perp$ is some typical transverse momentum.
(Recall that we assume all the transverse momenta
to be comparable; thus, to LLA, it makes no difference 
what is the precise value of $Q_\perp$.)
But the condition (\ref{strip-}) is better adapted to the
present background field problem, 
where we have homogeneity in time (so that we can work in
the $p^-$--representation), but strong
inhomogeneity in $x^-$ (cf. eqs.~(\ref{APM}) and (\ref{UTAF})).

A final subtlety refers to
the choice of a gauge condition in the LC-gauge propagator:
Recall that, even after imposing the condition $A^+=0$, 
one has still the freedom to perform $x^-$--independent 
gauge transformations. This ambiguity shows up as an unphysical 
pole at $p^+=0$ in the gluon propagator. Since in our calculations
with strip restriction on $p^-$, $p^+$ is not restricted anylonger,
an $i\epsilon$ prescription is needed to regulate this pole. 
This is important since, as our calculations show,
the final results for $\chi$ and $\sigma$ are generally dependent
upon this $i\epsilon$ prescription.
In fact, the discrepancy between our results below and those
reported in Refs. \cite{JKW99} may be partially
attributed to using different gauge-fixing prescriptions.

For consistency with the retarded boundary conditions 
(${\cal A}^i\to 0$ for $x^- \to -\infty$)
imposed on the classical solution (\ref{Atilde}),
we shall use a retarded $i\epsilon$ prescription also
in the LC-gauge propagator of the quantum fluctuations (see Appendix A).
With this choice, the induced source $\langle\delta \rho\rangle$
appears to have support only at {\it positive} $x^-$, as shown
eq.~(\ref{stripx-}). We have also verified that, by using advanced
prescriptions (both in the classical and the quantum calculations),
the support of $\langle\delta \rho\rangle$ would be rather at {\it negative}
$x^-$ (but with $|x^-|$ still constrained as in eq.~(\ref{stripx-})).
In both cases, however, the same results are finally
obtained for $\chi$ and $\sigma$ after integrating out the 
(prescription-dependent) longitudinal structure.
Thus, the RGE comes up the same with both retarded and advanced
prescriptions. On the other hand, 
we have not been able to give a sense to calculations with 
other prescriptions, like principal value or Leibbrandt-Mandelstam. 
(For instance, when computed with a principal value prescription,
the coefficients $\sigma$ and $\chi$ appear to be sensitive to the
internal structure of the source in $x^-$, which contradicts the
assumed separation of scales.)

\begin{figure}
\protect\epsfxsize=12.cm{\centerline{\epsfbox{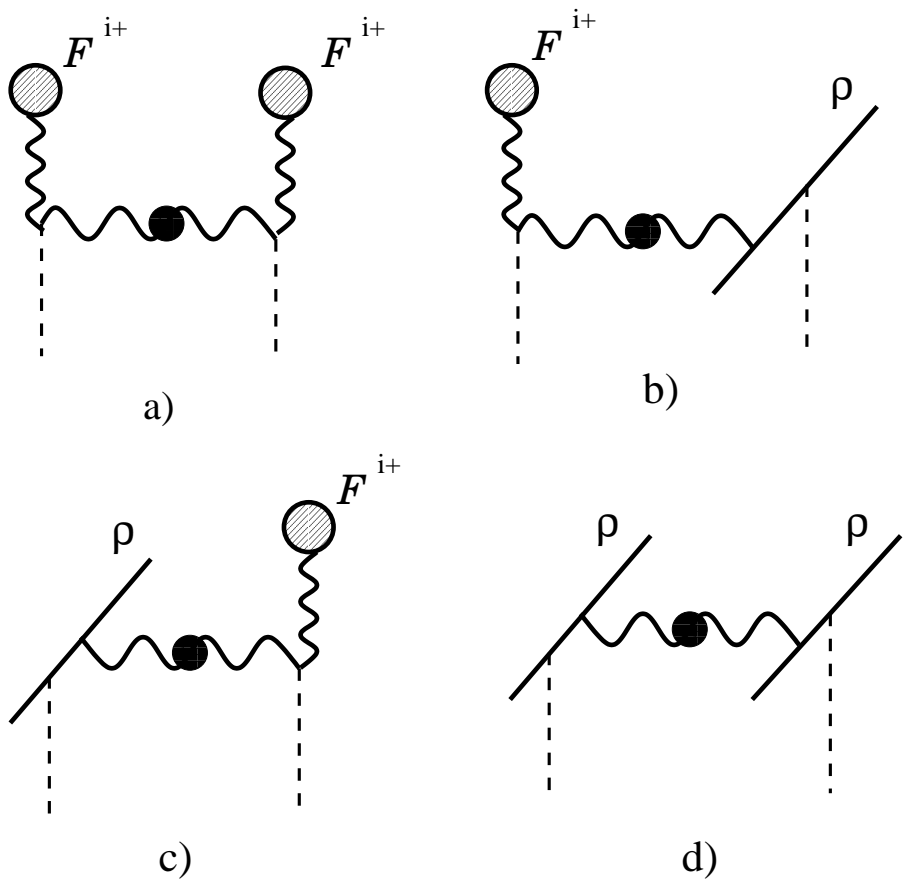}}}
         \caption{Feynman diagrams for the four contributions
to $\chi$ given in eq.~(\ref{chi2}).}
\label{CHIFIG}
\end{figure}

We conclude this introductory section by giving
explicit expressions for $\chi$ and $\sigma$ 
in terms of the background field
propagator $G^{\mu\nu}_{ab}(x,y)$ of the semi-fast gluons.
For $\chi$, we use eqs.~(\ref{sigperp}) and (\ref{chi0}) to deduce that
\be\label{chi1}
\alpha_s\ln{1\over b}\,
\chi_{ab}({\bf x}_\perp,{\bf y}_\perp)\,=\,
\int dx^- \int dy^-\,\hat\chi_{ab}(\vec x,\vec y),\ee
with (in matrix notations) 
\be\label{chi2}
\frac{1}{g^2}\,\hat \chi(\vec x, \vec y) &=&i\,
2{\cal F}^{+i}_x\, \langle x|G^{ij}|y\rangle \,2{\cal F}^{+j}_y \, +\, 
2{\cal F}^{+i}_x\,\langle x|G^{i-}\,{1 \over i\partial^-}|y\rangle \, \rho_y 
\nonumber\\&{}&\,\,\,\,-\,
\rho_x \, \langle x|{1 \over i\partial^-}\, G^{-i}|y\rangle \, 
2{\cal F}^{+i}_y\,+\,i
\rho_x \langle x|{1 \over i\partial^-} G^{--} {1 \over i\partial^-}
|y\rangle\,\rho_y.\ee
The equal-time limit is implicit here; it is achieved
by taking $y^+=x^++\epsilon$ within 
the time-ordered propagator (\ref{delAcorr})
\cite{PI}. Diagramatically, all the above contributions to $\chi$ 
are represented by {\it tree-like} Feynman graphs (no loops),
with vertices proportional to $\rho$ 
or ${\cal F}^{+i}$ (see Fig. \ref{CHIFIG}).
Since, physically, these are quantum corrections associated 
with the emission of a real (semi-fast) gluon, we shall sometimes
refer to $\chi$ or $\hat\chi$ as the {\it real correction}.
(In the weak field approximation, $\chi$ is responsible
for the real piece of the BFKL kernel \cite{JKLW97,PI};
see also Sect. \ref{sec:BFKL} below.)

\begin{figure}
\protect\epsfxsize=14.cm{\centerline{\epsfbox{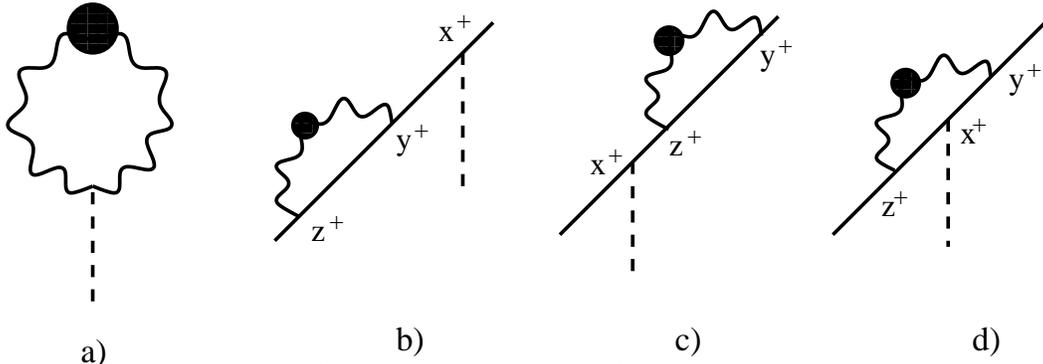}}}
         \caption{Feynman diagrams for $\hat\sigma_1$ (a) and
 $\hat\sigma_2$ (b,c,d). The wavy line with a blob denotes the
background field propagator of the semi-fast gluons;
the continuous line represents the source $\rho$; the precise
vertices can be read off eq.~(\ref{rho2}).}
\label{SIGFIG}
\end{figure}

To the same accuracy, the induced source $\langle\delta \rho\rangle$
involves only the terms in $\delta\rho$ of second order
in the fluctuations :
\be\labe{JIND}\,
\hat\sigma_a(\vec x)\,\equiv\,\langle\delta \rho_a(x)\rangle\,
\,=\,\langle\delta \rho_a^{(2)}(x)\rangle,\ee
with $\delta \rho_a^{(2)}$ given by eq.~(\ref{rho2}).
This yields 
\be\label{sigma0}
\alpha_s\ln{1\over b}\,\sigma_a ({\bf x}_\perp)\,=\,
\int dx^- \, {\rm Tr } \,(T^a \hat\sigma(\vec x)),\ee
where:
\be\label{sigma}
\hat\sigma(\vec x)&\equiv&-g \partial^+_y G^{ii}(x,y)\Big |_{x=y}\,
+\,ig^2\rho({\vec x})\Bigl\langle x\bigg|
 {1 \over i \partial^-} \,G^{--}  {1 \over i\partial^-}\bigg|x 
\Bigr\rangle\nn
&\equiv&\hat\sigma_1(\vec x)\,+\,\hat\sigma_2(\vec x).
\ee
In writing $\hat\sigma_2$ as above, we have used compact but
formal notations for the second contribution
to $\delta \rho_a^{(2)}$ in eq.~(\ref{rho2}), which is non-local
in time. Diagramatically, the two terms in eq.~(\ref{sigma})
are represented by the one-loop diagrams displayed 
in Fig. \ref{SIGFIG}. These are vertex and self-energy corrections
which in the weak field limit (i.e., to linear order in $\rho$)
generate the virtual piece of the BFKL kernel \cite{JKLW97,PI}.
Accordingly, we shall refer to $\sigma$ or $\hat\sigma$ as the
{\it virtual correction}.

%% file: S2.tex
\section{The induced charge-charge correlator, or ``real correction''}

We are now prepared for the explicit calculation of the 
coefficients in the RGE, to be completed 
in this and the following section. The one-loop quantum calculation
will be performed fully in the LC-gauge ($a^+_c={\cal A}^+_c=0$),
by using the retarded $i\epsilon$ prescription
discussed in Sect.~\ref{FRULES}. The corresponding background field
propagator is given explicitly in Appendix A.
The ensuing expressions for $\chi$ and $\sigma$, which
are gauge-covariant functionals of the background fields,
will be then rotated to the COV-gauge for the background fields
(cf. eq.~(\ref{tildesc})), which is the only gauge which
allows for explicit non-linear calculations.

In particular, the calculation of the ``real correction''
in this section will be organized 
as follows: We shall first compute $\hat \chi(\vec x, \vec y)$ by 
evaluating the matrix elements in the r.h.s. of eq.~(\ref{chi2}), 
then we shall derive $\chi({x}_\perp,{ y}_\perp)$
by integrating out the longitudinal structure of 
$\hat \chi(\vec x, \vec y)$, cf. eq.~(\ref{chi1}).
The result will be subsequently rotated to the (background) COV-gauge, 
to give $\tilde\chi({x}_\perp,{y}_\perp)$, cf. eq.~(\ref{tildesc}).
Finally, the coefficient  $\eta(x_\perp,y_\perp)$ in the RGE for
$W_\tau[\alpha]$ will be obtained according to  eq.~(\ref{etadef}).

\subsection{Non-linear effects and their dependence upon
the gauge-fixing  prescription}
\label{chinon}

By substituting the expressions (\ref{LCG}) for the gluon propagator 
in eq.~(\ref{chi2}) for $\hat \chi(\vec x, \vec y)$, one obtains,
after simple algebra,
\be\label{CHIFIN}
 {1 \over g^2} {\hat \chi} (\vec x,\vec y) & = &
i 2{\cal F}^{+i}_x\, \acute G^{ij}(x,y)\,2{\cal F}^{+j}_y\nonumber\\
&{}&
+(2 {\cal F}^{+i} {\cal D}^i+\rho)_x {1 \over i \partial^+}
 {\acute G}^{+j}(2 {\cal F}^{+j})_y
-(2 {\cal F}^{+i})_x {\acute G}^{i+} {1 \over i \partial^+}
 (2 {\cal D}^{\dagger j} {\cal F}^{+j}+\rho)_y \nonumber\\
 &{}&
+i\,
(2 {\cal F}^{+i} {\cal D}^i+\rho)_x
{1 \over i \partial^+} {\acute G}^{++} {1 \over i \partial^+}
(2{\cal D}^{\dagger j} {\cal F}^{+j}+\rho)_y \nonumber\\
&\equiv& {1 \over g^2}\Bigl( {\hat \chi}_1 (\vec x,\vec y)\,+\,
{\hat \chi}_2 (\vec x,\vec y)
\,+\,{\hat \chi}_3(\vec x,\vec y)\Bigr),\ee
where the equal-time limit $y^+=x^++\epsilon$ is implicit, and
 ${\hat \chi}_1 $, $ {\hat \chi}_2$, and ${\hat \chi}_3$
refer, respectively, to the terms in the first, second and third line.

In these expressions, ${\cal D}^i\equiv\partial^i -ig{\cal A}^i$ and
${\cal D}^{\dagger j}=\partial^{\dagger j} +ig{\cal A}^j$
(with the derivative $\partial^{\dagger j}$ acting on the 
function on its left) are covariant derivatives constructed
with the background field ${\cal A}^i$. Furthermore,
$\acute G^{\mu\nu}_{ab}(x,y)$ is the
gluon propagator in the {\it temporal} gauge $\acute a^-_c=0$, 
and is presented in Appendix A.3. It is a non-linear functional of
the background field $\alpha(\vec x)$, via the Wilson lines $V$ and 
$V^\dagger$ (see, e.g., eqs.~(\ref{G++c})--(\ref{G++n})).
What is however remarkable, and will be demonstrated in what
follows, is that all the non-linear 
effects encoded in $\acute G^{\mu\nu}$ drop out 
in the calculation of the matrix elements in eq.~(\ref{CHIFIN}).
That is, the final result for $\hat \chi(\vec x, \vec y)$ is
the same as obtained by evaluating
the r.h.s. of eq.~(\ref{CHIFIN}) with the {\it free} 
temporal-gauge propagator $\acute G^{\mu\nu}_0$.
This simplification is a consequence of our specific $i\epsilon$ 
prescription in the LC-gauge propagator, as we explain now.
Consider the following matrix element:
\be\label{me1}
\langle
x| {1 \over i\partial^+} 
{\acute G}^{++} 
{1 \over i\partial^+}|y\rangle
\ee
which enters the r.h.s. of eq.~(\ref{CHIFIN}). Our ``retarded''
prescription in $ G^{\mu\nu}$
implies that, on the left of ${\acute G}^{++}$ in eq.~(\ref{me1}),
${1 \over i \partial^+}\equiv {1 \over i \partial^+ + i \epsilon}$
should be retarded, while on the right
${1 \over i \partial^+}\equiv {1 \over i \partial^+ - i \epsilon}$
should rather be advanced (cf. eqs.~(\ref{delRA})--(\ref{+RA})).
Thus, eq.~(\ref{me1}) is the same as
\be\label{matriz2}
\int dz_1^-\, \int dz_2^-\,
\langle
x^-| {1 \over i \partial^+ + i \epsilon} |z_1^-
\rangle \,
\langle
x^+, z_1^-, x_{\perp} | \ {\acute G}^{++}  \ |y^+, z_2^-, y_{\perp}
\rangle \,
\langle
z_2^- | {1 \over i \partial^+ - i \epsilon}| y^-
\rangle
\nonumber\\
=
\int dz_1^-\, \int dz_2^-\,
\theta(x^- - z_1^-)\ {\acute G}^{++} (z_1^-,z_2^-)\
\theta(y^- - z_2^-)  \ ,
\ee
where in the second line
only the longitudinal coordinates have been shown explicitly.
In eq.~(\ref{CHIFIN}), this matrix element is sandwiched
between $\rho(\vec x)$ and $\rho(\vec y)$.  Since $\rho(\vec x)$
is localized at small$x^-$ ($x^- \simle x^-_\tau$, cf. eq.~(\ref{xtau})), 
while $G^{\mu\nu}(x,y)$
is relatively slowly varying as a function of $x^-$ and $y^-$
(since this is the propagator of the semi-fast gluons, with
$p^+ \ll \Lambda^+$), one can effectively
replace $x^-\simeq 0$ and $y^-\simeq 0$ in
eq.~(\ref{matriz2}) [and everywhere else in 
 eq.~(\ref{CHIFIN}); recall that the electric field
 ${\cal F}^{+i}$ is as localized as $\rho$].
This gives, for $x^+=y^+$,
\be
\label{matriz3}
\langle
x^-\simeq 0, \,x_{\perp}
| {1 \over i \partial^+} {\acute G}^{++} {1 \over i \partial^+}\,|
y^-\simeq 0, \,y_{\perp}
\rangle
=
\int dz_1^-\int dz_2^-\,
\theta(- z_1^-) {\acute G}^{++}(z_1^-,z_2^-) \theta(- z_2^-) \ ,
\ee
which shows that both external points $z_1^-, z_2^-$ in ${\acute G}$ are
negative, so ${\acute G}$ must be non-crossing.
(Recall that ``crossing'' and ``non-crossing'' refer to
whether the gluon has propagated or not accross the surface at $x^- =0$,
where the color source is located; see Sect. 6 of Paper I 
and Appendix A in this paper.)
Moreover, the non-crossing piece of ${\acute G(z_1^-,z_2^-)}$
at {\it negative}
$z_1^-, z_2^-$ is the same as the corresponding
piece of the {\it free} propagator. Thus, the matrix element
(\ref{matriz3}) singles out that particular piece of
${\acute G}$  which is not affected by the background field.

A similar conclusion applies to the other terms in 
the r.h.s. of eq.~(\ref{CHIFIN}), which involve the following
matrix elements
\be\label{me}
\langle x|\acute G^{ij}|y
\rangle,\qquad
\langle x| {1 \over i\partial^+} {\acute G}^{+i}|y
\rangle,\qquad \langle x| {\acute G}^{i+}{1 \over i\partial^+}|y
\rangle,
\ee
evaluated at $x^-\simeq y^-\simeq 0$ and $y^+=x^++\epsilon$.
For instance, in the first term above, one can use 
the continuity of ${\acute G}^{ij}$ at $x^-=0$ and $y^-=0$
(cf. Appendix A) to approach these points from $x^-<0$ and
$y^-<0$, i.e., from the domain where  ${\acute G}^{ij}$ 
coincides with the corresponding free propagator
${\acute G}^{ij}_0$. We thus deduce that
\be
\label{matriz4}
\langle 0, x_{\perp}
| \acute G^{ij}|  0, y_{\perp}
\rangle
=
\langle 0, x_{\perp}
| \acute G_0^{ij}|  0, y_{\perp}
\rangle\,,\ee
and similarly for the other matrix elements in eq.~(\ref{me}).

For the previous arguments, it has been essential that
the retarded prescription has been used systematically in 
the LC-gauge, both in the classical solution and in the
quantum propagator: ({\it a})
The retarded boundary condition on the classical solution has insured
that the background field ${\cal A}^i$ is non-vanishing only at $x^->0$,
or $z<t$. That is, the classical field sits behind its source,
the (fast degrees of freedom of the) hadron, which is located
at $z=t$. ({\it b}) The retarded $i\epsilon$ prescription
in the gluon propagator has implied that the semi-fast gluon
exchanged within $\hat\chi$ (cf. Fig. \ref{CHIFIG}) is confined at
negative $x^-$, where there is no background field.
Thus, this quantum gluon propagates freely
from $\vec y$ to $\vec x$, with $x^-\simeq y^-\simeq 0$.

Note, however, that this property of a free propagation
holds only for the {\it temporal} gauge
gluon $\acute a^\mu$ (with $\acute a^-=0$), which
does not couple directly to the singular color source $\rho$
\cite{PI}. Because of that, its propagator 
$\acute G^{\mu\nu}(x,y)$ is continuous at $x^-=0$ and $y^-=0$,
and the associated non-linear effects drop out in
the calculation of $\hat\chi$, as argued before.
By contrast, the {\it light-cone} gauge propagator $ G^{\mu\nu}(x,y)$ 
is sensitive also to the discontinuous gauge rotations at the
end points $x^-$ and  $y^-$, via the 
covariant derivatives ${\cal D}^i_x$ and ${\cal D}^{\dagger j}_y$
(which technically enter via the gauge rotation from the temporal
gauge to the LC gauge; cf. eq.~(\ref{GLC})).
As obvious on eq.~(\ref{CHIFIN}), the non-linear effects
associated with the fields within
${\cal D}^i_x$ and ${\cal D}^{\dagger j}_y$ do
subsist in the final result for $\hat \chi$.
Thus, while it is correct to replace the {\it temporal} gauge
propagator $\acute G^{\mu\nu}$ in eq.~(\ref{CHIFIN})
by its free counterpart, such a replacement would not be legitimate 
for the {\it light-cone} gauge propagator $G^{\mu\nu}$ 
in  eq.~(\ref{chi2}).

By inspection of the previous arguments, it should be
also clear that they would still hold, {\it mutas mutandis}, 
after replacing everywhere the retarded prescription
with an advanced one: The corresponding classical field
${\cal A}^i$ would have support only at negative $x^-$,
while the semi-fast gluon exchanged within $\hat\chi$
would freely propagate at positive $x^-$.

In Refs. \cite{MQ,K96,KM98}, where the
advanced prescription has been used extensively,
one has found similar simplifications in the calculation
of Feynman graphs for, e.g., the scattering of a
quark or a gluon off a hadronic target.
As shown there, it is only with this prescription
that one can ignore the final state interactions of
the struck quark (or gluon) in deep inelastic scattering,
which is essential if the produced jet is to be used as
an indicator of the hadron wavefunction.
Our present analysis corroborates the conclusions in
Refs. \cite{MQ,K96,KM98} that 
retarded and advanced LC-gauge prescriptions
are special in that they not only lead to
technical simplifications, but also allow for 
a more transparent physical interpretation of the results.
With more symmetrical prescriptions like principal
value PV$\,1\over p^+$ or Leibbrandt-Mandelstam,
one cannot avoid the overlap between the classical fields
and the quantum fluctuations, and thus neither the 
final/initial state interactions.

To summarize, when computing ${\hat \chi}$ with a retarded prescription,
one can replace ${\acute G}^{\mu\nu}$ in eq.~(\ref{CHIFIN})
by the corresponding {\it free} propagator  ${\acute G}^{\mu\nu}_0$.
This calculation will be presented in the following subsection.
For completness, in Appendix B we shall verify, on the example
of $\hat \chi_3$, that a lengthier calculation using the full propagator
${\acute G}^{\mu\nu}$ leads eventually to the same result.

\subsection{Explicit calculation of
 $\hat \chi(\vec x, \vec y)$}
\label{sect:comp_chi}

With  ${\acute G}^{ij}$ replaced by the free propagator
${\acute G}^{ij}_0=\delta^{ij}G_0$
the first term ${\hat \chi}_1 $ in the r.h.s. of eq.~(\ref{CHIFIN})
has been already computed in Sect. 5.2 of 
Paper I, with the following result:
\be\label{chi1final}
{\hat \chi}_1=
{g^2 \over \pi} \ln(1/b)\
{\cal F}^{+i}_x\ {\cal F}^{+j}_y\
\int {d^2p_\perp \over (2 \pi)^2}\,
{\rm e}^{ip_{\perp}\cdot(x_{\perp}-y_{\perp})}\,.
\ee
Consider now the matrix element (\ref{me1}) which enters
${\hat \chi}_3 $. With ${\acute G}^{++}\rightarrow {\acute G}^{++}_0$,
cf. eq.~(\ref{TGPROP}), this gives (for $y^+=x^+$, and 
$x^-\simeq y^-\simeq 0$) 
\be
\label{A00}
\langle 0, x_{\perp}| {1 \over i \partial^++ i \epsilon}\,
{\acute G}^{++}_0 {1 \over i \partial^+- i \epsilon}| 0, y_{\perp}
\rangle\,=\,\qquad\qquad\qquad\nn
\,=\,\int_{strip} {dp^-\over 2\pi}\int {dp^+ \over 2 \pi}
\int \frac{d^2 p_\perp}{(2\pi)^2}\,{\rm e}^{ip_\perp\cdot (x_\perp-y_\perp)}
\,\frac{1}{p^++ i \epsilon}\,\frac{2p^+}{p^-}\,G_0(p)\,
\frac{1}{p^+- i \epsilon}\,,\ee
where the integral over $p^-$ is restricted to the strip
(\ref{strip-}). The various factors of $p^+$ in the integrand 
can be combined into a PV prescription in $1/p^+$ :
\be
\frac{2p^+}{(p^++ i \epsilon)(p^+- i \epsilon)}\,=\,
\frac{1}{p^++ i \epsilon}\,+\,\frac{1}{p^+- i \epsilon}\,=\,
2{\rm PV}\,\frac{1}{p^+}\,.\ee
Then, the integral over $p^+$ is easily computed by contour techniques:
\be\label{intp+1}
\int {dp^+ \over 2 \pi}\,
{\rm PV}\,\frac{1}{p^+}\,
\frac{1}{2p^+p^- - p_\perp^2 +i\epsilon}\,=\,\frac{-i\epsilon(p^-)}
{2p_\perp^2}\,,\ee
where $\epsilon(p^-)\equiv \theta(p^-)-\theta(-p^-)$ is the
sign function. The restricted integral over $p^-$ generates
the expected logarithmic enhancement (below, $\Lambda\equiv
\Lambda^-$, cf. eq.~(\ref{strip-})) :
\be\label{LOGX}
\int_{strip} {dp^-\over 2\pi} \,\,\frac{\epsilon(p^-)}{p^-}\,\equiv\,
\left(\int_{-\Lambda/b}^{-\Lambda}+\int_{\Lambda}^{\Lambda/b}
\right){dp^- \over 2 \pi}\,{\epsilon(p^-)\over p^-}
\,=\,{1 \over \pi}\,\ln (1/b)\,.\ee
The final result reads
\be
\label{A00final}
\langle 0, x_{\perp}| {1 \over i \partial^++ i \epsilon}\,
{\acute G}^{++} {1 \over i \partial^+- i \epsilon}| 0, y_{\perp}
\rangle\,=\,{-i \over \pi} \ln(1/b)\,
\int {d^2p_\perp \over (2 \pi)^2}\,{1 \over p_{\perp}^2}\,
{\rm e}^{ip_{\perp}\cdot(x_{\perp}-y_{\perp})}\,,\ee
which immediately implies:
\be\label{chi3final}
{\hat \chi}_3=
{g^2 \over \pi} \ln(1/b)\ \int {d^2p_\perp \over (2 \pi)^2}\,
{1 \over p_{\perp}^2}\
(2 {\cal F}^{+i} {\cal D}^i+\rho)_x\
{\rm e}^{ip_{\perp}\cdot(x_{\perp}-y_{\perp})}\
(2{\cal D}^{\dagger j} {\cal F}^{+j}+\rho)_y \ .
\ee
A similar calculation yields:
\be\label{A0jfinal}
\langle 0, x_{\perp}| {1 \over i \partial^++ i \epsilon}\,
{\acute G}^{+i} | 0, y_{\perp}\rangle&=&
\langle 0, x_{\perp}|{\acute G}^{i+}
{1 \over i \partial^+- i \epsilon}| 0, y_{\perp}
\rangle\,\nn
&=&{-i \over 2\pi} \ln(1/b)
\int {d^2p_\perp \over (2 \pi)^2}\,{p^i \over p_{\perp}^2}\,
{\rm e}^{ip_{\perp}\cdot(x_{\perp}-y_{\perp})}\,,
\ee
which allows us to also compute ${\hat \chi}_2$ :
\be\label{chi2final}
{\hat \chi}_2=
{- g^2 \over \pi} \ln(1/b) \int {d^2p_\perp \over (2 \pi)^2}\,
{1 \over p_{\perp}^2}\
\biggl\{
(2 {\cal F}^{+i} {\cal D}^i+\rho)_x\
{\rm e}^{ip_{\perp}\cdot(x_{\perp}-y_{\perp})}\
(2 \partial^{\dagger j}{\cal F}^{+j})_y\,+\,
\nonumber\\\,\,\,\,
+\, (2 {\cal F}^{+i} \partial^i)_x\
{\rm e}^{ip_{\perp}\cdot(x_{\perp}-y_{\perp})}\
(2 {\cal D}^{\dagger j} {\cal F}^{+j}+\rho)_y
\biggr\}
\ee
The previous results for 
${\hat \chi}_1 $, $ {\hat \chi}_2$, and ${\hat \chi}_3$
are conveniently combined as:
\be\label{HATCHI}
{\hat \chi}(\vec x,\vec y)&=&
{g^2 \over \pi} \ln{1\over b}\,\Biggl\{
{\cal F}^{+i}_x\, \delta^{ij}_{\perp}(x_{\perp}-y_{\perp})\, {\cal F}^{+j}_y\,
+\, \nn &{}&\qquad \,\,\,\,+\,
\biggl[2 {\cal F}^{+i} \biggl({\cal D}^i - {\partial^i \over 2}\biggr)
+\rho\biggr]_x\,\langle
x_{\perp} | {1 \over -\grad_{\perp}^2} | y_{\perp}
\rangle\,
\biggl[2\biggl({\cal D}^{\dagger j} - {\partial^{\dagger j} \over 2}\biggr)
{\cal F}^{+j}+ \rho\biggr]_y\Biggr\}\ ,
\ee
which is our final expression for ${\hat \chi}$.
The following notations have been used:
\be\label{deltaT}
\delta^{ij}_{\perp}(x_{\perp}-y_{\perp})&=&
\int {d^2p_\perp \over (2 \pi)^2}\,
\left(\delta^{ij}- {p^i p^j \over p_{\perp}^2}\right)\
{\rm e}^{ip_{\perp}\cdot(x_{\perp}-y_{\perp})}\,,\\
\langle
x_{\perp} | {1 \over -\grad_{\perp}^2} | y_{\perp}
\rangle &=&
\int {d^2p_\perp \over (2 \pi)^2}\, {1 \over p_{\perp}^2}\
{\rm e}^{ip_{\perp}\cdot(x_{\perp}-y_{\perp})} \,.
\label{invdelta2}
\ee

\subsection{The two-dimensional correlator 
$\chi(x_\perp,y_\perp)$}

In eq.~(\ref{HATCHI}), both $\rho$ and ${\cal F}^{+i}$ have
support near the LC, at $x^- \simle x^-_\tau\equiv 1/\Lambda^+$. 
Thus, although induced by quantum modes with relatively large
longitudinal wavelengths $\Delta x^- \gg 1/\Lambda^+$,
the charge-charge correlator appears to be as localized in the
longitudinal direction as the original source at the scale $\Lambda^+$.
This is so because the vertices responsible
for this quantum effect are explicitly proportional to
$\rho$ or ${\cal F}^{+i}$ (cf. eq.~(\ref{rho10})).
Thus, $\hat\chi(\vec x, \vec y)$ is manifestly sensitive 
to the internal structure of the source, and also to the structure
of the background field at small distances $x^- \simle x^-_\tau$.
(Note, in particular, that eq.~(\ref{HATCHI}) involves
the product ${\cal F}^{+i}(\vec x){\cal A}^i(\vec x)$,
and thus the field ${\cal A}^i(\vec x)$ within the support
of $\rho$.)

If this was also true for the two-dimensional density
$\chi(x_\perp,y_\perp)$, which is the quantity
which enters the RGE (\ref{RGE}), this would spoil the separation
of scales assumed by the effective theory, and thus the validity 
of the latter. Note that, the simple fact that 
$\chi(x_\perp,y_\perp)$ is obtained from
$\hat\chi(\vec x, \vec y)$ after integrating out $x^-$ and $y^-$, 
cf. eq.~(\ref{chi1}), is by itself not sufficient
to guarantee that the physical information about small 
$x^-$ or $y^-$ is truly irrelevant. 
For instance, the following integrated quantity 
\be
\int dx^- {\cal F}^{+i}(\vec x){\cal A}^i(\vec x),\ee
although a function of $x_\perp$ alone, is nevertheless
sensitive to the values of ${\cal A}^i(\vec x)$ at 
$x^- \simle x^-_\tau$, since the electric field 
${\cal F}^{+i}(\vec x)$ has its support there (cf. eq.~(\ref{UTAF})).
On the other hand, the following  quantity
(recall that ${\cal A}^i(x^-\to\infty)={\cal A}^i_\infty$, while
${\cal A}^i(x^-\to -\infty)=0$) 
\be
\int dx^- {\cal F}^{+i}(\vec x)\,=\,
\int dx^- \partial^+{\cal A}^i\,=\,{\cal A}^i_\infty(x_\perp)\ee
is not sensitive to the
structure of ${\cal A}^i(\vec x)$ around the origin, but only to
its asymptotic value at large $x^-$.

As we shall demonstrate now, this latter situation applies
also to $\hat\chi(\vec x, \vec y)$, which, like the electric field
${\cal F}^{+i}(\vec x)$, is a total derivative with respect to
its longitudinal arguments $x^-$ and $y^-$:
\be\label{dervative}
{\hat \chi} (\vec x,\vec y) = \partial^+_x\partial^+_y\
{\cal C}(\vec x,\vec y) \ .
\ee
To see this, note the following identities, which hold
for an arbitrary function $\Phi(x_{\perp})$,
\be\label{derivatives}
(2 {\cal F}^{+i} {\cal D}^i+\rho)_x\ \Phi(x_{\perp})&=&
i \partial^+_x {\cal D}^2_x \Phi(x_{\perp}),\nn
\Phi(y_{\perp}) (2{\cal D}^{\dagger j} {\cal F}^{+j}+\rho)_y&=&
-i \Phi(y_{\perp}) {\cal D}^{\dagger 2}_y \partial^+_y\,\ee
(with $\partial^+_y$ in the second line acting on the
function on its left). These identities rely on the classical 
equation of motion (\ref{cleq0}), that is,
\be\label{eom}
\partial^i {\cal F}^{+i} - i g[{\cal A}^i,{\cal F}^{+i}]=\rho\,.\ee
For instance, one can obtain the first identity by writing:
\be
\partial^+ {\cal D}^2 \Phi(x_{\perp})&=&\partial^+ \Bigl((\partial^i
-ig{\cal A}^i)(\partial^i\Phi-ig{\cal A}^i\Phi)\Bigr)\nn
&=&-i({\cal F}^{+i}\Phi)-i{\cal F}^{+i}\partial^i\Phi
-ig({\cal F}^{+i}{\cal A}^i+{\cal A}^i{\cal F}^{+i})
\Phi\nn &=& -i\Bigl\{\partial^i{\cal F}^{+i}+ig
({\cal F}^{+i}{\cal A}^i-{\cal A}^i{\cal F}^{+i})+
2 {\cal F}^{+i} {\cal D}^i\Bigr\}\Phi\nn
&=&-i(\rho + 2 {\cal F}^{+i} {\cal D}^i)\Phi\,\ee
where in writing the second line we have used $\partial^+\Phi=0$
and $\partial^+{\cal A}^i={\cal F}^{+i}$, and the last line follows
from the previous one after using (\ref{eom}).

By using eqs.~(\ref{derivatives}) and (\ref{HATCHI}),
one deduces that $\hat\chi(\vec x, \vec y)$ is indeed
of the total-derivative form (\ref{dervative}), with
\be\label{Cderi}
{\cal C}(\vec x,\vec y)\,=\,{g^2 \over \pi} \ln{1\over b}\
\left\{
{\cal A}^i_x\ \delta^{ij}_{\perp}(x_{\perp}-y_{\perp})\ {\cal A}^j_y
+ ({\cal D}^i {\cal A}^i)_x
\langle
x_{\perp} | {1 \over -\grad_{\perp}^2} | y_{\perp}
\rangle
({\cal A}^j {\cal D}^{\dagger j})_y\right\} \ .
\ee
Then, the integrations over $x^-$ and $y^-$ in eq.~(\ref{chi1})
become trivial, and yield 
\be\label{CHIFINALDEVT}
{\chi}(x_{\perp},y_{\perp})= 4 
\biggl\{
{\cal A}^i_{\infty}(x_{\perp})\
\delta^{ij}_{\perp}(x_{\perp}-y_{\perp})
{\cal A}^j_{\infty}(y_{\perp})
+
({\cal D}^i_{\infty} {\cal A}^i_{\infty})_{x_{\perp}}
\langle
x_{\perp} | {1 \over -\grad_{\perp}^2} | y_{\perp}
\rangle
({\cal A}^j_{\infty} {\cal D}^{\dagger j}_{\infty})_{y_{\perp}}
\biggr\}
\ee
(with 
${\cal D}^i_{\infty} \equiv \partial^i - ig {\cal A}^i_{\infty}(x_{\perp})$)
which is sensitive only to the asymptotic fields.

The gauge rotation (\ref{tildesc}) of eq.~(\ref{CHIFINALDEVT})
to the background COV-gauge is straightforward.
By using ${\cal A}^i_\infty(x_\perp)
=(i/g)V_x \del^i V^\dagger_x$
(cf. eq.~(\ref{APM})), one immediately finds
\be\label{tchi}
\tilde \chi(x_\perp,y_\perp)\,=\,{4\over g^2}
\,\biggl\{\partial^iV^{\dagger}_x\,
\delta^{ij}_\perp(x_{\perp}-y_{\perp}) \partial^jV_y\,+\,
\partial^i_x\biggl(\Bigl(\partial^iV^{\dagger}\Bigr)_x
\langle x_\perp|\,\frac{1}{-\grad^2_\perp}\,|y_\perp\rangle
\Bigl(\partial^j V\Bigr)_y\biggr)\partial^{\dagger j}_y\biggr\},\,\,\ee
where $V^{\dagger}_x\equiv V^{\dagger}(x_\perp)$, cf. eq.~(\ref{v}),
and the derivatives not included in the brackets act on all the
functions on their right (or left). 

\subsection{From $\chi$ to $\eta$}
\label{etachi}

Eq.~(\ref{tchi}) provides $\tilde \chi(x_\perp,y_\perp)$ 
as an explicit functional of the field $\alpha(\vec x)$, 
via the asymptotic Wilson lines $V(x_\perp)$ and $V^\dagger(x_\perp)$. 
It is therefore more
convenient to write down the evolution equation in terms of
$\alpha$ rather than ${\tilde \rho}$. The appropriate transformation
$\tilde \chi \rightarrow \eta$ is shown in eq.~(\ref{etadef}),
and will be worked out in detail in what follows.

We shall do that in two steps, corresponding to the two terms
in the r.h.s. of eq.~(\ref{tchi}), which we write as
$\tilde \chi=\tilde \chi_1+\tilde \chi_2$. Correspondingly,
$\eta=\eta_1+\eta_2$ with (we omit the factor of
$4/g^2$ at intermediate steps)
\be\label{eta1}
\eta_1(x_\perp,y_\perp) &  \equiv &
\int d^2z_\perp \int d^2u_\perp
\langle x_\perp| \frac{1}{-\grad^2_\perp}\,|z_\perp\rangle
\,\Bigl(\partial^i_z \partial^j_u \left \{ V^{\dagger}_z
\delta^{ij}_{\perp}(z_{\perp}-u_{\perp}) V_u \right \}\Bigr)
\langle u_\perp|\frac{1}{-\grad^2_\perp}|y_\perp\rangle
\nonumber \\
& = & - \int d^2z_\perp \int d^2u_\perp
\langle x_\perp| \frac{\partial^i}{-\grad^2_\perp}|z_\perp\rangle
\,V^{\dagger}_z\,
\langle z_\perp|
\delta^{ij}- {\partial^i \partial^j \over \grad^2_\perp}
\,|u_\perp\rangle\, V_u \,
 \langle u_\perp| \frac{\partial^j}{-\grad^2_\perp} |y_\perp \rangle,
\ee
where the second line follows after some integrations by parts,
and we have written
\be\label{delog}
\langle x_\perp|\,\frac{\partial^i}{-\grad^2_\perp}\,|y_\perp\rangle\,
\equiv\,\partial^i_x\, \langle
x_\perp|\,\frac{1}{-\grad^2_\perp}\,|y_\perp\rangle\,
=\int {d^2p_\perp \over (2 \pi)^2}\, {-i\,p^i \over p_{\perp}^2}\,
{\rm e}^{ip_{\perp}\cdot(x_{\perp}-y_{\perp})}\, .
\ee
Similarly,
\be
\eta_2(x_\perp,y_\perp) &=  &
- \int d^2z_\perp \int d^2u_\perp
\langle x_\perp| \frac{\partial^i}{-\grad^2_\perp} |z_\perp\rangle
(\partial^i V^{\dagger}_z)
\langle z_\perp| \frac{1}{-\grad^2_\perp} |u_\perp\rangle
(\partial^j V_u)
\langle u_\perp| \frac{\partial^j}{-\grad^2_\perp}|y_\perp\rangle
\nonumber \\
\noindent
& = &  - \int d^2z_\perp \int d^2u_\perp
V^{\dagger}_z V_u \, \partial^i_z \partial^j_u
\langle x_\perp| \frac{\partial^i}{-\grad^2_\perp} |z_\perp\rangle
\langle z_\perp| \frac{1}{-\grad^2_\perp} |u_\perp\rangle 
\langle u_\perp| \frac{\partial^j}{-\grad^2_\perp} |y_\perp\rangle.
\ee
By using
\be
\partial^i_x
\langle x_\perp|\,\frac{\partial^i}{-\grad^2_\perp}\,|y_\perp\rangle\,
=\,-\delta^{(2)}(x_\perp-y_\perp), \ee
this can be further transformed as
\be
\eta_2(x_\perp,y_\perp)&=&
V^{\dagger}_x\,\langle
x_\perp|\,\frac{1}{-\grad^2_\perp}\,|y_\perp\rangle\,
V_y\,  \nonumber \\
&{}&\,\,+\, \int d^2u_\perp\, V^{\dagger}_x\,
\langle x_\perp|\,\frac{\partial^j}{-\grad^2_\perp}\,|u_\perp\rangle\,
V_u\,
\langle u_\perp|\,\frac{\partial^j}{-\grad^2_\perp}\,|y_\perp\rangle\,
\nonumber \\
&{}&\,\,+ \int d^2z_\perp\,
\langle x_\perp|\,\frac{\partial^i}{-\grad^2_\perp}\,|z_\perp\rangle\,
V^{\dagger}_z\,
\langle z_\perp|\,\frac{\partial^i}{-\grad^2_\perp}\,|y_\perp\rangle\,
V_y\,
\nonumber \\
&{}&\,\,+ \int d^2u_\perp\, \int d^2z_\perp\,
\langle x_\perp|\,\frac{\partial^i}{-\grad^2_\perp}\,|z_\perp\rangle\,
V^{\dagger}_z\,
\langle z_\perp|\,\frac{\partial^i
\partial^j}{-\grad^2_\perp}\,|u_\perp\rangle\,
V_u
\langle u_\perp|\,\frac{\partial^j}{-\grad^2_\perp}\,|y_\perp\rangle\, .
\ee
The last term in the above expression is non-local in both
$u$ and $z$, but
it cancels against the similar term that contributes to $\eta_1$,
eq.~(\ref{eta1}). So we arrive at
\be
\eta(x_\perp,y_\perp)&=&
(1\,+\,V^{\dagger}_x V_y)\, \langle
x_\perp|\,\frac{1}{-\grad^2_\perp}\,|y_\perp\rangle\,\nn
&{}&\,+\,\int d^2z_\perp\,
\langle x_\perp|\,\frac{\partial^i}{-\grad^2_\perp}\,|z_\perp\rangle\,
\langle z_\perp|\,\frac{\partial^i}{-\grad^2_\perp}\,|y_\perp\rangle\,
(V^{\dagger}_x V_z\, +\, V^{\dagger}_z V_y)\, .
\ee
By also using
\be
\int d^2z_\perp\,
\langle x_\perp|\,\frac{\partial^i}{-\grad^2_\perp}\,|z_\perp\rangle\,
\langle z_\perp|\,\frac{\partial^i}{-\grad^2_\perp}\,|y_\perp\rangle\,
=\,
-\langle x_\perp|\,\frac{1}{-\grad^2_\perp}\,|y_\perp \rangle\,
\ee
we finally express $\eta$ as (below, we reintroduce the overall
factor $4/g^2$)
\be\label{ETA}
\eta_{ab}(x_\perp,y_\perp)\,=\,{4\over g^2}
\int {d^2z_\perp\over (2\pi)^2}\,
\frac{(x^i-z^i)(y^i-z^i)}{(x_\perp-z_\perp)^2(y_\perp-z_\perp)^2 }
\Bigl\{1+ V^\dagger_x V_y-V^\dagger_x V_z - V^\dagger_z
V_y\Bigr\}_{ab}\, .
\ee
The kernel in this equation has been written in coordinate space
by using eq.~(\ref{delog}) and
\be\label{kernel}
 \int {d^2p_\perp \over (2 \pi)^2}
\int {d^2k_\perp \over (2 \pi)^2}\,\,
{p_{\perp} \cdot k_{\perp} \over p_{\perp}^2 k_{\perp}^2}\,\,
{\rm e}^{ip_{\perp}\cdot(x_{\perp}-z_{\perp})}\,
{\rm e}^{ik_{\perp}\cdot(z_{\perp}-y_{\perp})}\,=\,{1\over (2\pi)^2}\,
\frac{(x^i-z^i)(y^i-z^i)}{(x_\perp-z_\perp)^2(y_\perp-z_\perp)^2 }\,
.\ee
Eq.~(\ref{ETA}) is our final result for the ``real correction'', to
which we shall return in Sect. \ref{sect:PROP}.

%% file: S3.tex
\section{The induced color charge, of ``virtual correction''}
\label{sec:sigma}

We shall now similarly compute the coefficient $\nu$ in the RGE
(\ref{RGEA0}), or the ``virtual correction''. This calculation
will be organized as follows: By evaluating the r.h.s. of
eq.~(\ref{sigma}), we shall first obtain the three-dimensional color 
charge density $\hat\sigma(\vec x)$ in the LC gauge. The 
longitudinal structure of the result will be explored, 
and then integrated out, to yield the two-dimensional 
density $\sigma_a ({x}_\perp)$, cf. eq.~(\ref{sigma0}).
In a separate calculation, based on eq.~(\ref{tchi}) for
$\tilde \chi(x_\perp,y_\perp)$, we shall obtain the
classical polarization $\delta\sigma({x}_\perp)$ defined by
eq.~(\ref{sigclas}).  By combining the various results
as shown in eq.~(\ref{tildesig}), we shall deduce the
induced source $\tilde\sigma_a (x_\perp)$ 
in the (background-field) COV-gauge, from
which $\nu_a (x_\perp)$ will be finally extracted,
as shown in eq.~(\ref{nudef}).

At intermediate steps in this calculation we shall meet with
logarithmic ``tadpoles'', i.e.,
contributions to $\sigma$ or $\delta\sigma$
which are local in $x_\perp$ and proportional to
\be\label{tpole}
 \int {d^2p_\perp \over p_\perp^2}\,.\ee
Moreover, the momentum integrals giving the remaining,
non-local, contributions will generally involve
logarithmic ultraviolet (UV) divergences. After summing up
the various contributions, all UV divergences must go away. 
Indeed, our one-loop calculation cannot introduce other UV
divergences than those of the standard perturbation theory,
which, moreover, must be absent as well, since they are
not enhanced by the longitudinal logarithm $\ln(1/x)$.
(In other terms, the UV divergent terms are
purely ${\cal O}(\alpha_s)$ effects, which are a priori
ignored in our LLA.) 
To shorten the presentation, in this
section we shall simply ignore all the tadpoles. At the very end,
we shall verify that the final, non-local, result for
$\tilde\sigma_a$ (or $\nu_a$) is UV finite,
meaning that all the tadpoles must have cancelled among themselves.

The infrared finiteness, on the other hand, is a different issue, 
which, as in the corresponding BFKL problem, is related to
subtle compensations between real and virtual corrections.
We shall return to this 
issue in the discussion of the results in Sects. \ref{sect:RGE}
and \ref{sec:LIMITS}.

\subsection{Explicit calculation of $\hat\sigma(\vec x)$}
\label{sec:sigmaA}

Eq.~(\ref{sigma}) shows that there are two contributions to
the induced source: $\hat\sigma=\hat\sigma_1+\hat\sigma_2$.
The second contribution $\hat\sigma_2$, to be computed in
Appendix C, turns out to be a pure tadpole,
and as such it will be ignored in what follows.
The first contribution $\hat\sigma_1$, to be computed below,
is given by:
\be\label{sigma100}
{\hat \sigma}_1^a(\vec x)
\,=\,{\rm Tr } \,(T^a \hat\sigma_1(\vec x)),\qquad
{\hat \sigma}_1(\vec x)\,=\,
-g \partial^+_y G^{ii}(x,y)\Big |_{x=y}\,.\ee
It can be verified that, out of the four different terms in 
eq.~(\ref{Gij}) for $G^{ij}$, it is only the last one, namely
\be\label{G4}
G_4^{ii}(x,y)&\equiv& 
- \langle x|{\cal D}^i {1 \over \partial^+}\, \acute
G^{++}
\,{1 \over \partial^+}{\cal D}^{\dagger \,i}|y \rangle \\
&=& {\cal D}^i_x \int dz^-_1 dz^-_2
\langle
x^-| {1 \over i \partial^+ +i\epsilon}|z^-_1
\rangle
{\acute G}^{++}(z^-_1,x_{\perp};z^-_2,y_{\perp})
\langle
z^-_2|{1 \over i \partial^+-i\epsilon}|y^-
\rangle
{\cal D}^{\dagger \,i}_y\,,\nonumber\ee
that gives a non-vanishing contribution to $\hat\sigma_1$.
The derivative $ \partial^+_y$ in
eq.~(\ref{sigma100}) is readily performed as :
\be
 \partial^+_y\langle
z^-_2|{1 \over i \partial^+}|y^- \rangle\,\Big(\partial^{\dagger \,i}_y\,
+ ig {\cal A}^i(\vec y)\Bigr)\,=\,i\delta(z^-_2 -y^-)
{\cal D}^{\dagger \,i}_y\,+\,ig
\langle
z^-_2|{1 \over i \partial^+}|y^- \rangle\,{\cal F}^{+i}(\vec y)\,,\ee
thus giving the following two contributions to $\hat\sigma_1$ :
\be\label{sigma10}
{\hat \sigma}_1(\vec x)\,=
\,-ig^2 \langle x|{\cal D}^i
{1 \over i \partial^+} {\acute G}^{++} {1 \over i \partial^+}
|x\rangle \, {\cal F}^{+i}(\vec x) \,-\,
ig\langle x| {\cal D}^i {1 \over i \partial^+}
{\acute G}^{++} {\cal D}^{\dagger i}|x\rangle\,\equiv\,
{\hat \sigma}_{11}+{\hat \sigma}_{12}.
\ee
The first term ${\hat \sigma}_{11}$ 
involves the same matrix element (\ref{matriz3}) as $\hat \chi_3$ 
in eq.~(\ref{CHIFIN}). Indeed, since ${\cal F}^{+i}(\vec x)$
is localized at $x^- \simle x^-_\tau$, one can effectively
set $x^-\simeq 0$ in the above expression
${\hat \sigma}_{11}$, which leads us to the matrix element
evaluated in eq.~(\ref{A00final}). This gives
\be\label{sigma11nf}
{\hat \sigma}_{11} ({\vec{x}}) &=&
-\,{g^2 \over \pi} \ln (1/b)  \, {\cal D}^i_x
 \int {d^2p_\perp \over (2 \pi)^2}\,
\frac{{\rm e}^{ip_{\perp}\cdot(x_{\perp}-y_{\perp})}}
{p_{\perp}^2}\,{\cal F}^{+i} ({\vec y})\bigg |_{x=y}\nn
&=&
i\,{g^2 \over \pi} \ln (1/b) {\cal A}^i ({\vec x}){\cal F}^{+i} ({\vec x})
 \int {d^2p_\perp \over (2 \pi)^2}\,
\frac{1}
{p_{\perp}^2}\,,
\ee
which turns out to be a pure tadpole, and will be
therefore ignored in what follows.
As for the second term ${\hat \sigma}_{12}$ in eq.~(\ref{sigma10}),
that is,
\be\label{sigma120}
{\hat \sigma}_{12} \,=\,-
ig\langle x| {\cal D}^i {1 \over i \partial^++i\epsilon}\,
{\acute G}^{++} \,{\cal D}^{\dagger i}|x\rangle\,,
\ee
this is determined exclusively by the ``crossing'' piece
${\acute G}^{++(c)}$ of the propagator ${\acute G}^{++}$,
as given by eq.~(\ref{G++c}). (The corresponding ``non-crossing''
contribution is shown to vanish in Appendix C.)

To summarize, the whole non-trivial contribution to 
${\hat \sigma}_a(\vec x)$ comes from the
crossing piece of eq.~(\ref{sigma120}),
which after using eq.~(\ref{G++c}) can be rewritten as
(with ${\hat \sigma}_{12}$ simply denoted as ${\hat \sigma}$
from now on)
\be\label{sigma12}
{\hat \sigma} &=&-
ig\int_{strip} {dp^- \over 2 \pi}\, {2i \over p^-}\
\int {dp^+ \over 2 \pi} {dk^+ \over 2 \pi}\,
{{\rm e}^{-i(p^+ -k^+)x^-}  \over p^+ + i \epsilon} \nn
&{}&\quad\times\, {\cal D}^i_x \int d \Gamma_{\perp}\,
(p_{\perp} \cdot k_{\perp}) G_0(p) G_0(k)
\nn &{}&\quad\times 
\left \{
\theta(p^-) V(x_\perp)V^\dagger(z_\perp)
-\theta(-p^-) V(z_\perp)V^\dagger (y_\perp)
\right \}
{\cal D}^{\dagger i}_y{\Big |}_{y=x},
\ee
where it is understood that $k^-=p^-$, and we have used 
the following shorthand notation for the transverse
phase-space integrals (this is a function of
$x_\perp$ and $y_\perp$):
\be\label{dGamma}
\int d \Gamma_{\perp}\,\equiv\,  \int {d^2p_\perp \over (2 \pi)^2}\,
\int {d^2k_\perp \over (2 \pi)^2}\,\int d^2z_{\perp}\,
{\rm e}^{ip_{\perp}\cdot(x_{\perp}-z_{\perp})}\,
{\rm e}^{ik_{\perp}\cdot(z_{\perp}-y_{\perp})}.\ee
Thus, the transverse derivatives ${\cal D}^i_x$ and 
${\cal D}^{\dagger i}_y$ in eq.~(\ref{sigma12})
act on both $d \Gamma_{\perp}\,$ and the Wilson lines, and the 
limit $y_\perp\to x_\perp$ is taken at the very end.

The integrals over $p^+$ and $k^+$ in eq.~(\ref{sigma12})
are readily performed via contour techniques:
\be\label{intp+}
\int {dp^+ \over 2 \pi}\  e^{-ip^+x^-}\,
{1\over p^++i\epsilon}\
\frac{1}{2p^+p^- - p_\perp^2 +i\epsilon}\,=\,
\qquad\qquad\qquad\qquad\nn
\,=\,
{i \over p_{\perp}^2}
\left \{
\theta(x^-) -
\left [
\theta(x^-)\theta(p^-) -\theta(-x^-)\theta(-p^-) \right]
e^{-i {p_{\perp}^2 \over 2p^-}x^-}\right \},\ee
where the first term $\theta(x^-)$ in the accolades
is the contribution of the LC-gauge pole at $p^+=0$, 
while the other terms comes from the single-particle pole at 
$2p^+p^-=p_\perp^2 $. Similarly:
\be\label{G0x}
\int {dk^+ \over 2 \pi}\ e^{ik^+y^-} \,
\frac{1}{2k^+p^- - k_\perp^2 +i\epsilon}\,=\,
{-i \over 2 p^-}\,\left \{
\theta(p^-) \theta(-x^-) - \theta(-p^-) \theta(x^-)
\right \}
e^{i {k_{\perp}^2 \over 2p^-} x^-}\,.\ee
In the product of eqs.~(\ref{intp+}) and (\ref{G0x}),
the only term which survives is the one
proportional to $\theta(-p^-) \theta(x^-)$ survives. It yields
\be\label{sigint}
{\hat \sigma}({\vec x})\,=\,g \theta(x^-) {\cal D}^i_x
 \int_{strip}  {dp^- \over 2 \pi} \,{\theta(-p^-) \over (p^-)^2}
\int d \Gamma_{\perp} \,
{p_{\perp} \cdot k_{\perp} \over p_{\perp}^2}\,
 e^{i {k_{\perp}^2 \over 2 p^-}\ x^-}
V(z_\perp)V^\dagger (y_\perp)\,{\cal D}^{\dagger\, i}_y {\Big |}_{y=x},
\label{sigma12c2}
\ee
with the important consequence that the induced source
has support only at positive $x^-$. Clearly, this property is related
to our retarded gauge-fixing prescription, which is 
responsible for the $\theta(x^-)$ term in the r.h.s. of eq.~(\ref{intp+}).
If an advanced prescription $1/(p^+-i\epsilon)$
was used instead, eq.~(\ref{intp+}) would have rather generated 
a term in $\theta(-x^-)$, and thus a quantum correction
${\hat \sigma}({\vec x})$ with support at negative $x^-$.

To further explore the longitudinal structure of the induced
source, it is convenient to perform 
the restricted integration over $p^-$ in eq.~(\ref{sigint}).
One thus obtains:
\be\label{LOGX1}
\int_{strip} {dp^- \over 2 \pi}\,
{\theta(-p^-)\over (p^-)^2}\,e^{i {k_{\perp}^2 \over 2p^-}x^-}\,=\,
\frac{1}{k_{\perp}^2}\,\frac{e^{-ib\Lambda^+ x^-}-
e^{-i\Lambda^+ x^-}}{\pi i x^-}\,,\ee
where we have replaced $ k_{\perp}^2/2\Lambda^-\approx \Lambda^+$
to LLA (cf. the discussion after eq.~(\ref{strip-})). 
This shows that the $x^-$--dependence of $\hat\sigma$
is carried by the following ``form factor'' 
\be\label{FormF}
F(x^-)\,\equiv\,\theta(x^-)\,\frac{e^{-ib\Lambda^+ x^-}-
e^{-i\Lambda^+ x^-}}{x^-}\,,\ee
which has support at $1/\Lambda^+ \simle x^- \simle 1/b\Lambda^+\/$.
Indeed, $F(x^-)\approx 0$ both for small $x^-\ll
1/\Lambda^+$ (since in this case the two exponentials 
mutually cancel), and for large $x^-\gg 1/b\Lambda^+\/$
(where the two exponentials are individually small). 
For $x^- \simge 1/\Lambda^+$, ${\cal D}^i={\cal D}^i_\infty$,
with (cf. eq.~(\ref{APM}))
\be
{\cal D}^i_\infty \,\equiv\,
\partial^i -ig {\cal A}^i_{\infty}(x_{\perp}) =
\partial^i + V\partial^i V^\dagger.\ee
By also using
\be
\Big({\rm e}^{-ik_{\perp}\cdot y_{\perp}}
V^\dagger (y_\perp)\Big)\,{\cal D}^{\dagger\, i}_{\infty\,y}
\,=\,(\partial^i {\rm e}^{-ik_{\perp}\cdot y_{\perp}})V^\dagger (y_\perp)
\,=\,ik^i{\rm e}^{-ik_{\perp}\cdot y_{\perp}}V^\dagger (y_\perp),\ee
eq.~(\ref{sigint}) finally becomes (with
${\cal D}^i\equiv {\cal D}^i_{\infty}$)
\be\label{sig1x}
{\hat \sigma}({\vec x})\,=\,\frac{g}{\pi}\,F(x^-)\,{\cal D}^i_x
\int d \Gamma_{\perp}\,{p_{\perp} \cdot k_{\perp} 
\over p_{\perp}^2 k_{\perp}^2}
\,k^i\,V_z V^\dagger_y\Big |_{y_{\perp}=x_{\perp}},\ee
which shows explicitly that
the induced source ${\hat \sigma}({\vec x})$ has support 
at $1/\Lambda^+ \simle x^- \simle 1/b\Lambda^+\/$, as anticipated
in eq.~(\ref{stripx-}).
Moreover, the logarithmic enhancement is not yet manifest
on eq.~(\ref{sig1x}). This comes out only after the 
integration over $x^-\/$, via the following equation
\be 
\int dx^- F(x^-) \,=\,\ln {1 \over b}\,,\ee
which together with eq.~(\ref{sigma0}) implies:
\be\label{sigx}
g\sigma(x_\perp) \,=\,4\left({\cal D}^i_{x}
\int d\Gamma_{\perp}\,{p_{\perp} \cdot k_{\perp} \over p_{\perp}^2 k_{\perp}^2}
\,k^i\, V_z\right) V^\dagger_x.\ee
In this equation, the limit $y_\perp\to x_\perp$ within $d \Gamma_{\perp}$
can be taken also {\it before} acting with the derivative 
${\cal D}^i_x$. Indeed, it can be easily checked that the
additional term generated in this way cancels out after taking
the color trace with $T^a$. Thus, in eq.~(\ref{sigx}),
\be\label{dGamma1}
\int d \Gamma_{\perp}\,\equiv\,  \int {d^2p_\perp \over (2 \pi)^2}\,
\int {d^2k_\perp \over (2 \pi)^2}\,\int d^2z_{\perp}\,
{\rm e}^{i(p_{\perp}-k_{\perp})\cdot(x_{\perp}-z_{\perp})}\,.\ee
For what follows, it is useful to have the expression of 
$\sigma$ rotated to the covariant gauge. By using $V^{\dagger}
{\cal D}^i O=\partial^i(V^{\dagger} O)$, one obtains:
\be\label{tsigx}
g(V^{\dagger}\sigma V)(x_\perp)
 \,=\,{4}\,\partial^i_x
\int d\Gamma_{\perp}\,{p_{\perp} \cdot k_{\perp} \over p_{\perp}^2 k_{\perp}^2}
\,\,k^i\, V^\dagger_x V_z.\ee
Eqs.~(\ref{sigx}) and (\ref{tsigx}) represent our final
results in this subsection. But for what follows, it is important
to also keep in mind the longitudinal structure displayed
in eqs.~(\ref{FormF}) and (\ref{sig1x}). These equations show that,
unlike the real correction $\hat \chi(\vec x, \vec y)$ which
is as localized in the longitudinal direction as the original
source $\rho(\vec x)$, the virtual correction
$\hat\sigma(\vec x)$ is relatively delocalized, and has support
on top of the original source. As discussed in Sect.~\ref{QEVOL},
this justifies the peculiar
longitudinal structure of the RGE (\ref{RGE}) or (\ref{RGEA0}).

\subsection{From $\sigma$ to $\nu$}

According to eq.~(\ref{tildesig}), the induced source 
$\tilde\sigma_a (x_\perp)$ in the (background-field)
COV-gauge is obtained by  combining 
\be
V^{\dagger}_{ab}(x_\perp)\,\sigma_b(x_\perp)\,=\,
{\rm Tr}(T^a V^{\dagger}_x\,\sigma(x_\perp) V_x)\,,\ee
with the classical polarization term defined
by eq.~(\ref{sigclas}), that is,
\be \label{deltasig0}
\delta\sigma^a(x_\perp)&=& {\rm Tr}
\left ( T^a\, \delta \sigma(x_{\perp})
\right )\nn
\delta\sigma(x_\perp)&\equiv&i\,{g\over 2}\,
\int d^2y_\perp\,\,
\tilde\chi(x_\perp,y_\perp)\,
\langle y_\perp|\,\frac{1}{-\grad^2_\perp}\,|
x_\perp\rangle\,\equiv\,
\delta\sigma_1 + \delta\sigma_2,\ee
with the two pieces $\delta\sigma_1$ and $\delta\sigma_2$
corresponding to the two terms in the r.h.s. of eq.~(\ref{tchi})
for $\tilde \chi$. Their calculation is very similar to that
of $\eta_1$ and $\eta_2$ in Sect.~\ref{etachi}. The results are
\be
\label{deltasig1}
g\delta\sigma_1(x_\perp)&=&2 \int d \Gamma_{\perp}\,
{ p_\perp \cdot k_\perp \over p_{\perp}^2 k_{\perp}^2}\,p^i
(\partial^i V^\dagger)_x V_z\,,\nonumber\\
g\delta\sigma_2(x_\perp)&=&2 \int d \Gamma_{\perp}\,
{ (p_\perp - k_\perp) \cdot k_\perp \over p_{\perp}^2 k_{\perp}^2}
\left \{
p^i (\partial^i V^\dagger)_x V_z + i (\grad^2_\perp V^\dagger)_x V_z
\right \},
\ee
with $d \Gamma_{\perp}$ defined as in eq.~(\ref{dGamma1}).
After adding these two expressions together, and using
\be
\int d \Gamma_{\perp}\,
{ p_\perp \cdot k_\perp \over p_{\perp}^2 k_{\perp}^2}\,
\Bigl
(2p^i(\partial^i V^\dagger)_x V_z+
i (\grad^2_\perp V^\dagger)_x V_z\Bigr)\,=\,i\partial^i_x
\int d \Gamma_{\perp}\,
{ p_\perp \cdot k_\perp \over p_{\perp}^2 k_{\perp}^2}\,
(\partial^i V^\dagger)_x V_z,\ee
(which follows from the symmetry of the integrand under 
the exchange $p^i\leftrightarrow -k^i$) one obtains:
\be
\label{deltasig4}
g\delta\sigma(x_\perp)\,=\,2\left \{i \partial^i_x \int d
\Gamma_{\perp}\,
{p_{\perp} \cdot k_{\perp} \over p_{\perp}^2 k_{\perp}^2}
(\partial^i V^\dagger)_x V_z\,-\,
i (\grad^2_\perp V^\dagger)_x V_x \int {d^2p_\perp \over (2 \pi)^2}
{1 \over p_{\perp}^2}\right \},
\ee
where the second term within the braces is a pure tadpole
and thus will be neglected here, to keep in line with our previous
strategy. Our final result for the classical polarization 
term is therefore:
\be\label{classpol}
g\delta\sigma_a(x_\perp)
\,=\,{2i}\,\partial^i_x
\int d\Gamma_{\perp}\,{p_{\perp} \cdot k_{\perp} \over p_{\perp}^2 k_{\perp}^2}
\,\partial^i_x\,{\rm Tr}\Bigl(T^a V^\dagger_x V_z\Bigr).\ee
As a matter of fact, the transverse integrations in eq.~(\ref{classpol})
develop a logarithmic ultraviolet divergence, which in the original
expression (\ref{deltasig4}) was precisely cancelled by the explicit 
tadpole there. Still, neglecting this tadpole is a
legitimate procedure since the UV divergence in eq.~(\ref{classpol})
is actually needed to cancel a corresponding divergence
in eq.~(\ref{tsigx}) for $V^\dagger\sigma V$ (see below).
This simply shows that in a more complete calculation where
all the tadpoles would have been kept explicitly,
the tadpole in eq.~(\ref{deltasig4}) would have cancelled
against similar tadpoles emerging in the calculation of
$\sigma$, and which have been neglected in arriving at
eq.~(\ref{tsigx}). 

To conclude, the final result for the induced source 
$\tilde\sigma_a (x_\perp)$ 
in the COV-gauge is obtained by combining
eqs.~(\ref{tsigx}) and (\ref{classpol}) according to 
eq.~(\ref{tildesig}), and reads:
\be\label{tsigma}
g\tilde\sigma_a (x_\perp)\,=\,-\grad_\perp^2\left\{2i\,
\int d\Gamma_{\perp}\,{p_{\perp} \cdot k_{\perp} \over p_{\perp}^2 k_{\perp}^2}
\,{\rm Tr}\Bigl(T^a V^\dagger_x V_z\Bigr)\right\}
\,\equiv\,-g\,\grad_\perp^2\nu_a (x_\perp).\ee
In order to obtain this simple form, it has been convenient to rearrange
the integrand in eq.~(\ref{tsigx}) as follows:
\be
\int d\Gamma_{\perp}\,\cdots\,k^i\,=\,(1/2)\int d\Gamma_{\perp}\,
\cdots\,(k^i-p^i)\,=\,(-i/2)\partial^i_x \int d\Gamma_{\perp}\,
\cdots\,,\ee
which relies again on the symmetry of the integrand
under $p^i\leftrightarrow -k^i$.

As anticipated, the expression (\ref{tsigma}) is UV finite.
This can be verified as follows:
Let $q_{\perp}\equiv p_{\perp}-k_{\perp}$ be the
momentum conjugate to the transverse distance $x_{\perp}-z_{\perp}$.
Then
\be\label{sigUV}
\int d\Gamma_{\perp}\,{p_{\perp} \cdot k_{\perp} \over p_{\perp}^2 k_{\perp}^2}
\,{\rm Tr}\Bigl(T^a V^\dagger_x V_z\Bigr)&=&
\int d^2z_{\perp} \int {d^2p_\perp \over (2 \pi)^2}
\int {d^2q_\perp \over (2 \pi)^2}\,\,
{\rm e}^{iq_{\perp}\cdot(x_{\perp}-z_{\perp})}\,
{\rm Tr}\Bigl(T^a V^\dagger_x V_z\Bigr)\nn
&{}&\quad \times\,
\frac{1}{2}\left\{
\frac{1}{p_{\perp}^2}\,+\,\frac{1}{(p_\perp-q_\perp)^2}
\,-\,\frac{q_\perp^2}{p_{\perp}^2(p_\perp-q_\perp)^2}\right\},\ee
where the following identity has been used:
\be\label{UVID}
{p_{\perp}\cdot(p_\perp-q_\perp) \over p_{\perp}^2
(p_\perp-q_\perp)^2}\,=\,\frac{1}{2}\left\{
\frac{1}{p_{\perp}^2}\,+\,\frac{1}{(p_\perp-q_\perp)^2}
\,-\,\frac{q_\perp^2}{p_{\perp}^2(p_\perp-q_\perp)^2}\right\}.
\ee
Since $q_\perp$ is kept finite by the non-locality in
$x_{\perp}-z_{\perp}$, potential UV problems can occur only
from the limit $p_{\perp}\to \infty$ at fixed $q_\perp$.
In this limit, the last term within the braces in 
eq.~(\ref{sigUV}) is manifestly UV finite (since it decreases
like $1/p_{\perp}^4$), while the first two terms give
logarithmic tadpoles $\int d^2p_\perp/p_{\perp}^2$.
But in these two terms, the integration over $q_\perp$
sets $x_{\perp}=z_{\perp}$, so that $ V^\dagger_x V_z \to 1$.
Thus, the potential tadpoles go away after taking the color
trace with $T^a$.
Incidentally, the above discussion shows that eq.~(\ref{tsigma})
can be equivalently rewritten as:
\be\label{tsigma1}
g\tilde\sigma_a (x_\perp)\,=\,i\grad_\perp^2
\int d^2z_{\perp} \int {d^2p_\perp \over (2 \pi)^2}
\int {d^2q_\perp \over (2 \pi)^2}\,\,
{\rm e}^{iq_{\perp}\cdot(x_{\perp}-z_{\perp})}\,
\frac{q_\perp^2}{p_{\perp}^2(p_\perp-q_\perp)^2}\,
{\rm Tr}\Bigl(T^a V^\dagger_x V_z\Bigr),\ee
a form which will be useful later on.

From eq.~(\ref{tsigma}), the coefficient $\nu_a (x_\perp)$ in the
COV-gauge RGE (\ref{RGEA0}) can be trivially identified,
according to eq.~(\ref{nudef}). It reads:
\be\label{NU}
g\nu^a (x_\perp)&=&
2i\int {d^2z_\perp\over (2\pi)^2}\,\frac{1}{(x_\perp-z_\perp)^2}
\,{\rm Tr}\Bigl(T^a V^\dagger_x V_z\Bigr),
\ee
where eq.~(\ref{kernel}) has also been used.
Physically, the quantity $\nu_a (x_\perp)$ 
is the induced color field in the COV-gauge, that is, the modification
in the original field $\alpha_a$ induced by quantum corrections
\cite{PI}.

%% file: S4.tex
\section{The Renormalization Group Equation}
\label{sect:RGE}

We are finally in a position to write down the renormalization group 
equation explicitly and study some of its properties.
In the $\alpha$--representation, this is given by eq.~(\ref{RGEA0})
with the coefficients $\eta$ and $\nu$ from eqs.~(\ref{ETA}) and
(\ref{NU}), respectively. 
We summarize here these equations for convenience:
\be\labe{RGEA}
{\del W_\tau[\alpha] \over {\del \tau}}\,=\,
\left\{ {1 \over 2} {\delta^2 \over {\delta
\alpha_\tau^a(x_\perp) \delta \alpha_\tau^b(y_\perp)}} 
[W_\tau\eta_{xy}^{ab}] - 
{\delta \over {\delta \alpha_\tau^a(x_\perp)}}
[W_\tau\nu_{x}^a] \right\}\,,
\ee
where 
\be\label{eta}
\eta^{ab}(x_\perp,y_\perp)
&=&{1\over \pi}\int {d^2z_\perp\over (2\pi)^2}\,
\frac{(x^i-z^i)(y^i-z^i)}{(x_\perp-z_\perp)^2(y_\perp-z_\perp)^2 }
\Bigl\{1+ V^\dagger_x V_y-V^\dagger_x V_z - V^\dagger_z V_y\Bigr\}^{ab},
\\
\nu^a (x_\perp)&=&{ig\over 2\pi}
\int {d^2z_\perp\over (2\pi)^2}\,\frac{1}{(x_\perp-z_\perp)^2}
\,{\rm Tr}\Bigl(T^a V^\dagger_x V_z\Bigr).
\label{nu}\ee
Note that, as compared with the original equations (\ref{RGEA0}),
(\ref{ETA}) and (\ref{NU}), we have rescalled here the coefficients
$\eta$ and $\nu$ to absorb the overall factor $\alpha_s=g^2/4\pi$.
The equations above represent our main result in this and the
accompanying paper \cite{PI}. 
They govern the flow with $\tau=\ln(1/x)$ of the probability density
$W_\tau[\alpha]$ for the stochastic color field $\alpha_a(\vec x)$
which describes the CGC in the COV-gauge. We now turn to a systematic
discussion of the properties of these equations.

\subsection{General properties}
\label{sect:PROP}

{\sf i) The coefficients $\eta$ and $\nu$ are real quantities.}
This is so since the Wilson lines in the
adjoint representation are real color matrices: 
$V^*_{ab}=V_{ab}$, and therefore $V^\dagger_{ab} = V_{ba}$.
By using this same property, one can further verify that:

{\sf ii) The 2-point function 
$\eta^{ab}(x_\perp,y_\perp)$ is symmetric and positive semi-definite: }
\be\label{etasym}
\eta^{ab}(x_\perp,y_\perp)&=&\eta^{ba}(y_\perp,x_\perp)\nn
&=&{1\over \pi}\int {d^2z_\perp\over (2\pi)^2}\,
\frac{(x^i-z^i)(y^i-z^i)}{(x_\perp-z_\perp)^2(y_\perp-z_\perp)^2 }
\Bigl(1- V^\dagger_z V_x\Bigr)^{ca}
\Bigl(1- V^\dagger_z V_y\Bigr)^{cb}.\ee
This guarantees that the solution $W_\tau[\alpha]$ is 
positive semi-definite (as it should) at any $\tau$ 
provided it was like that at the initial  ``time'' $\tau_0$.

{\sf iii) The RGE preserves the
correct normalization of the weight function:} 
\be\label{norm1}
\int {\cal D}\alpha\, \,W_\tau[\alpha]\,=\,1\qquad
{\rm at\,\, any}\,\, \tau.\ee
Indeed, the r.h.s. of eq.~(\ref{RGEA})
is a total derivative with respect to $\alpha$. Thus, if
eq.~(\ref{norm1}) is satisfied by the initial condition at
time $\tau_0$, then it will be automatically true at any $\tau>\tau_0$.

{\sf iv) The solution $W_\tau[\alpha]$ encodes the information about the
longitudinal support of $\alpha^a(x^-,x_\perp)$ and its evolution with
$\tau$.} From the calculation of quantum corrections, we know that
the quantum evolution up to $\tau$ generates a non-trivial color field 
$\alpha^a(\vec x)$ only within the range $0\le x^-\le x^-_\tau$,
with  $x^-_\tau=x^-_0{\rm e}^{\tau}$ and $x^-_0= 1/P^+$. Thus, as a functional
of $\{\alpha^a(x^-,x_\perp)\,|\,0\le x^-< \infty\}$, $W_\tau[\alpha]$ must
have the following structure:
\be\label{longitW}
W_\tau[\alpha]\,=\,\delta[\alpha_>]\,
{\cal W}_\tau[\alpha_<].\ee
Here, $\alpha_<$ ($\alpha_>$) is the function $\alpha(\vec x)$
for $x^-< x^-_\tau$ ($x^-> x^-_\tau$),
\be
\alpha(\vec x)\,\equiv\,\theta(x^-_\tau-x^-)\alpha_<(\vec x)\,+\,
\theta(x^- - x^-_\tau)\alpha_>(\vec x),\ee
the  {\it functional} $\delta$--function $\delta[\alpha_>]$
should be understood
in terms of some discretization of the longitudinal axis,
as the product of ordinary $\delta$--functions at all
the points $x^->x^-_\tau$ :
\be
\delta[\alpha_>]\,\equiv\,\prod_{x^->x^-_\tau}
\delta(\alpha_{x^-})\ee
(this simply shows that $\alpha^a(x^-,x_\perp)=0$
with probability one for any $x^->x^-_\tau$),
and the new functional ${\cal W}_\tau[\alpha_<]$ involves 
$\alpha^a(x^-,x_\perp)$ only within the restricted
range $0\le x^-\le x^-_\tau$.

The structure (\ref{longitW}) can be used to simplify the
average over $\alpha$ when computing observables:
\be\label{OBSERV}
\langle O[\alpha] \,\rangle_\tau&=&
\int_{0\le x^-< \infty} \,{\cal D}\alpha\,O[\alpha] \,W_\tau[\alpha]\nn
&=&\int_{0\le x^-\le x^-_\tau}\,{\cal D}\alpha\,O[\alpha] \,
{\cal W}_\tau[\alpha],\ee
where in the first line the functional integral
runs over fields with support at $x^- \ge 0$, while in the second
line the support is restricted to $x^-\le x^-_\tau$. The
expression in the first line is nevertheless more convenient to
derive an evolution equation for $\langle O[\alpha] \,\rangle_\tau$,
since there the whole dependence on $\tau$ is carried by the weight 
function:
\be\labe{evolO}
{\del \over {\del \tau}}\langle O[\alpha] \,\rangle_\tau\,=\,\int
{\cal D}\alpha\,O[\alpha] \,{\del W_\tau[\alpha] \over {\del \tau}},\ee
with $\del W_\tau/\del \tau$ determined by the RGE (\ref{RGEA}).

{\sf v) Within the RGE,
the Wilson lines $V^\dagger$ and $V$,
which were a priori defined as path-ordered
integrals along the whole $x^-$ axis
(cf. eq.~(\ref{v})), can be effectively replaced by:}
\be\labe{vtau}
V^\dagger(x_{\perp})\,\rightarrow\,U^\dagger_\tau(x_{\perp}),\qquad
U^\dagger_\tau(x_{\perp})\,\equiv\,{\rm P} \exp
 \left \{
ig \int_{0}^{x^-_\tau} dx^-\,\alpha (x^-,x_{\perp})
 \right \}.\ee
This obvious consequence of eq.~(\ref{longitW}) is interesting since
it shows that the functional derivatives within the RGE (\ref{RGEA})
act on Wilson lines as derivatives with respect to the color field 
at the {end} point $x^-=x^-_\tau$. Explicitly:
\be\label{DIFFU}
{\delta U_\tau^\dagger (x_{\perp})\over \delta \alpha^a_\tau(z_\perp)}\,=\,
ig\delta^{(2)}(x_{\perp}-z_\perp)\,T^a U_\tau^\dagger (x_{\perp}),\quad
{\delta U_\tau(y_{\perp})\over \delta \alpha^a_\tau(z_\perp)}\,=\,
-ig \delta^{(2)}(y_{\perp}-z_\perp)\, U_\tau(y_{\perp})T^a\ee
which should be compared with more general formulae like:
\be\label{DIFFU1}
{\delta V^\dagger (x_{\perp})\over \delta \alpha^a_\tau(z_\perp)}\,=\,
ig\delta^{(2)}(x_{\perp}-z_\perp)\,U_{\infty,\tau}^\dagger(x_{\perp}) 
T^a U_\tau^\dagger (x_{\perp})\,,\ee
where the integral over $x^-$ within $U_{\infty,\tau}^\dagger$
runs from $x^-_\tau$ up to $\infty$.
Clearly, eq.~(\ref{DIFFU1}) reduces to the first  eq.~(\ref{DIFFU})
for a field $\alpha$ with support at $x^-\le x^-_\tau$.

{\sf vi) The longitudinal coordinate and the evolution time
get identified by the quantum evolution.} The previous arguments
at points {iv) -- vii)}  reveal a strong correlation
between the longitudinal coordinate $x^-$ and the rapidity $\tau$
which eventually allows us to identify these two variables. 
To make this identification more precise, let us 
introduce, in addition to the 
{\it momentum}  rapidity $\tau\equiv\ln(P^+/\Lambda^+) = \ln(1/x)$,
also the {\it space-time} rapidity ${\rm y}\,$, 
defined as (for positive $x^-$ and $x^-_0= 1/P^+$):
\be\label{etarapdef} {\rm y}\,\equiv\,\ln(x^-/ x^-_0)\,,
\qquad -\infty < {\rm y} < \infty\,.\ee
Then the field 
\be\label{alphatau}
\alpha_{{\rm y}}^a(x_\perp)
\,\equiv\, \alpha^a(x^-=x^-_{\rm y},x_\perp),\qquad
x^-_{\rm y}\,\equiv\,x^-_0{\rm e}^{{\rm y}},\ee
at {\it positive}\footnote{The field $\alpha_{{\rm y}}$ at
negative ${\rm y}$, i.e., at $x^-<x^-_0$, is rather a part of the
initial conditions, in that it exists independently of the quantum evolution.}
 ${\rm y}$ (i.e., at $x^->x^-_0$) has been
generated by the quantum evolution from $\tau'=0$ up to $\tau$,
with a one-to-one correspondence between ${\rm y}$ and $\tau$.
That is, the field $\alpha_{{\rm y}'}$ within the {\it space-time} rapidity 
layer ${\rm y}\le {\rm y}' \le {\rm y}+\Delta{\rm y}$ has been
obtained by integrating out the quantum modes within the 
{\it momentum}  rapidity layer $\tau\le \tau'\le \tau+\Delta\tau$
with $\tau={\rm y}$ and $\Delta\tau=\Delta{\rm y}$.
This leads us to treat ${\rm y}$ and $\tau$ on the same
footing, as an ``evolution time''. Then $\{\alpha_{{\rm y}}^a(x_\perp)\,|\,
-\infty < {\rm y}<\infty\}$ is conveniently interpreted as a {\it trajectory}
in the functional space spanned by the {\it two-dimensional} color fields
$\alpha^a(x_\perp)\,$; this trajectory effectively ends up at ${\rm y}=\tau$.
With this interpretation, eq.~(\ref{RGEA}) describes effectively a
2+1 dimensional field theory (the two transverse coordinates plus the
evolution time), which is however {\it non-local} in
both $x_\perp$ and ${\rm y}$
(since Wilson lines likes (\ref{vtau}) involve integrals over all
${\rm y}\in (-\infty,\tau)$).

{\sf vii) The infrared and ultraviolet behaviours of the RGE.} 
These are determined by the corresponding 
behaviours of the kernel $\eta^{ab}(x_\perp,y_\perp)$ (we shall
shortly see that $\nu$ is related to $\eta$), and, more precisely,
by the behaviour of the integrand in eq.~(\ref{eta}) for
fixed $x_\perp$ and $y_\perp$ (since these are the ``external'' 
points at which we probe correlations in the system). 
For the infrared, we need this integrand in the 
limit where $z_\perp$ is much larger than both $x_\perp$ and $y_\perp$.
Then the products of Wilson lines involving $z_\perp$ are expected to
be small (since, e.g., 
$\langle V^\dagger_x V_z \rangle \to 0$ 
as $|z_\perp-x_\perp|\to \infty$ \cite{SAT}),
so it is enough to study the large--$z_\perp$ behaviour of
\be\label{Kxyz}
{\cal K}(x_\perp,y_\perp,z_\perp)\,\equiv\,
\frac{(x^i-z^i)(y^i-z^i)}{(x_\perp-z_\perp)^2(y_\perp-z_\perp)^2}\,.\ee
For $z_\perp \gg x_\perp,y_\perp$, this gives
${\cal K}_{IR} \sim 1/z_\perp^2$ and the ensuing integral 
$(d^2z_\perp/z_\perp^2)$ has a logarithmic infrared divergence.
Thus, there is potentially an IR problem in the RGE. This is
not necessarily a serious difficulty, since IR problems are expected
to be absent only for the {\it gauge-invariant} observables.
In this context, we would expect the IR divergences to cancel out
when the RGE is used to derive evolution equations for 
gauge-invariant observables. Although a general proof in this sense 
is still missing, we shall nevertheless see some explicit examples
where such cancellations take place indeed.

Coming now to the ultraviolet, or short-range,
behaviour, it is easy to see on eqs.~(\ref{eta}) or
(\ref{etasym}) that no UV problem is to be anticipated.
For instance, the would-be linear pole of ${\cal K}(x_\perp,y_\perp,z_\perp)$
at $|z_\perp-x_\perp|\to 0$ is actually cancelled by the
factor $1- V^\dagger_z V_x$ which vanishes in the same limit.

\subsection{The RGE in Hamiltonian form}

The RGE (\ref{RGEA}) is a functional Fokker-Planck equation,
that is, a diffusion equation for a (functional) probability density.
It portrays the quantum evolution towards small $x$ as
a random walk (in time $\tau$) in the functional space spanned
by the two-dimensional color fields $\alpha^a(x_\perp)$.
The term involving second order derivatives in this equation is
generally interpreted as the ``diffusion term'', while that 
with a single derivative is rather the ``force term''.
But this interpretation is not unique for our RGE where the
2-point function $\eta^{ab}(x_\perp,y_\perp)$ --- the
analog of the diffusion ``constant'' --- is itself a 
functional of the field variable $\alpha^a(x_\perp)$, so that
eq.~(\ref{RGEA}) may be as well rewritten as
\be\labe{RGEAnew}
{\del W_\tau[\alpha] \over {\del \tau}}\,=\,
{\delta \over {\delta \alpha_\tau^a(x_\perp)}}
\left\{ {1 \over 2} \eta_{xy}^{ab} \,{\delta W_\tau\over 
{\delta \alpha_\tau^b(y_\perp)}}  + \left({1 \over 2} \,
{\delta \eta_{xy}^{ab}\over {\delta \alpha_\tau^b(y_\perp)}}
 -\nu_{x}^a\right)W_\tau \right\}\,,
\ee
which rather features ${1 \over 2} 
({\delta \eta/{\delta \alpha_\tau}}) -\nu$ as the
effective ``force term''.
An important property, which is full of consequences, is that
this effective ``force term'' is precisely zero.

Specifically, the following functional relation holds between
the coefficients of the RGE:
\be\label{sigchi}
{1 \over 2} \int d^2y_\perp 
{\delta \eta^{ab}(x_\perp,y_\perp)\over {\delta \alpha_\tau^b(y_\perp)}}
\,=\,\nu^a(x_\perp)\,.\ee
It is straightforward to prove this relation by computing the
effect of $\delta/\delta \alpha_\tau^b(y_\perp)$ on the terms 
involving Wilson lines in eq.~(\ref{eta}) for $\eta^{ab}(x_\perp,y_\perp)$.
For instance,
\be\label{anti}
{\delta\over {\delta \alpha_\tau^b(y_\perp)}}\,(V^\dagger_x V_y)^{ab}
&=& {\delta V^{\dagger\,ac}_x\over {\delta \alpha_\tau^b(y_\perp)}}\,V_y^{cb}
\,+\,V^{\dagger\,ac}_x{\delta V_y^{cb}\over {\delta \alpha_\tau^b(y_\perp)}}\nn
&=& ig\delta^{(2)}(x_{\perp}-y_\perp)\Bigl(T^bV^\dagger_y\Bigr)_{ac}V_y^{cb}
-ig\delta^{(2)}(0_{\perp})V^{\dagger\,ac}_x\Bigl(V_yT^b\Bigr)_{cb}\,=\,0\,,\ee
where each of the two terms vanishes because of the antisymmetry of the
color group generators in the adjoint representation (e.g.,
$(T^b)_{ab}=0$). One can similarly show that:
\be
-{\delta\over {\delta \alpha_\tau^b(y_\perp)}}\,(V^\dagger_x V_z)^{ab}
&=&-ig\delta^{(2)}(x_{\perp}-y_\perp)\Bigl(T^b V^\dagger_x V_z\Bigr)_{ab}
\,=\,ig\delta^{(2)}(x_{\perp}-y_\perp) 
{\rm Tr}\Bigl(T^a V^\dagger_x V_z\Bigr),\nn
-{\delta\over {\delta \alpha_\tau^b(y_\perp)}}\,(V^\dagger_z V_y)^{ab}&=&0.\ee
The only non-vanishing contribution is that in the first line above,
and this precisely reproduces eq.~(\ref{nu}) after integration over $y_\perp$,
since (cf. eq.~(\ref{Kxyz}))
\be\label{Kxxz}
 {\cal K}(x_\perp,x_\perp,z_\perp)\,=\,\frac{1}{(x_\perp-z_\perp)^2}\,.\ee
By using eqs.~(\ref{RGEAnew}) and (\ref{sigchi}), the RGE
can be finally brought into the Hamiltonian form:
\be\labe{RGEH}
{\del W_\tau[\alpha] \over {\del \tau}}\,=\,-\,H W_\tau[\alpha], 
\ee
with the following Hamiltonian:
\be\labe{H}
H&\equiv&
{1 \over 2}
\int d^2x_\perp\int d^2y_\perp\,
{\delta \over {\delta
\alpha_\tau^a(x_{\perp})} }\,\eta_{xy}^{ab}\,
{\delta \over {\delta \alpha_\tau^b(y_{\perp})}}\,=\,
\int {d^2 z_\perp\over 2\pi }\,J^i_a(z_\perp)\,J^i_a(z_\perp),\nn
J^i_a(z_\perp)&\equiv& \int {d^2 x_\perp\over 2\pi }\,
\frac{z^i-x^i}{(z_\perp-x_\perp)^2}\,(1 - V^\dagger_zV_x)_{ab}\,
{i \delta \over {\delta
\alpha_\tau^b(x_\perp)} }\,,\ee
which is Hermitian (since $\eta_{xy}^{ab}$ is real and symmetric)
and positive semi-definite (since the ``current'' $J^i_a(z_\perp)$
is itself Hermitian). 

At a first sight, the interpretation of eqs.~(\ref{RGEH})--(\ref{H})
as a Hamiltonian system may look compromised by the fact that
the operator (\ref{H}) appears to be time-dependent, e.g.,
via the $\tau$-dependence of the differentiation point 
$\alpha_\tau^a(x_\perp)$, and even non-local in time,
via the Wilson lines $V$ and $V^\dagger$ (which involve integrals
over all times ${\rm y}\le\tau$; cf. the discussion
after eq.~(\ref{alphatau})). But this rather means that we have not
yet identified the correct canonical variables: these are not the
fields $\alpha_\tau^a$,  but rather the Wilson lines 
ending at $\tau$. Specifically, the
canonically conjugate variables and momenta are 
$\{U^{ab}_\tau(x_{\perp}),\,\Pi^c_\tau(y_\perp)\}$, with 
$U_\tau(x_{\perp})$ as defined in eq.~(\ref{vtau}), and
$\Pi^a_\tau(x_\perp)$ acting on $U_\tau$ or $U^\dagger_\tau$
as the functional derivative w.r.t. the field at the end point:
\be\labe{CANPI}
\Pi^a_\tau(x_\perp)\,\equiv\,\frac{1}{ig}\,{ \delta \over {\delta
\alpha_\tau^a(x_\perp)} }\,.\ee
The associated Poisson brackets, or equal-time commutation
relations, are easily inferred from eq.~(\ref{DIFFU}):
\be\label{PB}
\Big[\Pi^a_\tau(x_\perp),\,U^\dagger_\tau(y_{\perp})\Big]&=&
T^a U_\tau^\dagger (y_{\perp})\delta^{(2)}(x_{\perp}-y_\perp),\quad
\Big[\Pi^a_\tau(x_\perp),\,U_\tau(y_{\perp})\Big]\,=\,-
 U_\tau(y_{\perp})T^a\delta^{(2)}(x_{\perp}-y_\perp),\nn
\Big[\Pi^a_\tau(x_\perp),\,\Pi^b_\tau(y_{\perp})\Big]&=&
if^{abc}\Pi^c_\tau(x_\perp)\delta^{(2)}(x_{\perp}-y_\perp),
\qquad\Big[U_\tau^{ab}(x_{\perp}),\,U_\tau^{cd}(y_{\perp})\Big]
\,=\,0.\ee
(Non-shown commutators like $[ U_\tau^\dagger, U_\tau^\dagger]$
or $[ U_\tau^\dagger,U_\tau]$ vanish as well. This should not be 
confused with the fact that, as color matrices, $U_\tau(x_{\perp})$
and $U_\tau(y_{\perp})$ do not commute with each other at different
points $x_{\perp}\ne y_{\perp}$.)

Eqs.~(\ref{vtau}) and (\ref{CANPI}) provide explicit
representations for the canonical variables in terms 
of the gauge field $\alpha_{\rm y}^a(x_\perp)$. But we can extend
these relations to more abstract definitions, which hold
independently of any representation. 
The generic canonical ``coordinates'' are 
the group-valued 2-dimensional fields $V^{ab}(x_{\perp})$, while
the  canonical ``momenta'' $\Pi^a(x_\perp)$ are Lie derivatives
w.r.t. these fields, whose action is {\it defined} by eqs.~(\ref{PB}).
These momenta satisfy the commutation relation of the Lie algebra,
as they should (cf. the second line of eq.~(\ref{PB})).
In terms of these variables, the Hamiltonian (\ref{H}) takes the more
standard form of a second-quantized Hamiltonian for a 2-dimensional
field theory:
\be\labe{H2q}
H[\Pi,V]\,=\,
{1 \over 2}
\int d^2x_\perp\int d^2y_\perp\,\Pi^a(x_\perp)\,\Big(g^2
\eta_{xy}^{ab}[V,V^\dagger]\Big)\,\Pi^b(y_\perp)\,.\ee
Note that this is a purely kinetic Hamiltonian (no potential).
It describes free motion on a non-trivial manifold (here, a 
functional group manifold), 
with the kernel $g^2\eta^{ab}_{xy}[V,V^\dagger]$ playing the role
of the metric on that manifold. Then, (\ref{RGEH}) is like 
the evolution equation for this Hamiltonian in imaginary time.
Actually, giving the probabilistic interpretation of $W_\tau[\alpha]$,
the best analogy is not with the (functional) Schr\"odinger equation
in imaginary time, but rather with the Fokker-Planck equation.
We shall return to this analogy by the end of this section (see also
\cite{BIW}).

The Hamiltonian  (\ref{H2q}) has been first obtained by Weigert
\cite{W} in a rather different context, namely from an analysis
of Balitsky's equations. These are coupled evolution equations
for Wilson-line operators that have been derived by Balitsky \cite{B} 
via the operator product expansion of high-energy QCD scattering
in the target rest frame.
Weigert has subsequently recognized that all these equations,
which form an infinite hierarchy, can be generated from a
rather simple functional evolution equation with the Hamiltonian
(\ref{H2q}). The fact that we have come across the same Hamiltonian means
that our RGE too generates Balitsky's equations,
which we shall verify explicitly in the next section on the
example of the 2-point function 
$\langle {\rm tr}\big(V^\dagger(x_{\perp}) V(y_{\perp})\big)\rangle_\tau$.


Weigert has been also the first one to notice the remarkable
relation (\ref{sigchi}) between the coefficients in the functional
evolution equation. Let us devote the end of this section to a more
thourough discussion
of this relation and some of its consequences.

As a relation between one- and two-point functions, eq.~(\ref{sigchi})
is reminiscent of a Ward identity, and is most probably a consequence
of gauge symmetry, although we have not been able to demonstrate 
this convincingly. It can be checked that a corresponding relation holds
between the coefficients $\tilde\chi$ and $\tilde\sigma$ 
in the RGE for $W_\tau[\tilde\rho]$ :
\be\label{sigchiCOVG}
{1 \over 2} \int d^2y_\perp 
{\delta \tilde\chi^{ab}(x_\perp,y_\perp)\over 
{\delta \tilde\rho_\tau^b(y_\perp)}}
\,=\,\tilde\sigma^a(x_\perp)\,\ee
(this is a rather obvious,
consequence of eqs.~(\ref{sigchi}) and (\ref{nudef})) and, more
significantly, also between the coefficients $\chi$ and $\sigma$
of the original RGE (\ref{RGE}) for $W_\tau[\rho]$ :
\be\label{sigchiLC}
{1 \over 2} \int d^2y_\perp 
{\delta \chi^{ab}(x_\perp,y_\perp)\over {\delta \rho_\tau^b(y_\perp)}}
\,=\,\sigma^a(x_\perp)\,.\ee
This last relation is more significant in that it sheds some more light
on the classical polarization term $\delta\sigma$ : this is such as
to make the two relations (\ref{sigchiCOVG}) and (\ref{sigchiLC})
consistent with each other, and with the $\rho$--dependence of the
gauge rotation (\ref{tildechi}) from $\chi$ to $\tilde\chi$. 
That is, $\delta\sigma$ has precisely
the right value to compensate the functional derivatives of
the Wilson lines $V$ and $V^{\dagger}$ in eq.~(\ref{tildechi}).

There are at least two important consequences of the relation
(\ref{sigchi}). The first one refers to the cancellation of
infrared divergences in the evolution equations for gauge-invariant
quantities. We shall see in Sect. \ref{sec:BK} that,
when the Balitsky--Kovchegov equation is derived directly 
from the RGE in Hamiltonian form (\ref{RGEH}) --- where eq.~(\ref{sigchi}) 
has been already used ---, the IR divergences are automatically absent.
But if one rather starts with the original RGE (\ref{RGEA}), 
where both $\eta$ and $\nu$ are still present, then each of these 
terms will individually give rise to IR singularities, which 
will cancel only in the final sum. We expect similar
cancellations to work for all gauge-invariant observables,
but a general proof in this sense is still lacking. This would be
the non-linear generalization of the standard cancellation of
IR divergences between virtual and real corrections.

The second consequence refers to the behaviour of the weight function
$W_\tau[\alpha]$ at large ``times'' $\tau$, which is what
governs the high energy limit of gluon correlations and,
ultimately, of hadron cross-sections. To better appreciate the connection
between eq.~(\ref{sigchi}) and the large time behaviour, it is useful to 
consider first the simpler example of the Brownian motion, 
which is governed too by a diffusion equation similar to eq.~(\ref{RGE}),
but for which an explicit solution can be readily worked out.
Consider thus a particle suspended in a highly viscous liquid,
and in the presence of some external force. The particle performs
a random walk because of its random collisions off the molecules
in the liquid. The relevant quantity 
--- the analog\footnote{This analogy will be developed in more 
detail somewhere else \cite{BIW}.} of $W_\tau[\alpha]$ --- is 
$P(x,t)$ ($\equiv$ the probability density
to find the particle at point $x$ at time $t$), which obeys
the following (Fokker-Planck) equation \cite{ZJ,Parisi}:
\be\label{FPBM}
{\del P(x,t)\over {\del t}}\,=\,D{\del^2\over \del x^i\del x^i}\,P(x,t)\,-\,
{\del\over \del x^i}\Bigl(F^i(x) P(x,t)\Bigr),\ee
where $D$ is the diffusion constant (for simplicity, we assume
this to be truly a constant, i.e., independent of $x$ or $t$),
and $F^i(x)$ is the ``external force'' (actually, the force divided
by the viscosity). 
If $F^i=0$, the solution is immediate, and reads (with the
initial condition $P(x,0)=\delta^{(3)}(x)$):
\be P(x,t)\,=\,{1\over (4\pi Dt)^{3/2}}\,\,
{\rm exp}\left\{-\frac{x^2}{4Dt}\right\}.\ee
This is normalized to unity, as it should, at any $t$
($\int d^3x P(x,t)  =1$), and goes smoothly to zero at any $x$ when
$t\to \infty$ (runaway solution). Thus, for sufficiently large time,
the probability density is quasi-homogeneous (no $x$ dependence),
but it is  everywhere close to zero (to cope with the normalization
condition). The particle simply diffuses in all the available space.

This situation changes, however, if the motion of the particle
is biased by an external force. Assume this force can be derived from
a potential:
\be F^i\,=\,-{\del V\over \del x^i}\,.\ee
Then the time-independent distribution $P_0(x)\sim
{\rm exp}[-\beta V(x)]$ is a stationary solution to eq.~(\ref{FPBM})
provided $\beta D=1$. Of course, this solution is acceptable as a probability
density only if it is normalizable, which puts some constraints on the form 
of the potential. But assuming this to be the case, then $P_0(x)\sim
{\rm e}^{-\beta V}$ represents an equilibrium distribution which is
(asymptotically) reached by the system at large times. Once this is
done, all the correlations become independent of time.

Returning to our quantum evolution, a stationary distribution would 
correspond to a non-trivial fixed point of the RGE (\ref{RGE}) (a solution
${\cal W}_0[\alpha]$ which is normalizable and independent of time).
If such a distribution existed, then the high energy limit of QCD
scattering would be trivial: at sufficiently high energies,
all the cross sections would become independent of the energy.
The relation (\ref{sigchi}) between the coefficients in the RGE
guarantees, however, that this cannot happen: The effective force 
in eq.~(\ref{RGEAnew}) vanishes, and the corresponding evolution
Hamiltonian (\ref{H}) is a purely kinetic operator, which
describes diffusion on the group manifold.
We do not expect any non-trivial fixed
point\footnote{Of course, a constant distribution 
${\cal W}_0=const.$ (no dependence on $\alpha$) would be a stationary
point for the Hamiltonian, but this should scale as $1/V$ with $V=$
the volume of the manifold --- to be normalizable ---, 
and thus would vanish as $V\to \infty$.}
for this Hamiltonian, just runaway solutions, and this is
indeed the case for the approximate solutions to eq.~(\ref{RGEH})
found in Ref. \cite{SAT}.

%% file: S5.tex
\section{Recovering some known equations}
\label{sec:LIMITS}

If $\langle O[\alpha] \,\rangle_\tau$ is any observable which
can be computed as an average over $\alpha$, as in eq.~(\ref{OBSERV}),
then it satisfies an evolution equation given by (\ref{evolO}), namely,
\be\labe{evolOBS}
{\del \over {\del \tau}}\langle O[\alpha] \,\rangle_\tau\,=\,
\left\langle {1 \over 2}\,{\delta \over {\delta
\alpha_\tau^a(x_{\perp})} }\,\eta_{xy}^{ab}\,
{\delta \over {\delta \alpha_\tau^b(y_{\perp})}}\,O[\alpha]\right
\rangle_\tau\,,\ee
where we have also used the RGE in Hamiltonian form, eq.~(\ref{RGEH}),
and we have integrated twice by parts within the
functional integral over $\alpha$. In what follows we shall apply
this equation for two choices of the operator $O[\alpha]$ : 

\noindent
--- The unintegrated gluon distribution function\footnote{This is
 independent of $k^+$ since the electric field
$ {\cal F}^{i+}(\vec x)$ is almost a $\delta$-function in $x^-$;
cf. eq.~(\ref{UTAF}).} (i.e., the integrand of eq.~(\ref{GCL})):
\be\label{varphi}
\varphi_\tau (k^2_{\perp})\,\equiv\,
k^2_{\perp}\,
\Bigl\langle\,|{\cal F}^{i+}_a(k^+,k_\perp)|^2\Bigr\rangle_\tau\,\ee
in the weak field approximation; we shall thus recover the BFKL
equation, as expected.

\noindent
--- The following two-point function of the Wilson lines
(see below for its interpretation):
\be\label{Stau}
S_\tau(x_{\perp},y_{\perp})\,\equiv\,
\Big\langle {\rm tr}\big(V^\dagger(x_{\perp}) V(y_{\perp})\big)
\Big\rangle_\tau,\ee
for which we shall recover the Balitsky--Kovchegov equation
\cite{B,K}.

\subsection{The BFKL limit}
\label{sec:BFKL}

The BFKL equation is obtained in the weak field (and source)
approximation, in which one can expand the Wilson lines in the
Hamiltonian to lowest non-trivial order. For instance, the BFKL
Hamiltonian in the present formalism is obtained by replacing,
in eqs.~(\ref{H}),
\be
1 - V^\dagger_zV_x\,\longrightarrow\,ig\Big(\alpha^a(x_\perp)
- \alpha^a(z_\perp)\Big)T^a,\qquad \alpha^a(x_\perp)
\equiv \int dx^-\, \alpha^a(x^-,x_\perp)\,.\ee
Clearly, this is formally the same as the 
perturbative expansion (i.e., the expansion in powers of $g$) 
of the Hamiltonian to lowest order. After this expansion, the
Hamiltonian takes the generic form
\be\label{HBFKL} H_{BFKL}\,\sim\,\alpha\,{i \delta \over {\delta
\alpha} }\,\alpha\,{i \delta \over {\delta\alpha} }\,,\ee
which is like the second-quantized Hamiltonian for a
non-relativistic many-body problem: In this weak field limit,
the quantum evolution is diagonal in the number of fields,
i.e., it couples only correlators $\langle
\alpha(1) \alpha(2)\cdots\alpha(n)\rangle_\tau$ with the same number $n$
of fields. The BFKL equation is the corresponding evolution equation
for the 2-point function (\ref{varphi}).

To write down this equation, we first observe that, for
such weak fields, the solution to the Yang-Mills equations
(\ref{cleq0}) can be linearized as well:
\be {\cal F}^{+i}_a(k)\,\approx \,
i(k^i/k^2_\perp)\,\rho_a(k),\ee
which implies an approximate expression for the
unintegrated gluon distribution (\ref{varphi}):
\be\label{varphiWF}
\varphi_\tau (k^2_{\perp})\,\approx\,
\Big\langle\rho_a(k_\perp)\rho_a(-k_\perp)\Big\rangle_\tau,\ee
where
\be \rho_a(k_\perp)\,=\,\int d^2x_\perp\,
{\rm e}^{-ik_\perp\cdot x_\perp}\,\rho_a(x_\perp),\qquad
\rho_a(x_\perp)\,\equiv\,\int dx^-\, \rho_a(x^-,x_\perp).\ee
We thus need the equation satisfied by the charge-charge
correlator $\langle\rho\rho\rangle_\tau$, which follows quite
generally from the RGE (\ref{RGE}) in the $\rho$-representation:
\be\labe{RGE2pLIN}
{d\over {d\tau}}\,
\Big\langle\rho_a(x_\perp)\rho_b(y_\perp)\Big\rangle_\tau\,=\,\alpha_s
\Big\langle\sigma_a(x_\perp)\rho_b(y_\perp)
\,+\,\rho_a(x_\perp)\sigma_b(y_\perp)\,+\,
\chi_{ab}(x_\perp,y_\perp)\Bigr\rangle_\tau\,.\ee
For generic, strong, fields and
sources, the r.h.s. of this equation involves $n$-point correlators 
$\langle\rho(1) \rho(2)\cdots\rho(n)\rangle_\tau$ of arbitrarily high
order $n$. But in the weak field limit, where $\sigma$ is linear in
$\rho$ and $\chi$ is quadratic, this becomes a closed equation for
the 2-point function, in agreement with the general argument after
eq.~(\ref{HBFKL}). The coefficients $\chi$ and $\sigma$ in this
limit will be now obtained by expanding the general formulae
(\ref{CHIFINALDEVT}) and (\ref{sigx}) to lowest order in $\rho$.

Consider $\chi^{(0)}$ first. To the order of interest in $\rho$, one
can replace, in eq.~(\ref{CHIFINALDEVT}),
\be
{\cal A}^i_{\infty}(x_{\perp})\,\approx\,-\partial^i\alpha(x_{\perp})
\,\approx\,(\partial^i/\grad_\perp^2)\rho(x_{\perp}),\qquad
{\cal D}^i_{\infty}\,\approx\,\partial^i,\ee
which after simple algebra yields (in matrix notations:
$(\rho_x\rho_y)^{ab}=\rho_x^{ac}\rho_y^{cb}$, etc.) :
\be\label{CHIBFKL}
\chi^{(0)}_{ab}(x_\perp,y_\perp)&=&4
\int {d^2p_\perp \over (2 \pi)^2}\,
\frac{{\rm e}^{ip_{\perp}\cdot(x_{\perp}-y_{\perp})}}
{p_{\perp}^2}\nonumber\\
&{}&\,\,\,\,\times\left\{\rho_x\rho_y-i\rho_x(p^i\partial^i\alpha)_y
+i(p^i\partial^i\alpha)_x\rho_y+ p^2_\perp (\partial^i\alpha)_x
(\partial^i\alpha)_y\right\}^{ab},
\ee
or, after a Fourier transform,
\be\label{chiBFKL}
\chi^{(0)}_{aa}(k_\perp,-k_\perp)\,=\,4N_ck_\perp^2
\int \frac{d^2 p_\perp}{(2\pi)^2}\,\frac{\rho_a(p_\perp)\rho_a(-p_\perp)}
{p^2_{\perp} (k_{\perp}-p_{\perp})^2}\,.\ee
Consider similarly $\sigma^{(0)}$: to linear order in $\rho$,
this can be extracted from any of the expressions (\ref{sigx}),
(\ref{tsigx}), or (\ref{tsigma}), so let us choose the latter
expression, for convenience. To this aim, it is enough to replace,
in eq.~(\ref{tsigma1}),
\be
{\rm Tr}\Bigl(T^a V^\dagger_x V_z\Bigr)\,\approx\,igN_c[\alpha^a(x_\perp)
-\alpha^a(z_\perp)],\ee
as appropriate for weak fields, and thus get, after simple algebra,
\be\label{sigmaq}
\sigma^{(0)}_a (k_\perp)
\,=\,-N_c \,\rho_a(k_\perp)
\int \frac{d^2 p_\perp}{(2\pi)^2}\,{k^2_{\perp} \over p_{\perp}^2 
(p_{\perp}- k_{\perp})^2}\,.\ee

By inserting eqs.~(\ref{chiBFKL}) and (\ref{sigmaq}) into 
the evolution equation (\ref{RGE2pLIN}), and using (\ref{varphiWF}),
one finally obtains:
\begin{eqnarray}
 {\partial \varphi_\tau (k^2_{\perp}) \over \partial \tau} & = & \,\,\,
{\alpha_s N_c \over \pi^2}\,
\int d^2 p_{\perp}
 {k^2_{\perp} \over p^2_{\perp} (k_{\perp}-p_{\perp})^2}\,
 \varphi_\tau(p^2_{\perp}) \nonumber \\
& & -\,
{\alpha_s N_c \over2 \pi^2}\,
\int d^2 p_{\perp}
 {k^2_{\perp} \over p^2_{\perp} (k_{\perp}-p_{\perp})^2}\,
 \varphi_\tau(k^2_{\perp})\,,
\label{BFKL}
\end{eqnarray}
which coincides, as anticipated, with the BFKL equation \cite{BFKL}.
The first term in the r.h.s., which here is generated by 
$\chi^{(0)}$, is the {\it real} BFKL kernel, 
while the second term, coming from $\sigma^{(0)}$,
is the corresponding {\it virtual} kernel.

\subsection{The Balitsky--Kovchegov equation}
\label{sec:BK}

Note first that, although written in a specific gauge --- namely,
the covariant gauge where the color field of the hadron reads
$A^\mu_a=\delta^{\mu+}\alpha_a(\vec x)$ ---, the quantity (\ref{Stau})
has a gauge invariant meaning, as the COV-gauge expression of the
gauge-invariant operator
\be\label{StauP}
S_\tau(x_{\perp},y_{\perp})\,\equiv\,
\Big\langle {\rm tr} \,L (x_{\perp},y_{\perp})\Big\rangle_\tau,
\ee
where $L (x_{\perp},y_{\perp})$ is the Wilson loop obtained by
closing the two Wilson lines along $x^-$ (at $x_{\perp}$ and
$y_{\perp}$, respectively) with arbitrary paths joining
$x_{\perp}$ and $y_{\perp}$ in the transverse planes at $x^-=\infty$
and $x^-=-\infty$.

Physically, $S_\tau(x_{\perp},y_{\perp})$
is the S-matrix element for the scattering of a ``color
dipole'' off the hadron in the eikonal approximation. To clarify this, 
consider DIS in a special frame (the "dipole frame") where most of the
energy is put in the hadron which moves along the positive $z$ direction
---  in this respect, this is like the infinite momentum frame, 
so one can use the CGC effective theory for the hadron wavefunction ---,
but the virtual photon is itself energetic enough for the scattering to
proceed as follows: 
The photon first fluctuates into an energetic 
quark--antiquark pair (a ``color dipole") which then propagates in the
negative $z$ (or positive $x^-$) direction, i.e., towards the hadron,
against which it scatters eventually, with a cross section
(see, e.g., \cite{K,W,Levin} for more details)
\be\label{sigmadipole}
\sigma_{dipole}(\tau,r_\perp)\,=\,2\int d^2b_\perp\,\frac{1}{N_c}
\Big\langle {\rm tr}\Big(1- V^\dagger(x_{\perp}) V(y_{\perp})\Big)
\Big\rangle_\tau.\ee
Here $x_{\perp}$ and $y_{\perp}$ are the transverse coordinates
of the quark and antiquark in the pair,
$r_\perp=x_{\perp}-y_{\perp}$ is the size of the dipole, and
$b_\perp=(x_{\perp}+y_{\perp})/2$ is the impact parameter.
$V^\dagger_x$ and $V_y$ are Wilson lines in the $x^-$ direction,
defined as in eq.~(\ref{v}) but in the fundamental representation.
They describe the eikonal coupling of 
the quark and antiquark to the color field $A^+_a=\alpha_a$ of the hadron. 
The average over the hadronic target in eq.~(\ref{sigmadipole}) 
is here understood as
an average over $\alpha$ in the sense of eq.~(\ref{OBSERV}).
When the scattering energy increases (i.e., $\tau$ increases),
the weight function $W_\tau[\alpha]$ for this average changes due to quantum corrections, and so does the scattering cross section. 
This change is governed by eq.~(\ref{evolOBS}) with 
$O[\alpha]=S_\tau(x_{\perp},y_{\perp})$, that we shall construct
now explicitly.

The final outcome of this calculation, to be detailed in Appendix D,
is an evolution equation for $S_\tau(x_{\perp},y_{\perp})$
that we write down here for a color dipole in some arbitrary 
representation $R$, since this is not more difficult:
\be\labe{evolVR}
{\del \over {\del \tau}}\Big\langle {\rm tr}_R (V^\dagger_x V_y)
\Big\rangle_\tau=-{\alpha_s\over \pi^2}\int d^2z_\perp
\frac{(x_\perp-y_\perp)^2}{(x_\perp-z_\perp)^2(y_\perp-z_\perp)^2 }
\left\langle C_R{\rm tr}_R (V^\dagger_x V_y)-
{\rm tr}_R (V^\dagger_zt^a V_z V^\dagger_xt^aV_y)
\right\rangle_\tau.\nn\ee
In this equation, all the Wilson lines 
$V^\dagger$, $V$, and the color group generators $t^a$, $t^b$
belong to the generic representation $R$, and $C_R=t^a t^a$.

In particular, for the fundamental representation, the
following Fierz identity:
\be\label{Fierz}
t_a^{ij}\,t_a^{kl}\,=\,{1 \over 2}\,\delta^{il}\delta^{jk}\,-\,
{1 \over 2N_c}\, \delta^{ij}\delta^{kl}\,\ee
allows one to simplify the color trace in the last term in 
eq.~(\ref{evolVR}). By also using $C_R=(N^2_c-1)/2N_c$, one
finally obtains (with $R=F$ kept implicit)
\be\labe{evolV}
{\del \over {\del \tau}}\langle {\rm tr}(V^\dagger_x V_y)
\rangle_\tau=-{\alpha_s\over 2 \pi^2}\int d^2z_\perp
\frac{(x_\perp-y_\perp)^2}{(x_\perp-z_\perp)^2(y_\perp-z_\perp)^2 }
\left\langle N_c {\rm tr}(V^\dagger_x V_y)
- {\rm tr}(V^\dagger_x V_z){\rm tr}(V^\dagger_z V_y)\right\rangle_\tau,\ee
which coincides with the equation derived by Balitsky
within a rather different formalism \cite{B} : via the operator
expansion of high-energy scattering in the target rest frame.

The first observation is that the above equations are
not closed. They relate a 2-Wilson-line correlation 
function to a 4-line function, for which one can derive an
evolution equation too \cite{B}, but one may already
guess that this would be not the end of the story: 
The 4-line function will be in turn coupled, via its evolution 
equation, to a 6-line function, and so on, so that one is really
dealing with an infinite hierarchy of coupled equations 
of which eq.~(\ref{evolV}) is just the first equation.

But a closed equation can still be obtained in
the large $N_c$ limit, since in this limit the 4-line correlation function
in eq.~(\ref{evolV}) factorizes:
\be
\left\langle {\rm tr}(V^\dagger_x V_z)\,
{\rm tr}(V^\dagger_z V_y)\right\rangle_\tau
\longrightarrow 
\left\langle {\rm tr}(V^\dagger_x V_z)\right\rangle_\tau\,
\left\langle{\rm tr}
(V^\dagger_z V_y)\right\rangle_\tau\quad {\rm for}\,\,N_c\to\infty,
\ee
and eq.~(\ref{evolV}) reduces to a closed 
equation\footnote{A different closed equation has been recently
obtained in Ref. \cite{KW01} by evaluating the 4-line correlator
in eq.~(\ref{evolV}) with a Gaussian weight function for finite
$N_c$. Note however that the general solution to the RGE
(\ref{RGEH}) is {\it not} a Gaussian \cite{BIW} and, moreover,
eq.~(\ref{evolV}) is not even consistent with a Gaussian, or
``mean field'', approximation to the general RGE \cite{SAT,SATL}.}
for the quantity ${\cal N}(x_\perp,y_\perp)\equiv 
\langle{\rm tr}(1-V^\dagger_x V_y)\rangle_\tau$ ($=$
the forward scattering amplitude of the color dipole 
with the hadron) : 
\be\labe{evolN}
{\del \over {\del \tau}} {\cal N}_{xy}
={\alpha_s\over 2 \pi^2}\int d^2z_\perp
\frac{(x_\perp-y_\perp)^2}{(x_\perp-z_\perp)^2(y_\perp-z_\perp)^2 } 
\left\{ N_c\Bigl({\cal N}_{xz} + {\cal N}_{zy} - {\cal N}_{xy}\Bigr)
- {\cal N}_{xz}{\cal N}_{zy}\right\}.\,\,\,\,\,\ee
This coincides with the evolution equation obtained by Kovchegov \cite{K}
within still a different formalism\footnote{A similar non-linear equation
describing the multiplication of pomerons was suggested in Ref. \cite{GLR}
and proved in \cite{MQ} in the double-log approximation. More recently,
Braun has reobtained this 
equation by resumming ``fan'' diagrams \cite{B00}.}, 
namely the dipole model by  Mueller \cite{AM3}.

The second observation is that the Wilson lines in the adjoint 
representation which were originally present in the Hamiltonian
(\ref{H}) have melted together with the Wilson lines in the generic
representation of the external dipole to generate the color traces
in the r.h.s. of eq.~(\ref{evolVR}). This melting has been made possible
by relations like:
\be\label{melt}
\tilde V^{\dagger}_{ab}t^b\,=\,Vt^aV^{\dagger},\ee
which allows one to trade a group element $\tilde V^{\dagger}$ in
the adjoint representation (that we denote here with a tilde, for more
clarity) for a pair of group elements $V$ and $V^{\dagger}$ in the
generic representation with generators $t^a$, $t^b$.
Note also the simple correspondence between the two terms
within the braces in the r.h.s. of eq.~(\ref{evolVR}) and the four terms
in eq.~(\ref{eta}) for $\eta$ :
The first two terms, $1$ and $\tilde V^{\dagger}_x \tilde V_y$, in eq.~(\ref{eta}) have contributed equally to the first term, involving
$\langle{\rm tr}_R (V^\dagger_x V_y)\rangle$, in eq.~(\ref{evolVR}),
while the other two terms, $\tilde V^{\dagger}_x \tilde V_z$ and
$\tilde V^{\dagger}_z \tilde V_y$, have given identical contribution to 
the remaining piece, involving the 4-line function, in eq.~(\ref{evolVR}).

The third important observation refers to the transverse kernel
in eq.~(\ref{evolVR}), or (\ref{evolV}). A brief comparison shows that
this is not the same as the original kernel
${\cal K}(x_\perp,y_\perp,z_\perp)$, eq.~(\ref{Kxyz}), of the RGE.
Rather, this has been generated as (see the explicit calculation
in Appendix C)
\be\label{decay}
{\cal K}(x_\perp,x_\perp,z_\perp)+
{\cal K}(y_\perp,y_\perp,z_\perp)- 2
{\cal K}(x_\perp,y_\perp,z_\perp)\,=\,
\frac{(x_\perp-y_\perp)^2}{(x_\perp-z_\perp)^2(y_\perp-z_\perp)^2 }\,.\ee
The final result, also known as ``the dipole kernel'' (since
it is proportional to the probability for the decay of a dipole of size
$x_\perp-y_\perp$ into two dipoles of sizes $x_\perp-z_\perp$
and $y_\perp-z_\perp$, respectively), has the remarkable feature to 
show a better IR behaviour than the original kernel (\ref{Kxyz}).
When $z_\perp \gg x_\perp,\,y_\perp$, the r.h.s. of eq.~(\ref{decay})
decreases like $(x_\perp-y_\perp)^2/z_\perp^4$, so its integral
$\int d^2z_\perp$ is actually finite. This guarantees, in particular,
that the evolution equation (\ref{evolN}) in the large $N_c$ limit
is free of IR problems, and it strongly suggest a similar property for
the exact equation (\ref{evolV}), although a rigorous proof
in this latter case would still require a knowledge of 
the large--$z_\perp$ behaviour of the 4-line function
$\langle
{\rm tr}(V^\dagger_x V_z){\rm tr}(V^\dagger_z V_y)\rangle$.
(All of the kernels occuring in the evolution equation for this 4-line 
correlation function are of the dipole type too \cite{B}.)

The dipole kernel (\ref{decay}) has ultraviolet poles 
at $z_\perp=x_\perp$ and $z_\perp=y_\perp$, but these are inocuous,
since the accompanying Wilson-line operators
in eqs.~(\ref{evolVR}) or (\ref{evolV}) cancel each other
at these points.

%% file: S6.tex
\section{Conclusions and perspectives}

In this paper we have completed the construction of the 
renormalization group equation for the Color Glass Condenstate
by explicitly computing the coefficients in this equation,
the one-loop real and virtual
contributions, to all orders in the color glass field.
The resulting expressions appear to be quite simple --- given
especially the complexity of the calculation --- and to exhibit
some remarkable properties with important consequences. We have
started the exploration of these properties and of some of their
consequences, but there is much room left for further studies.
In particular, the deep reason for some of these properties, 
like the functional relation between the real and virtual 
contributions, is not yet fully understood.

Our results turn out to be equivalent, at least as far as the
evolution of Wilson line operators is concerned, with previous
results obtained by Balitsky \cite{B} and (in the
large $N_c$ limit) by Kovchegov \cite{K} within quite different
approaches: the operator expansion of high-energy scattering
amplitudes in Refs. \cite{B}, and the Mueller dipole
model in the case of Ref. \cite{K}. In fact, our RGE (\ref{RGEA})
appears to be the same (up to some change of variables)
as the functional equation written down by Weigert to summarize 
in closed form Balitsky's infinite hierarchy of equations
 This agreement between our respective
results --- in spite of the profound differences between the various
formalisms staying behind them --- gives us even more confidence
that these are indeed the right equations to describe non-linear
evolution at small $x$ in the leading logarithmic approximation.

On the other hand, our results appear to disagree from those
reported in previous attempts \cite{JKW99} to compute the 
coefficients in the RGE. There are several
noticeable differences between our calculations and those in Refs. 
\cite{JKW99} --- chiefly among them, the different constructions
of the background field propagator (especially in relation with
the gauge-fixing prescription),  the different ways to treat
the longitudinal extent of the color source, and the role
of the classical polarization term, which has been missed in
Refs. \cite{JKW99} --- which may explain the different results, 
but a deep understanding of these differences requires more study. 

Now that the RGE is known explicitly, the next objective is, of course,
to try and solve it. Significant progress has been already achieved
in this direction, with interesting consequences. In Ref. \cite{SAT},
some approximate solutions have been obtained in a mean
field approximation in which the coefficients in the RGE have
been replaced by their expectation values, self-consistently 
determined from the ensuing approximation to the weight function.
The results thus obtained distinguish between two regimes:\\
---  A short-distance, or high (transverse) momentum regime
 $k_\perp \gg 1/Q_s(\tau)$, where one recovers the original Ansatz
of the McLerran--Venugopalan model, namely that of a system of
independent color sources whose density increases with $\tau$.
This is a low density regime, where the linear evolution equation is
valid.\\
--- A long-distance, or low momentum ($k_\perp \ll 1/Q_s(\tau)$), regime,
where the system of color charges rather forms a {\it Coulomb gas}
with long-range correlations in the transverse plane. This is a high-density
regime which exhibits {\it gluon saturation}, with a maximal gluon
phase-space density of order $1/\alpha_s$.

The separation scale $Q_s(\tau)$ between these two regimes (the 
``saturation scale'') is estimated as \cite{GLR,MQ,LT99,SAT} 
(with $R\,=$ the hadron radius)
\be
R^2 Q_s^2(\tau)\,\sim\,\alpha_s\,\frac{N_c}{N_c^2-1}\,xG(x,Q_s^2(\tau))
\,\propto\,{\rm e}^{C\alpha_s\tau}\,,
\ee
where
the last estimate (with $C$ a pure number) is based on the linear
evolution equation in the double logarithmic approximation and
with a fixed coupling constant $\alpha_s$. If one rather
uses the one-loop running coupling constant $\alpha_s(Q^2)$ 
with $Q^2=Q_s^2(\tau)$, then the saturation scale rises more slowly, 
as $Q_s^2(\tau)\sim {\rm e}^{\sqrt{b\tau}}$, with $b$ some number.

It is interesting to note that these
predictions of the RGE are in a good agreement with
recent investigations of the Kovchegov equation (\ref{evolN}), for which
various solutions have been obtained via a combination of 
analytical and numerical methods \cite{K,LT99,B00,LL}.
Although the large $N_c$ limit (in which Kovchegov' equation 
becomes applicable) is a priori not the same as the mean field
approximation of Ref. \cite{SAT}, it is nevertheless clear that
these two approximations are closely related, which explains the
similar results. Thus, more effort is required to relax these
approximations, and compute $1/N_c$ corrections. A possibility
to do so is via lattice simulations. The exact solution to the
RGE (\ref{RGEH}) can be given a path integral representation,
to be presented somewhere else \cite{BIW}. Another path integral
has been recently proposed by Balitsky \cite{Balitsky2001},
and provides an exact solution to his coupled equations. All
these path integrals are those of a quantum field theory in three
Euclidean dimensions [two transverse coordinates plus 
``imaginary time'' $\equiv\tau$ (rapidity)], and thus are well
suited for lattice calculations.

A better control of the solutions to the RGE would allow one to
extract its phenomenological consequences, notably in relation with
high-energy deep inelastic scattering and ultrarelativistic heavy
ion collisions. Possible signatures of saturation in DIS
and heavy ion collisions have been already discussed in the literature
\cite{GBW99,KV00,KOV00,LM00,KN01,MSB01,MD01,KKL01,GP01,KL01,Larry01}.
It would be interesting to see how these and related predictions
are affected by the details of the effective theory at small $x$.

\bigskip
\bigskip
\bigskip
{\large{\bf Acknowledgements}}

The authors gratefully acknowledge conversations with 
Ian Balitsky, Jean-Paul Blaizot, Kazu Itakura, Al Mueller, 
Eugene Levin, Yuri Kovchegov, Raju Venugopalan, and Heribert Weigert.

This manuscript has been authorized under Contract No. DE-AC02-98H10886 
with the U. S. Department of Energy.

%% file: A1.tex
\section{The background field gluon propagator}

The propagator $ G^{\mu\nu}_{ab}(x,y)$
of the semi-fast gluons in the background
of the tree-level field (\ref{Atilde}) and in the LC-gauge
($a^+_c={\cal A}^+_c=0$) has been obtained
in Sect. 6 of Paper I (see also Refs. \cite{AJMV95,HW98})
via the following three steps:
{\it i)} First one has constructed the background field 
propagator $G(x,y)$ of a {\it scalar} field. {\it ii)} Then, 
one has computed the gluon propagator $\acute G^{\mu\nu}_{ab}(x,y)$
for a quantum gluon in the {\it temporal} gauge $\acute a^-_c=0$ 
(but with the background field ${\cal A}^\mu_c=\delta^{\mu i}
{\cal A}^i_c$ still in the LC-gauge). {\it iii)} Finally,
the  LC-gauge propagator $ G^{\mu\nu}_{ab}$
has been obtained from $\acute G^{\mu\nu}_{ab}$ via
a gauge rotation. 

Below, we shall present the final results of these calculations.
For convenience, we start with the free propagator.

\subsection{The free LC-gauge propagator}

Consider first the free propagator (no background field)
$G^{\mu\nu}_{0\,ab}=\delta_{ab}G^{\mu\nu}_0$.
With the retarded prescription
advocated in Sect. 1.3, this reads, in momentum space:
\be\label{LCPROP}
G_0^{i-}(p)&=&{p^i\over p^++i\epsilon}\,G_0(p),\quad
G_0^{-i}(p)={p^i\over p^+-i\epsilon}\,G_0(p),\nn
G_0^{ij}(p)&=&\delta^{ij}G_0(p),\qquad
G_0^{--}(p)=\,{\rm PV}\, {2p^-\over p^+}\,G_0(p)\,,\ee
where $G_0(p)=1/(2p^+p^--p_\perp^2+i\epsilon)$ [the free
propagator of a massless scalar field], and PV denotes
the principal value prescription:\be
{\rm PV}\ {1 \over p^+} \equiv {1 \over 2}
\left ( 
{1 \over p^+\ -i \varepsilon} + {1 \over p^+\ +i \varepsilon}
\right ) \,.
\labe{PV}\ee
Note that, strictly speaking, the gauge-fixing prescription 
in eq.~(\ref{LCPROP}) is ``retarded''
only as far as the component $G^{i-}_0$ is concerned; by hermiticity,
the corresponding prescription in $G_0^{-i}$ is
advanced, while in $G_0^{--}$ it is a principal value.

Other prescriptions that are referred to in the main text are 
the ``advanced'' prescription (see, e.g., Refs. \cite{MQ,K96,KM98}) 
which is obtained by changing the sign of $i\epsilon$ for the axial 
poles in eq.~(\ref{LCPROP}) [that is, this is advanced for the
component $G^{i-}_0$], and the
PV-prescription, for which $1/p^+ \equiv {\rm PV}(1/p^+)$
in all the components of the propagator.

For further reference, we note also the expression of the scalar
propagator in the $x^-$ representation:
\be\label{scprop0}
         G_0(x^-,p^-,{ p}_\perp)&=&
\int {dp^+ \over 2 \pi}\  e^{-ip^+x^-}\,
{1 \over {2p^+p^--p_\perp^2 + i\epsilon}}\,\nn
&=& - {i\over {2p^-}}\, \left\{ \theta (x^-) \theta (p^-)
- \theta (-x^-) \theta (-p^-) \right\}
{\rm e}^{ -i\frac{p_{\perp}^2}{2p^-}x^-}\,.\ee

\subsection{ The background field propagator of a scalar field}

Since the background fields are static, and thus we have
homogeneity in time, it is convenient to work in the
$p^-$--representation, that is, to construct the propagator
$G(\vec x,\vec y, p^-)$ for a given $p^-$. This also makes it
easy to implement the strip restriction (\ref{strip-}).
The propagator is known only for the
discontinuous background field
${\cal A}^i(\vec x)=\theta(x^-){i\over g}V\del^i V^\dagger$
in eq.~(\ref{APM}),
which is how the actual field of eq.~(\ref{Atilde}) is ``seen''
by the semi-fast gluons\footnote{With $p^-$ constrained 
as in eq.~(\ref{strip-}), the typical momentum scale
$p^+\sim p^2_\perp/p^-$ for longitudinal dynamics is
relatively small, $p^+\ll \Lambda^+$, so that the 
semi-fast gluons are unable to discriminate the internal 
structure of the ``$\theta$''-function in eq.~(\ref{APM}).}.
The final result for $G(\vec x,\vec y, p^-)$ can be written as:
\be\label{GSCALAR}
G(\vec x,\vec y, p^-)&=&G_0(\vec x-\vec y, p^-)
\Bigl\{\theta (x^-)\theta (y^-)V(x_\perp)V^\dagger (y_\perp)
 + \theta (-x^-)\theta (-y^-)\Bigr\}\nn
&+& 2ip^-\int d^3\vec z\,\, G_0(\vec x-\vec z, p^-)
\, \delta (z^-)\,G_0(\vec z-\vec y, p^-) \nn
&{}&\times\,\left\{ \theta (x^-)\theta (-y^-)
V(x_\perp)V^\dagger(z_\perp) - \theta (-x^-) \theta(y^-)
V(z_\perp)V^\dagger (y_\perp)\right\}.\ee
It can be easily verified that this function is
continuous at both $x^- = 0$ and $y^- = 0$.

On eq.~(\ref{GSCALAR}), we distinguish two type of contributions, 
``crossing'' and ``noncrossing'', corresponding to trajectories which
cross, or do not cross, the plane at $x^-=0$ where is
located the singularity of the background field (or
its source). Consider ``noncrossing'' trajectories first, for
which $x^-$ and $y^-$ are of the same sign:
When they are both negative, the propagation takes place
 in a domain where the field vanishes; this is therefore
free propagation. When $x^-$ and $y^-$ are both positive,
the gluon propagates in a background field which is just a gauge
rotation, ${\cal A}^i={i\over g}V\del^i V^\dagger\,$;
thus, the net effect comes from the difference
between the gauge rotations at the end points. If the trajectory
crosses the discontinuity at  $x^-=0$, say in going from $y^-<0$
to $x^->0$, then there is also a gauge factor $V^\dagger(z_\perp)$
associated with this crossing (at some arbitrary $z_\perp$).

We thus write, with obvious notations, 
$G = G^{(n)} + G^{(c)}$. In the crossing piece $G^{(c)}$, it is
further possible to replace:
\be\label{xyp}
\theta (x^-)\theta (-y^-)\longrightarrow \theta (p^-),\qquad
\theta (-x^-) \theta(y^-)\longrightarrow \theta (-p^-),\ee
because of the correlation between the sign of 
$p^-$ and the direction of propagation in $x^-$, as manifest
on eq.~(\ref{scprop0}).

\subsection{ The gluon propagator in the temporal gauge}

This has the following non-zero components ($G\,=\,$ the scalar
propagator in  eq.~(\ref{GSCALAR})):
\be\label{acuteG}
\acute G^{ij}(\vec x,\vec y, p^-)&=&\delta^{ij} G(\vec x,\vec y, p^-),\nn
\acute G^{+i}(\vec x,\vec y, p^-)&=&{i \over p^-}\, {\cal D}_x^i 
G(\vec x,\vec y, p^-),\qquad
\acute G^{i+} (\vec x,\vec y, p^-)\,=\,-\,{i \over p^-}\,
G(\vec x,\vec y, p^-)\,{\cal D}^{\dagger \,j}_y\,,
\nonumber\\
\acute G^{++}(\vec x,\vec y, p^-)&=&{1 \over (p^-)^2}\,\left\{{\cal D}_x^i
G(\vec x,\vec y, p^-){\cal D}^{\dagger \,i}_y\,+\,\delta^{(3)}(\vec x-\vec y)
\right\}\quad.
\ee
Note that, because of the strip restriction (\ref{strip-}) on $p^-$,
the operator $1/p^-$ is never singular.

In the above equations, ${\cal D}^i\equiv\partial^i -ig{\cal A}^i$ and
${\cal D}^{\dagger j}=\partial^{\dagger j} +ig{\cal A}^j$
with the derivative $\partial^{\dagger j}$ acting on the function on its left.
By using the Wilson lines in eq.~(\ref{GSCALAR}), it is possible to
convert these covariant derivatives into ordinary derivatives.
This relies on the following identities, valid
for any function $O(x)$ (cf. eq.~(\ref{Atilde})) :
\be
{\cal D}^i \Bigl[U(\vec x)O(x)\Bigr]\,=\,U(\vec x)\,\partial^i O(x),
\quad \Bigl[O(x)U^{\dagger}(\vec x)\Bigr]{\cal D}^{\dagger \,i}\,=\,
\Bigl(\partial^i O(x)\Bigr)U^{\dagger}(\vec x).\ee
This gives the following expressions for the crossing and 
non-crossing pieces of, e.g., $\acute G^{++}$:
\be\label{G++c}
{\acute G}^{++(c)}(\vec x,\vec y; p^-)&=&
{2i \over p^-} \int d^2z_{\perp}\
\partial^i_x G_0(x^-,x_{\perp}-z_{\perp})\,
\partial^i_y G_0(-y^-,z_{\perp}-y_{\perp})\nonumber\\
&{}&\,\times\,\left \{
\theta(p^-) V(x_\perp)V^\dagger(z_\perp) -
\theta(-p^-) V(z_\perp)V^\dagger (y_\perp)
\right \},
\ee
and
\be\label{G++n}
{\acute G}^{++(n)}(\vec x,\vec y; p^-)&=&
\Bigl\{\theta(x^-)\theta(y^-) V(x_\perp)V^\dagger (y_\perp)
+\theta(-x^-)\theta(-y^-)\Bigr\}
\nonumber\\&{}&\,\,\times\,{1 \over (p^-)^2}\,
\partial^i_x\partial^i_y G_0(\vec x-\vec y; p^-)\,
+\,\delta^{(3)}(\vec x-\vec y){1 \over (p^-)^2}\, .
\ee
In the crossing piece we have also performed the replacement 
(\ref{xyp}). 

In the zero field limit, the above equations yield the free
temporal-gauge propagator in the expected form:
\be\label{TGPROP}
{\acute G}_0^{ij}(p)=\delta^{ij}G_0(p),\quad
{\acute G}_0^{i+}(p)={\acute G}_0^{+i}(p)={p^i\over p^-}\,G_0(p),
\quad{\acute G}_0^{++}(p)=\,{2p^+\over p^-}\,G_0(p)\,,\ee
vwhere $G_0(p)=1/(2p^+p^--p_\perp^2+i\epsilon)$.

\subsection{ The gluon propagator in the LC gauge}

This is related to the temporal-gauge propagator presented
above via the relation:
\be\label{GLC}
iG^{\mu\nu}_{bc}(x,y)\,\equiv\,
\langle {\rm T}\,a_b^\mu(x) a_c^\nu(y)\rangle\,=\,
\left\langle {\rm T}\Bigl(
\acute a^\mu-{\cal D}^\mu{1 \over {\partial^+}} \,\acute a^+ \Bigr)_x^b
\Bigl(\acute a^\nu-{\cal D}^\nu{1 \over {\partial^+}} \,\acute a^+
\Bigr)_y^c\right\rangle,\ee
or, more explicitly,
\begin{mathletters}
\be\label{Gij}
G^{ij} & = &
\acute G^{ij} - {\cal D}^i {1 \over \partial^+}\, \acute G^{+j}
+ \acute G^{i+} {1 \over \partial^+} {\cal D}^{\dagger j}-
{\cal D}^i {1 \over \partial^+}\,{\acute G}^{++} 
{1 \over \partial^+} {\cal D}^{\dagger j}\,,\\\label{G-i}
G^{-i} &= &- \,{\partial^- \over \partial^+}\, \acute G^{+i} -
{\partial^- \over \partial^+} {\acute G}^{++}
 {1 \over \partial^+} {\cal D}^{\dagger i}\,,\\
\label{Gi-}
G^{i-} & =&   \,\acute G^{i+}\, {\partial^- \over \partial^+} - \,
{\cal D}^i \,{1 \over \partial^+}\, {\acute G}^{++} 
{\partial^- \over \partial^+ }\,,\\
\label{G--}
G^{--} & = &- \,{\partial^- \over \partial^+}
{\acute G}^{++} {\partial^- \over \partial^+}\,,
\ee
\label{LCG}
\end{mathletters}
where we use the convention that the 
derivatives written on the right act on the functions on their left;
e.g., $\partial^- F \,\partial^-\equiv \partial^-_x\partial^-_y F(x,y)$.

The axial poles at $p^+=0$ in the expressions above are
regularized according to the ``retarded'' prescription 
in eq.~(\ref{LCPROP}). This implies that the operators $1/\partial^+$
written on the left (right) of $\acute G^{\mu\nu}$ in eq.~(\ref{LCG})
are regularized with a retarded (advanced) prescription.
For instance,
\be \label{delRA}
{1 \over \partial^+}\,{\acute G}^{++} 
{1 \over \partial^+}\,\equiv \,{1 \over \partial^+_R}\,{\acute G}^{++} 
{1 \over \partial^+_A}\,,\ee
where
\begin{mathletters}
\be\label{delret}
\langle x^-|{1 \over {i\partial^+_R}}|y^-\rangle&\equiv&
\int {dp^+\over 2\pi} \,\frac{{\rm e}^{ip^+(x^--y^-)}}
{p^++i\epsilon}\,=\,-i\theta(x^--y^-)\,,\\
\langle x^-|{1 \over {i\partial^+_A}}|y^-\rangle&\equiv&\,
\int {dp^+\over 2\pi} \,\frac{{\rm e}^{ip^+(x^--y^-)}}
{p^+-i\epsilon}\,=\,i\theta(y^--x^-)\,.\label{deladv}
\ee\label{+RA}
\end{mathletters}
In going from eq.~(\ref{GLC}) to eq.~(\ref{LCG}),
we have also used :
\be \langle x^-|{1 \over {i\partial^+_A}}|y^-\rangle\,=\,-
\langle y^-|{1 \over {i\partial^+_R}}|x^-\rangle\,.\ee

%% file: A2.tex
\section{Alternative calculation of ${\hat \chi}_3$}

In this Appendix we shall verify that a complete calculation
of ${\hat \chi}_3$ using the full temporal gauge propagator 
${\acute G}^{++}$ leads to the same result
(\ref{chi3final}) as the simplified calculation with
${\acute G}^{++}$ replaced by its free counterpart ${\acute G}^{++}_0$. 

According to its definition in eq.~(\ref{CHIFIN}), ${\acute G}^{++}$ 
involves the following  matrix element (the equal-time limit
$x^+=y^+$ is understood)
\be
{\cal M}(x_{\perp},y_{\perp})\,\equiv\,
\langle x^-=0,x_{\perp}
| {1 \over i \partial^+_R} {\acute G}^{++} 
{1 \over i \partial^+_A}| y^-=0,y_{\perp}
\rangle \,\equiv {\cal M}^{(c)}\,+\,{\cal M}^{(n)}\,,
\label{matrizn1}\ee
with both crossing (${\cal M}^{(c)}$) and non-crossing (${\cal M}^{(n)}$)
contributions. The crossing piece is evaluated with the propagator
${\acute G}^{++(c)}$ in eq.~(\ref{G++c}), which yields
\be
{\cal M}^{(c)}&=&
\int_{strip} {dp^-\over 2\pi}\int {dp^+ \over 2 \pi}
\int {dk^+ \over 2 \pi}\,\frac{1}{p^++i\epsilon}\,\frac{1}{k^+-i\epsilon}
\,{\acute G}^{++(c)}(p^+,k^+;x_{\perp},y_{\perp}; p^-)\nn
&=&\int_{strip} {dp^-\over 2\pi}\,\frac{2i}{p^-}\,
\int {dp^+ \over 2 \pi}
\int {dk^+ \over 2 \pi}\,\frac{1}{p^++i\epsilon}\,\frac{1}{k^+-i\epsilon}\,
\int d \Gamma_{\perp}\,(p_{\perp} \cdot k_{\perp})\nn
&{}&\,\,G_0(p)G_0(k)\left \{
\theta(p^-) V(x_\perp)V^\dagger(z_\perp) -
\theta(-p^-) V(z_\perp)V^\dagger (y_\perp)
\right \}.
\ee
(We have also used here the notation (\ref{dGamma}) for the integrals
over the transverse phase-space.) This vanishes since
the integrals over $p^+$ and $k^+$ generate mutually excluding
$\theta$ functions :
\be
\int {dp^+ \over 2 \pi} {1\over p^++i\epsilon}\,G_0(p)\,=\,
{-i \over p_{\perp}^2}\,\theta(-p^-)\,,\qquad
\int {dk^+ \over 2 \pi} {1\over k^+-i\epsilon}\,G_0(k)\,=\,
{i \over k_{\perp}^2}\,\theta(p^-)\,.\ee
Thus, ${\cal M}^{(c)}=0$, in agreement with the arguments
following eq.~(\ref{matriz3}).

As for the non-crossing contribution, this is most
easily evaluated as (cf. eqs.~(\ref{matriz3}) and (\ref{G++n}))
\be\label{matrizn2}
{\cal M}^{(n)}
= \int\, dz^-_1 \int\, dz^-_2 \theta(-z^-_1) \theta(-z^-_2)
{\acute G}^{++(n)}(z_1^-,x_{\perp},z_2^-,y_{\perp})
\nonumber\\
= \int\, dz^-_1 \int\, dz^-_2 \theta(-z^-_1) \theta(-z^-_2)
\int {d p^- \over 2 \pi}\ {1 \over (p^-)^2}
\int {dp^+ \over 2 \pi}\  {\rm e}^{ip^+ (z^-_1 - z^-_2)}
\nonumber\\
\times\
\int {d^2p_\perp \over (2 \pi)^2}\
{\rm e}^{ip_{\perp}\cdot(x_{\perp}-y_{\perp})}\ \Big(p_{\perp}^2 \ G_0(p)
+1\Big)
\nonumber\\
= \int\, dz^-_1 \int\, dz^-_2 \theta(-z^-_1) \theta(-z^-_2)
\int {d p^- \over 2 \pi}\ {2 \over p^-}
\int {dp^+ \over 2 \pi}\  p^+ {\rm e}^{ip^+ (z^-_1 - z^-_2)}
\nonumber\\
\times\
\int {d^2p_\perp \over (2 \pi)^2}\
{\rm e}^{ip_{\perp}\cdot(x_{\perp}-y_{\perp})}\ G_0(p)
\ee
The integrations over $z_{1,2}^-$ are performed using
\be
\int\, dz^- \theta(-z^-) {\rm e}^{\mp i p^+ z^-}
\, = \, {\pm i \over  p^+ \mp i \epsilon}
\ee
which gives 
\be
\label{matrizn3}
{\cal M}^{(n)}=
\int {d p^- \over 2 \pi}\
\int {dp^+ \over 2 \pi}\
\int {d^2p_\perp \over (2 \pi)^2}\
{\rm e}^{ip_{\perp}\cdot(x_{\perp}-y_{\perp})}\
{1 \over p^+ + i \epsilon}\ {2 p^+ \over p^-}\ {1 \over p^+ - i
\epsilon}\
G_0(p).
\ee
This is the same expression (\ref{A00})
as obtained by using directly the free propagator
${\acute G}^{++}_0$. From now on, the calculation proceeds
as in Sect. \ref{sect:comp_chi} and leads
to eq.~(\ref{chi3final}) for ${\hat \chi}_3$, as anticipated.

%% file: A3.tex
\section{More details on the virtual correction}

In this Appendix we collect some calculations
pertinent to the virtual correction $\hat\sigma(\vec x)$
which have not been included in the main text. 

{\bf (a) Computing $\hat\sigma_2$}

We first compute the second contribution $\hat\sigma_2$ to
eq.~(\ref{sigma}) for the induced source and show that this
is indeed a pure tadpole, as stated in Sect. \ref{sec:sigmaA}.
This is the quantum expectation value of the second piece
in eq.~(\ref{rho2}) for $\delta \rho_a^{(2)}$, that is:
\be\label{sigma2}
\hat\sigma^a_2(\vec x)&=&-i\,\frac{g^2}{N_c}
  \int dy^+ \int dz^+ \,
{\cal M}^{abcd}(x^+,y^+,z^+)\,\rho_{b}({\vec x})\,
G^{--}_{cd}(y^+,{\vec x}\,;z^+,{\vec x}),\nn
{\cal M}^{abcd}(x^+,y^+,z^+)&\equiv&
\theta (x^+ -z^+) \theta (z^+ -y^+) {\rm Tr} (T^a T^b T^c T^d)+
\theta (z^+ -y^+)\theta (y^+ -x^+) {\rm Tr} (T^a T^c T^d T^b)\nn
&{}&+\,
\theta (z^+ -x^+) \theta (x^+ -y^+) {\rm Tr}(T^a T^d T^b T^c).\ee
By using eq.~(\ref{G--}) for $G^{--}$ together with
time homogeneity, one can write:
\be\label{G--sig2}
G^{--}_{cd}(y^+,{\vec x}\,;z^+,{\vec x})&=& \int
{dp^- \over 2 \pi}\, {\rm e}^{-ip^-(y^+ -z^+)}\,
G^{--}_{cd}({\vec x},{\vec x};p^-)\nn
&=& \int{dp^- \over 2 \pi}\, {\rm e}^{-ip^-(y^+ -z^+)}\,(p^-)^2\,
\langle \vec x | {1 \over i \partial^+_R} {\acute G}^{++}_{cd}(p^-)\, 
{1 \over i \partial^+_A}| \vec x \rangle.\ee
In eq.~(\ref{sigma2}), this is multiplied by $\rho_{b}({\vec x})$,
so one can replace $x^-\simeq 0$.
Then (\ref{G--sig2}) is essentially the same matrix element
as in eq.~(\ref{matriz3}), except that this is now evaluated for
arbitrary time variables $y^+$ and $z^+$. But the arguments in 
Sect. \ref{chinon}, according to which this matrix element can be
equivalently evaluated with the {\it free} propagator 
$G^{--}_{0\,cd}=\delta_{cd}G^{--}_0$, are still valid, since
they rely just on the longitudinal structure of the propagator
together with our axial prescriptions.
Thus, for $x^-=0$ eq.~(\ref{G--sig2}) is the same as
(cf. eqs.~(\ref{A00})--(\ref{intp+1}))
\be\label{G--0}
\delta_{cd}G^{--}_0(y^+,0^-,x_\perp\,;z^+,0^-,x_\perp)
\,=\,-i\delta_{cd}
\int_{strip}{dp^- \over 2 \pi}\,{\rm e}^{-ip^-(y^+ -z^+)}\,
p^-\epsilon(p^-) \int {d^2p_\perp \over (2 \pi)^2}\,
{1 \over p_{\perp}^2}\,,\ee
which is the same matrix element that would enter at
BFKL level. Because of that, also the final result for $\hat\sigma_2$
is the same as at BFKL level, and is therefore well known \cite{JKLW97}.
For completeness, let us nevertheless finish its calculation.

With the trivial colour structure of eq.~(\ref{G--0}),
the colour traces in eq.~(\ref{sigma2}) become straightforward:
\be\label{Mab}
{\cal M}^{abcc}(x^+,y^+,z^+)&=&\Big\{
\theta (x^+ -z^+) \theta (z^+ -y^+) +
\theta (z^+ -y^+)\theta (y^+ -x^+)\Big\} {\rm Tr} (T^a T^b T^c T^c)+\nn
&{}&+\,\theta (z^+ -x^+) \theta (x^+ -y^+) {\rm Tr}(T^a T^c T^b T^c)\nn
&=&\theta (z^+ -x^+) \theta (x^+ -y^+) {\rm Tr}(T^a[T^c, T^b] T^c)
+ \theta (z^+ -y^+){\rm Tr} (T^a T^b T^c T^c)\nn
&=& -\delta^{ab}\frac{N_c^2}{2}\theta (z^+ -x^+) \theta (x^+ -y^+)
\,+\,\delta^{ab}N_c^2\theta (z^+ -y^+).\ee
In going from the first to the second line above, we have used
the identity:
\be
\theta (x^+ -z^+) \theta (z^+ -y^+) +
\theta (z^+ -y^+)\theta (y^+ -x^+)=-\theta (z^+ -x^+) \theta (x^+ -y^+)
+\theta (z^+ -y^+)\,.\ee
To obtain $\hat\sigma_2$, one still has to perform the integrations
over $y^+$, $z^+$ and $p^-$. After these operations, we get a non-vanishing
contribution only from the first term (the term
 involving two $\theta$-functions) in the last line of
eq.~(\ref{Mab}). This involves:
\be
\int dy^+ \int dz^+ \,\theta (z^+ -x^+) \theta (x^+ -y^+)\,
{\rm e}^{-ip^-(y^+ -z^+)}\,=\,\frac{-1}{(p^-)^2}\,,\ee
where the prescription at $p^-=0$ is irrelevant since $p^-$
is anyway restricted to the strip (\ref{strip-}). This restricted
integration over $p^-$ then generates the 
 expected logarithmic enhancement, cf. eq.~(\ref{LOGX}),
so that the final result reads:
\be\label{sigma2f}
\hat\sigma^a_2(\vec x)\,=\,-\frac{g^2N_c}{2\pi}\,\ln \frac{1}{b}\,
\rho^a(\vec x)\int {d^2p_\perp \over (2 \pi)^2}\,
{1 \over p_{\perp}^2}\,,\ee
which is linear in $\rho$, and a pure tadpole, as anticipated. 

{\bf (b) The ``non-crossing'' contribution to $\hat\sigma_1$ vanishes}

Let us now check that the ``non-crossing'' contribution to
eq.~(\ref{sigma120}) for $\hat\sigma_1$ vanishes indeed.
This involves the propagator ${\acute G}^{++(n)}$
of eq.~(\ref{G++n}), and is therefore proportional to
\be
{\hat \sigma}_{12}^{(n)} &\propto &
{\cal D}^i_x  \int dz^-\,
\theta(x^- - z^-)\ {\acute G}^{++(n)} (z^-,x^-;x_\perp,y_\perp)\,
{\cal D}^{\dagger i}_y \Big |_{x=y}\\
 &\propto &\int dz^-\,
\theta(x^- - z^-) {\cal D}^i_x \Bigl\{
\theta(z^-)\theta(x^-) V_x V^\dagger_y
+\theta(-z^-)\theta(-x^-)\Bigr\}\Bigl(
\partial^i_x\partial^i_y G_0\Bigr)\,
{\cal D}^{\dagger i}_y \Big |_{x=y}\nn
&=&\int dz^-\, \theta(x^- - z^-)\Bigl\{
\theta(z^-)\theta(x^-) + +\theta(-z^-)\theta(-x^-)\Bigr\}
\grad_\perp^4  G_0(z^- -x^-,x_\perp-y_\perp)
\Big |_{x_\perp=y_\perp},\nonumber\ee
which is trivial in color ($\propto \delta^{ab}$), and
therefore gives no contribution after taking the color trace
with $T^a$ (cf. eq.~(\ref{sigma100})). In going from the second
to the third line in the equation
above, we have used (cf. eq.~(\ref{APM}))
\be\label{calD}
 {\cal D}^i_x\,=\,\theta(-x^-)\partial^i_x + \theta(x^-)
 {\cal D}^i_\infty,\qquad  {\cal D}^i_\infty =
\partial^i -ig {\cal A}^i_{\infty}(x_{\perp}) =
\partial^i + V\partial^i V^\dagger,\ee
to deduce that, for any function $O(x_\perp)$,
\be\label{DVO}
{\cal D}^i_\infty (VO)\,=\,V\partial^i O,
\qquad (OV^{\dagger}){\cal D}^{\dagger \,i}_\infty\,=\,
(\partial^i O)V^{\dagger}.\ee
This has allowed us to convert the covariant derivatives
into ordinary derivatives
acting on $G_0\,$. Finally, in the local limit
$x_\perp\to y_\perp$,  $V_x V^\dagger_y\to 1$,
and all the color structure has disappeared.

%% file: A4.tex
\section{Deriving the Balitsky--Kovchegov equation}

We present here some of the calculations leading
to eq.~(\ref{evolV}) in the main text. The starting point is
eq.~(\ref{evolOBS}) with $O[\alpha]=S_\tau(x_{\perp},y_{\perp})$
given by eq.~(\ref{Stau}). This gives (the colour group
representation $R$ is arbitrary):
\be\labe{evolS0}
{\del \over {\del \tau}}\Big\langle {\rm tr}_R (V^\dagger_x V_y)
\Big\rangle_\tau\,=\,
{1 \over 2}
\int d^2u_\perp\int d^2v_\perp\,\left\langle
{\delta \over {\delta
\alpha_\tau^a(u_{\perp})} }\,\eta_{uv}^{ab}\,
{\delta \over {\delta \alpha_\tau^b(v_{\perp})}}\,
{\rm tr}_R (V^\dagger_x V_y)\right\rangle_\tau\,\ee
where (cf. eqs.~(\ref{eta}) and (\ref{Kxyz})) : 
\be\label{etauv}
\eta^{ab}_{uv}
&=&{1\over \pi}\int {d^2z_\perp\over (2\pi)^2}\,
{\cal K}(u_\perp,v_\perp,z_\perp)
\Bigl\{1+ \tilde V^\dagger_u \tilde V_v-\tilde V^\dagger_u \tilde V_z -
 \tilde V^\dagger_z \tilde V_v\Bigr\}^{ab}
\,.\ee
The first derivative in eq.~(\ref{evolS0}) is immediately
performed with the help of eq.~(\ref{DIFFU}) (with ${\rm tr}
\equiv {\rm tr}_R$)
\be\labe{diff1}
{\delta \over {\delta \alpha_\tau^b(v_{\perp})}}\,
{\rm tr} (V^\dagger_x V_y)\,=\,
ig\,{\rm tr} (t^bV^\dagger_x V_y)(\delta_{xv}-\delta_{yv}),\ee
where $\delta_{xv}\equiv \delta^{(2)}(x_{\perp}-v_\perp)$.
To perform similarly the second derivative in (\ref{evolS0}), 
we consider separately the contributions of the four
terms within $\eta^{ab}_{uv}$, eq.~(\ref{etauv}).

The first term, involving the unit matrix $\delta^{ab}$, is
immediate:
\be\labe{diff21}
{\delta \over {\delta
\alpha_\tau^a(u_{\perp})} }\,\delta^{ab}\,
{\delta \over {\delta \alpha_\tau^b(v_{\perp})}}\,
{\rm tr} (V^\dagger_x V_y)\,=\,-g^2C\,
{\rm tr} (V^\dagger_x V_y)\,(\delta_{xv}-\delta_{yv})\,
(\delta_{xu}-\delta_{yu})\,.\ee
We have used here $t^at^a=C_R$ together with the cyclic
invariance of the trace.

For the second term,  involving 
$(\tilde V^\dagger_u \tilde V_v)^{ab}$, we use the
identity to (\ref{melt}) to first write:
\be
(\tilde V^\dagger_u \tilde V_v)^{ab}\,{\rm tr} (t^bV^\dagger_x V_y)
\,=\,(\tilde V^\dagger_u)^{ac}\,
{\rm tr} (V^\dagger_v t^c V_v V^\dagger_x V_y)
\,=\,(\tilde V^\dagger_u)^{ac}{\rm tr} (V^\dagger_xt^c  V_y),\ee
where the second identity holds in the presence of the
factor $(\delta_{xv}-\delta_{yv})$. The first factor 
$(\tilde V^\dagger_u)^{ac}$ in this equation commutes with the derivative
$\delta/{\delta \alpha_\tau^a(u_{\perp})}$, since $(T^a)_{ab}=0$
(this is similar to eq.~(\ref{anti})). Thus, the contribution of the
second term in eq.~(\ref{etauv}) can be evaluated as:
\be\labe{diff22}
ig\,(\delta_{xv}-\delta_{yv})\,(\tilde V^\dagger_u)^{ac}\,
{\delta \over {\delta
\alpha_\tau^a(u_{\perp})} }\,{\rm tr} (V^\dagger_xt^c  V_y)
&=&-g^2\,(\delta_{xv}-\delta_{yv})\,(\delta_{xu}-\delta_{yu})\,
(\tilde V^\dagger_u)^{ac}\,{\rm tr} (t^aV^\dagger_xt^c  V_y)\nn
&=&-g^2C_R\,
{\rm tr} (V^\dagger_x V_y)\,(\delta_{xv}-\delta_{yv})\,
(\delta_{xu}-\delta_{yu})\,,\ee
where in writing the second line we have used eq.~(\ref{melt})
once again, together with the $\delta$-functions at $u=x$ and
$u=y$. Remarkably, the final result above is exactly the same
as the first contribution in eq.~(\ref{diff21}).

Consider similarly the third term
$(-\tilde V^\dagger_u \tilde V_z)^{ab}$ in eq.~(\ref{etauv}).
It is easy to check that this commute with the derivative
$\delta/{\delta \alpha_\tau^a(u_{\perp})}$ (as in eq.~(\ref{anti})),
so its contribution reads:
\be\labe{diff23}
{\delta \over {\delta
\alpha_\tau^a(u_{\perp})} }(-\tilde V^\dagger_u \tilde V_z)^{ab}
{\delta \over {\delta \alpha_\tau^b(v_{\perp})}}
{\rm tr} (V^\dagger_x V_y)&=&-ig\,(\delta_{xv}-\delta_{yv})\,
(\tilde V^\dagger_u \tilde V_z)^{ab}{\delta \over {\delta
\alpha_\tau^a(u_{\perp})} }\,{\rm tr} (t^bV^\dagger_x  V_y)\nn
&=&g^2\,(\delta_{xv}-\delta_{yv})(\tilde V^\dagger_u)^{ac}
\tilde V_z^{cb}\left[{\rm tr} 
(t^bt^aV^\dagger_x  V_y)\delta_{xu} -{\rm tr} 
(t^at^bV^\dagger_x  V_y)\delta_{yu}\right]\nn
&=&g^2\,(\delta_{xv}-\delta_{yv})\,(\delta_{xu}-\delta_{yu})\,
\tilde V_z^{cb}\,{\rm tr} (t^bV^\dagger_x t^c V_y)\,,\ee
after by now standard manipulations.

A similar calculation shows that
last term $(-\tilde V^\dagger_z \tilde V_v)^{ab}$ in eq.~(\ref{etauv})
gives exactly the same contribution as in eq.~(\ref{diff23}).
By putting together the previous results, it is now 
straightforward to obtain  eq.~(\ref{evolVR}).
In particular, the two integrals over $u_\perp$ 
and $v_\perp$ in eq.~(\ref{evolS0}) are trivially
performed with the help of the $\delta$--functions
 $(\delta_{xv}-\delta_{yv})(\delta_{xu}-\delta_{yu})$, thus generating
the dipole kernel (\ref{decay}).